\documentclass[fleqn,usenatbib,useAMS]{mnras}

\usepackage{graphicx}	
\usepackage{amsmath}	
\usepackage{amssymb}	
\usepackage{multicol}
\usepackage{bm}		
\usepackage{pdflscape}	
\usepackage{tabularx}
\usepackage{subcaption}
\usepackage{xfrac}

\usepackage[T1]{fontenc}
\usepackage{ae,aecompl}

\usepackage{newtxtext,newtxmath}

\title[Membership of open clusters based on GDR3] {The membership of stars, density profile and mass segregation in open clusters using a new machine learning-based method}

\author[M. Noormohammadi]{
M. Noormohammadi,$^{1}$\thanks{E-mail: mnoormohammadi@aut.ac.ir}
M. Khakian Ghomi,$^{1}$\thanks{E-mail: khakian@aut.ac.ir}
and H. Haghi,$^{2}$\thanks{E-mail: haghi@iasbs.ac.ir}
\\
$^{1}$Physics and Energy Engineering Department, Amirkabir University, Tehran, IRAN\\
$^{2}$Department of Physics, Institute for Advanced Studies in Basic Sciences (IASBS), Zanjan, IRAN\\
}

\date{Accepted XXX. Received YYY; in original form ZZZ}

\pubyear{2022}

\begin{document}
\label{firstpage}
\pagerange{\pageref{firstpage}--\pageref{lastpage}}
\maketitle
\color{black}
\begin{abstract}
\noindent
A combination of two unsupervised machine learning algorithms, DBSCAN and GMM are used to find members with a high probability of twelve open clusters, M38, NGC2099, Coma Ber, NGC752, M67,  NGC2243, Alessi01, Bochum04, M34, M35, M41, and M48, based on Gaia DR3. These clusters have different ages, distances, and numbers of members which makes a suitable cover of these parameters situation to analyze this method.
We have identified 752, 1725, 116, 269, 1422, 936, 43, 38, 743, 1114, 783, and 452, probable and possible members with a higher probability than 0.8 for M38, NGC2099, Coma Ber, NGC752, M67, NGC2243, Alessi01, Bochum04, M34, M35, M41, and M48, respectively. Moreover, we obtained the tidal radius, core radius, and clear evidence of mass segregation in ten clusters. From an examination of the high-quality color-magnitude data of the cluster, we obtained one white dwarf for each of NGC752, Coma Ber and M67.
In the young open cluster M38, we found all members inside the tidal radius however in the older clusters we found some members outside of the tidal radius, indicating that the young open clusters had not enough time to form clear tidal tails. It is seen that mass segregation occurs at a higher rate in older clusters than the younger ones.

\end{abstract}


\section{INTRODUCTION} \label{section.intro}

The open clusters (OCs) are considered ideal systems for studying stellar astrophysical properties and the dynamics of stellar systems as their stars lie roughly at the same distance with the same age and chemical abundance, but different masses. In this regard, identifying the member stars is of great importance in determining the physical characteristics of star clusters. Therefore, identifying the member stars of OCs is of great importance in determining their astrophysical parameters (reddening, distance and age), structural parameters (core, cluster and tidal radii), overall masses, mass function (MF), relaxation times, and evolutionary parameters. Moreover, the members of the OCs undergo internal and external perturbations such as stellar evolution, two-body relaxation, mass segregation, and Galactic tidal field. In this regard, the ongoing ESA space-based mission Gaia \citep{perryman2001,gaia2016} is building the largest, most precise three-dimensional map of our Galaxy by surveying nearly two billion objects. Hence the deep observational study of star clusters is receiving a new impetus.\\
\noindent
There are two main approaches to identify cluster membership including the astronomical methods (which are based on the position and motion of the stars) and photometric methods (that derive from the color and brightness of stars). Since cluster stars follow the same evolutionary path, they can also be identified photometrically using color–magnitude diagrams by separating the cluster stars from the background stars using color and brightness information.\\ 
Using Gaia data release with its precise astrometric and photometric parameters, considerable efforts have been expended in the research of cluster stars membership such as: \cite{galah}, \cite{2019A&A}, \cite{2019A&A630A.119M}, \cite{2020A&A633A.146A} and \cite{2019A&A631A57M}.\\
\noindent
To process such an extensive database, the use of neural networks and machine-learning algorithms in different areas of astronomy is mandatory. For example, the machine learning algorithms have been used to detect reliable star cluster members, using the K-means algorithm by \cite{65alaziz} and \cite{Cantat-Gaudin2018}, DBSCAN by \cite{Gao_2014}, \cite{XINHUA2014}, \cite{BHATTACHARYA2017} and \cite{Wilkinson2018}, spectral clustering by \cite{Gao_spectral_clustering}, Random Forest and Spectral clustering by \cite{73}, GMM and Random Forest by \cite{Gao_gmm_rf_peraesepe}, \cite{Gao_m67_gmm_rf}, \cite{Gao_gmm_rf_Pleiades}, KNN and GMM by \cite{ML/mnras/stab118}.\\
The latest release of Gaia data (EDR3, Gaia Collaboration 2021)~\citep{gaiadr3} with improved precision, offers the opportunity to re-visit the OC membership and physical characteristics allow the application of machine-learning methods to search for the new members that would go unnoticed with traditional methods. Here in this paper, we investigate the efficient method by a combination of suitable unsupervised machine learning algorithms DBSCAN and GMM for cluster membership and apply them to twelve OCs. The combination of these two algorithms covers the weakness of each method with the strength of the other one. The paper is organized as follows. In Section \ref{section.Data} we explain the observational data. In Section \ref{section.Method}  the combined method of DBSCAN and GMM is described. This is followed by a presentation of the results in Section \ref{section.Results}. We compared our method with other methods in Section \ref{section.compare}. And finally a discussion of the results and conclusions are presented in Sections \ref{section.discus} and \ref{con.Data}.

\section{DATA} \label{section.Data}

The data used in this paper for membership determination of twelve OCs is Gaia DR3~\citep{gaiadr3}. Gaia DR3 is the first delivery of the third data release, which provides about 1.81 billion objects of which 1.47 billion data with fully astrometric and photometric parameters, observations with an improved precision with respect to Gaia DR2. The high accuracy in parallax and proper motion parameters are excellent features of Gaia data which is essential for star cluster membership investigation. Moreover, the accuracy of the data is limited by the level of magnitude.\\
\noindent
Stars are selected in this work if they have full astronomical data, including the position in $RA$ and $DEC$, parallax and proper motions in $RA$ and $DEC$; and photometrical data, including color index $B-R$ and magnitude in band G for twelve open clusters M38, NGC2099, NGC2243, M67, Coma Ber, NGC752, Alessi01, Bochum04, M34, M35, M41, and M48. These clusters cover a wide range of ages, distances, and numbers of stars. We selected all data around 150 arcmin radius from the cluster center for M38, NGC2099, M67, NGC2243, Alessi01, M34, M35, M41, M48, 80 arcmin for Bochum04, and 400 arcmin for the nearest one, Coma Ber.  In order to see the tidal effect we select 5 degrees around the less populated cluster NGC752.\\
These radii correspond to 50.61 (M38), 68.53 (NGC2099), 9.93 (Coma Ber), 38.14 (NGC752), 37.97 (M67), 159.64 (NGC2243), 30.22 (Alessi01), 30.06 (Bochum04), 21.99 (M34), 39.42 (M35), 32.80 (M41) and 33.56 (M48) pc.\\
\noindent
Two categories of the selected data around each star cluster are eliminated to improve the results. The first one is the negative parallax and the second one are stars with magnitudes fainter than 20. The data for each cluster was normalized by scale function from scikit-learn library(\url{https://scikit-learn.org/stable/modules/generated/sklearn.preprocessing.scale.html}) before machine learning analysis.

\section{METHOD} \label{section.Method}

Our method includes two steps: in the first step, DBSCAN algorithm is used in 3 astronomical dimensions (two proper motions and parallax) to find the best candidate around the cluster center. In the second step, the membership probability for each star is calculated using the GMM algorithm in five dimensions (two positions, two proper motions, and one parallax). In this method, the objects with higher probability are selected as the star cluster members. In follow, the details of the different steps are described.
\subsection{DBSCAN algorithm} \label{section.Method.DBSCAN}

The DBSCAN algorithm which is a density-based clustering algorithm is used in the first step to find candidate members of clusters among the field stars \citep{dbscan}. Two essential parameters are required in the algorithm: (1) Eps which is the value given to the sphere radius that is considered in the sample space, and (2) MinPts which determines how many data points in the sphere should exist to consider the sphere center as a cluster member.\\
\noindent
The algorithm considers a sphere with a radius of Eps to the center of every single data point and counts the number($n$) of data points inside the sphere. If the number is more than MinPts, the sphere center is considered as a core point, and the algorithm goes to the next point. All points in the sphere are considered as the core neighbors. If any of the neighbors in their sphere, have $n$ more than the MinPts, it is considered as another core point. Otherwise, it is considered as a border point. The data that is not the core point and not the border point is considered as noise. All of the core and border points are considered to be members of the cluster by the DBSCAN algorithm.\\
\noindent
Although DBSCAN is a perfect algorithm to work with large data sets in multi-dimension, it needs two free parameters of Eps and MinPts which can cause mistakes regarding the efficiency of finding cluster members.\\
\noindent
Some cluster members are considered field stars when running the algorithm by taking strict constraint values for Eps and MinPts. On the contrary, some field stars are considered cluster members when selecting soft constraint values for the free parameters. This defect is transformed into a strong point in this study by combining this algorithm and another machine learning algorithm (GMM algorithm).
\subsection{GMM algorithm} \label{section.Method.GMM}

GMM is a machine learning algorithm that finds similar data points that shows clear Gaussian distribution and indicates a probable value for each data \citep{gmm2002_Figueiredo}. Since star clusters show precise Gaussian distribution in astrometric parameters, GMM counts as an appropriate algorithm to identify cluster members in star fields. However, as mentioned by \citet{ML/mnras/stab118}, three conditions are vital in using GMM algorithm, and when these conditions are not met, GMM cannot give acceptable results:\\
1. The data in the current study should be set to have enough level of accuracy to separate from each other. \\
2. The number of cluster members among the field stars must be significant.\\
3. The distribution of star clusters must be rather different from field stars.\\
The first condition is met by using Gaia DR3, which has the best accuracy in astrometric data. Using DBSCAN in the first step meets the other two conditions, which have been explained in the following. The GMM applies Gaussian distribution to each data with Equation~\ref{Eq gaussian.Eq}
\begin{equation}\label{Eq gaussian.Eq}
  d(G)=\sum_{j=1}^{k}w_{j}(-2\pi)^{-\frac{n}{2}}|\Sigma_{j}|^{-\frac{1}{2}}exp(-\frac{1}{2}(X-\mu_{j})^{T}\Sigma_{i}^{-1}(X-\mu_{j}))
  \end{equation}
In Equation~\ref{Eq gaussian.Eq} $d(G)$ is the probability distribution, $X=(x_{1},x_{2},...x_{i})$ is the data points, $n$ is the dimension of the data sets, $\mu_{j}$ and $\Sigma_{j}$ are the mean value and a covariant matrix of the Gaussian distribution, $j$ determines the number of clusters and $w_{j}$ is the weight distribution of the $j^{th} $ Gaussian component. In this work $j=2$ (cluster and field).\\
It is obvious that:\\
\begin{equation}
\sum_{j=1}^{2}(w_{j})=1 
\end{equation}
The GMM calculates the probability for every single data by EM procedure \citep{EM_Dempster} that shows every data belongs to which Gaussian distribution.

\subsection{DG Method} \label{section.Method.Our Mthod}

Here, we propose a new method that combines both DBSCAN and GMM to identify cluster members from a large number of sample sources. To show the capability of this method, it is applied to twelve open clusters with a different number of stars, ages, and distances.\\
\noindent
First of all, the data are normalized regarding the position, proper motion, and parallax. In the next step, DBSCAN is applied to the sample source, and the best values for MinPts and Eps are found. It is clear that the cluster members show central points in the proper motion and parallax among the star fields which means, DBSCAN can identify the probable candidate members of the cluster. Eps and MinPts were not considered with strict constraints, because the GMM  cleans the data from field stars contamination in the next step.\\
\noindent
Nevertheless, DBSCAN met two conditions of the GMM algorithm, 2 and 3 during the first step. MinPts and Eps have been selected in such a way that the rate of star clusters among star fields was significant and the distribution of star clusters rather different from star fields for the GMM algorithm. In the third step, GMM was applied in five parameters, two positions($RA$,$DEC$), two proper motions($pmRA$,$pmDEC$), and parallax to stars that DBSCAN determined. Although without GMM, the existence of two values for DBSCAN causes the problem that has been mentioned before, here they are very useful for the GMM condition, and the best data sample to be provided for GMM.\\
\noindent

\section{RESULTS} \label{section.Results}

\begin{figure}
\centering
\captionsetup[subfigure]{labelformat=empty}
\begin{subfigure}{0.44\textwidth}
        \centering
           \includegraphics[width=\textwidth]{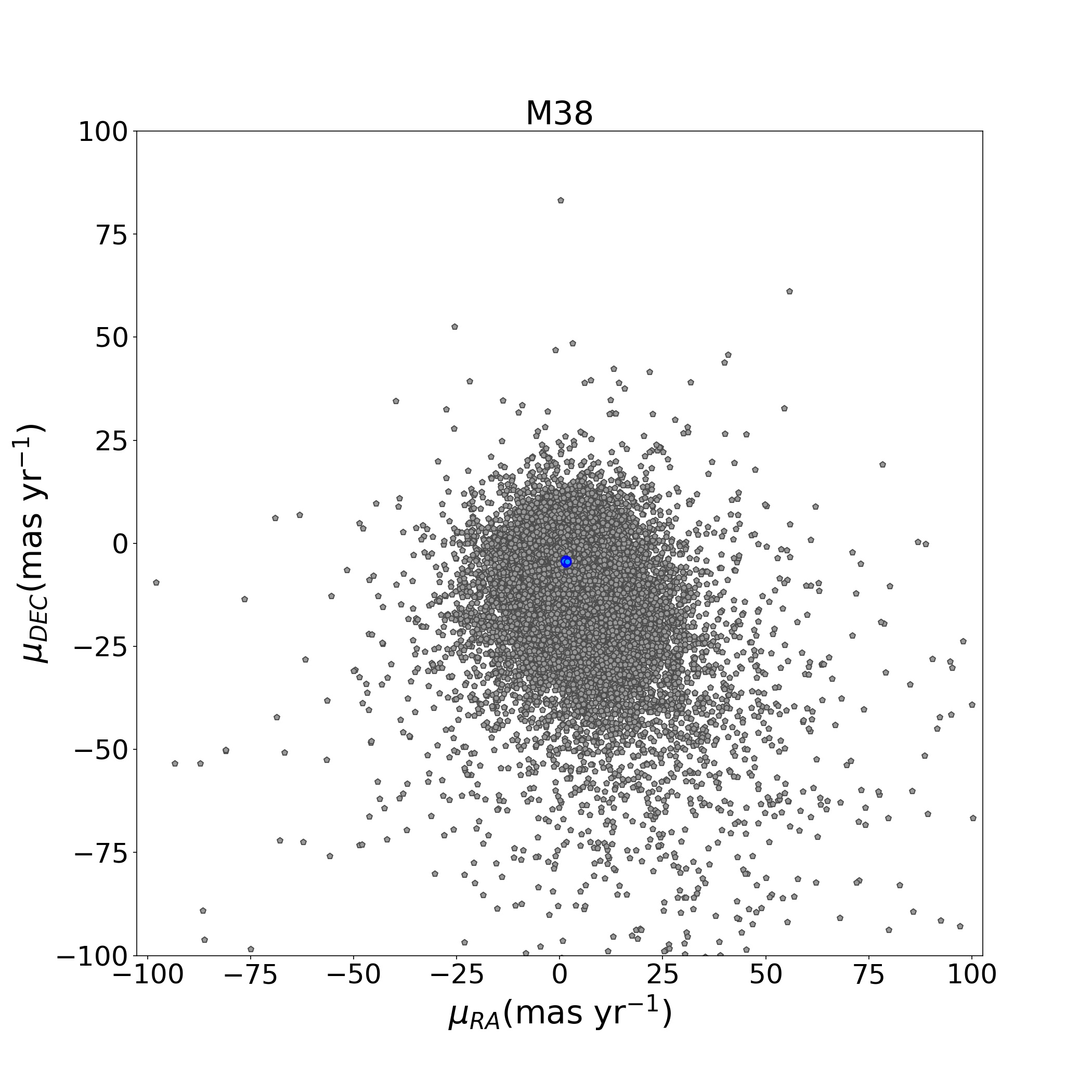}

        \end{subfigure}
        \begin{subfigure}{0.44\textwidth}

                \centering
                \includegraphics[width=\textwidth]{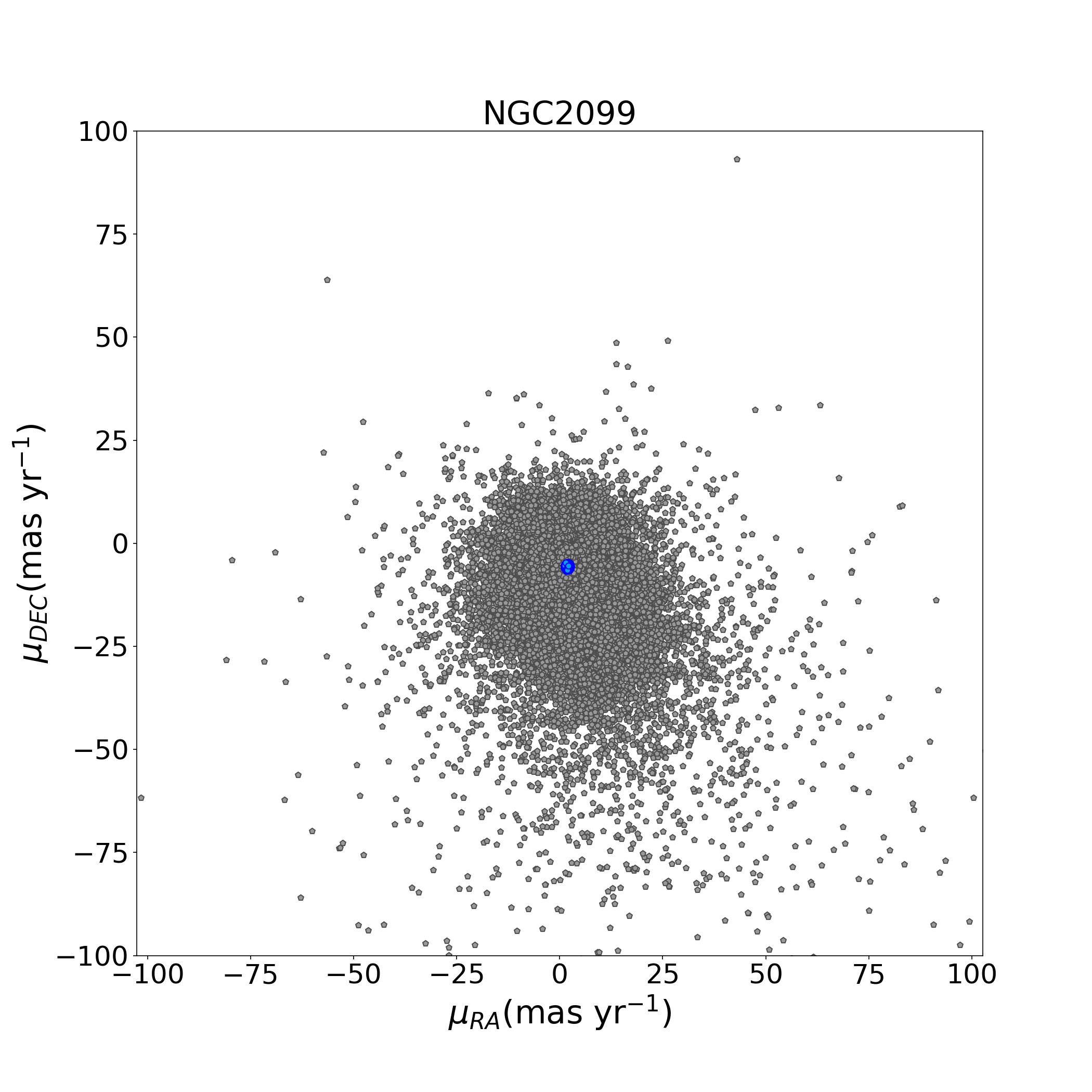}

        \end{subfigure}
        \begin{subfigure}{0.44\textwidth}
                \centering
           \includegraphics[width=\textwidth]{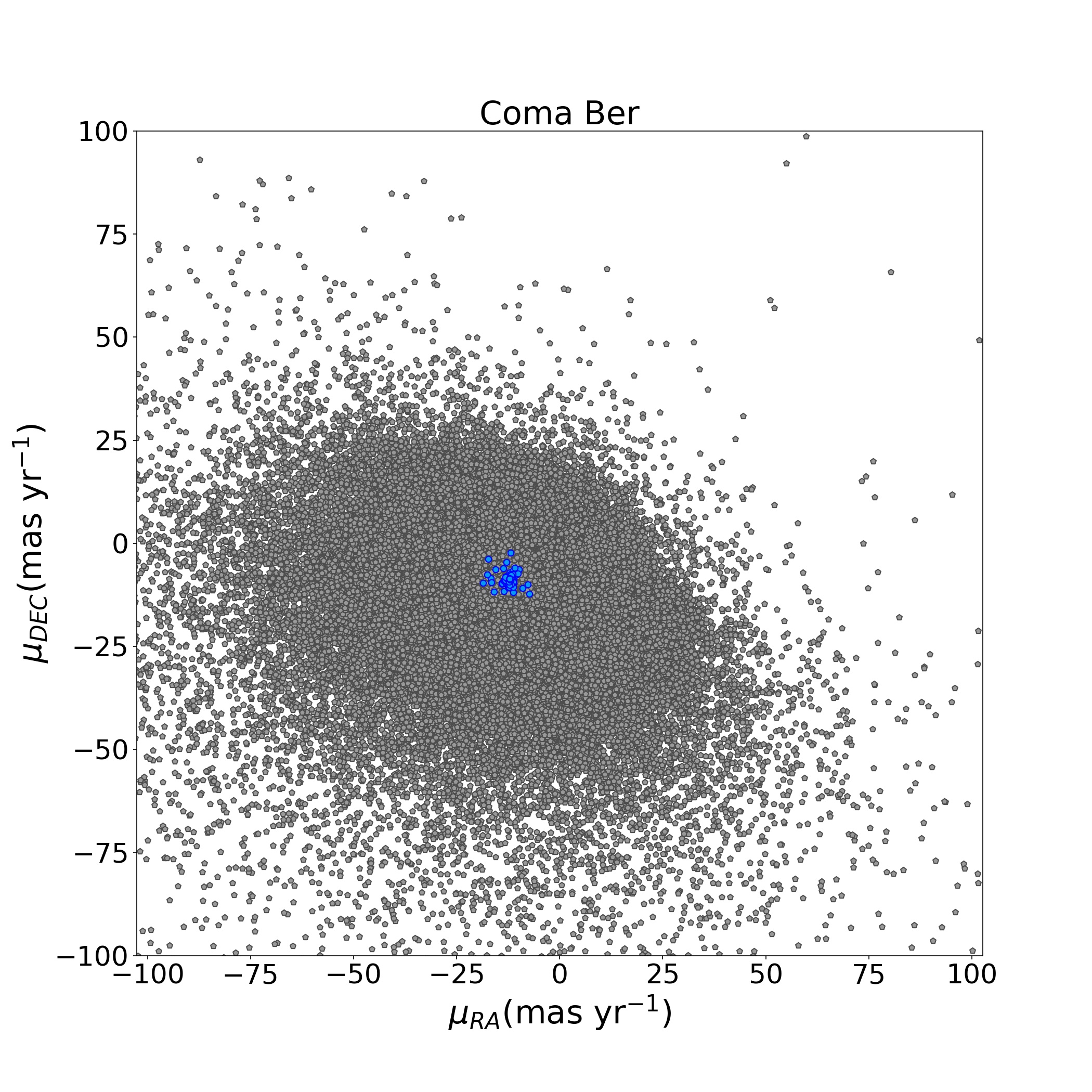}

        \end{subfigure}
        \begin{subfigure}{0.44\textwidth}
                \centering

                \includegraphics[width=\textwidth]{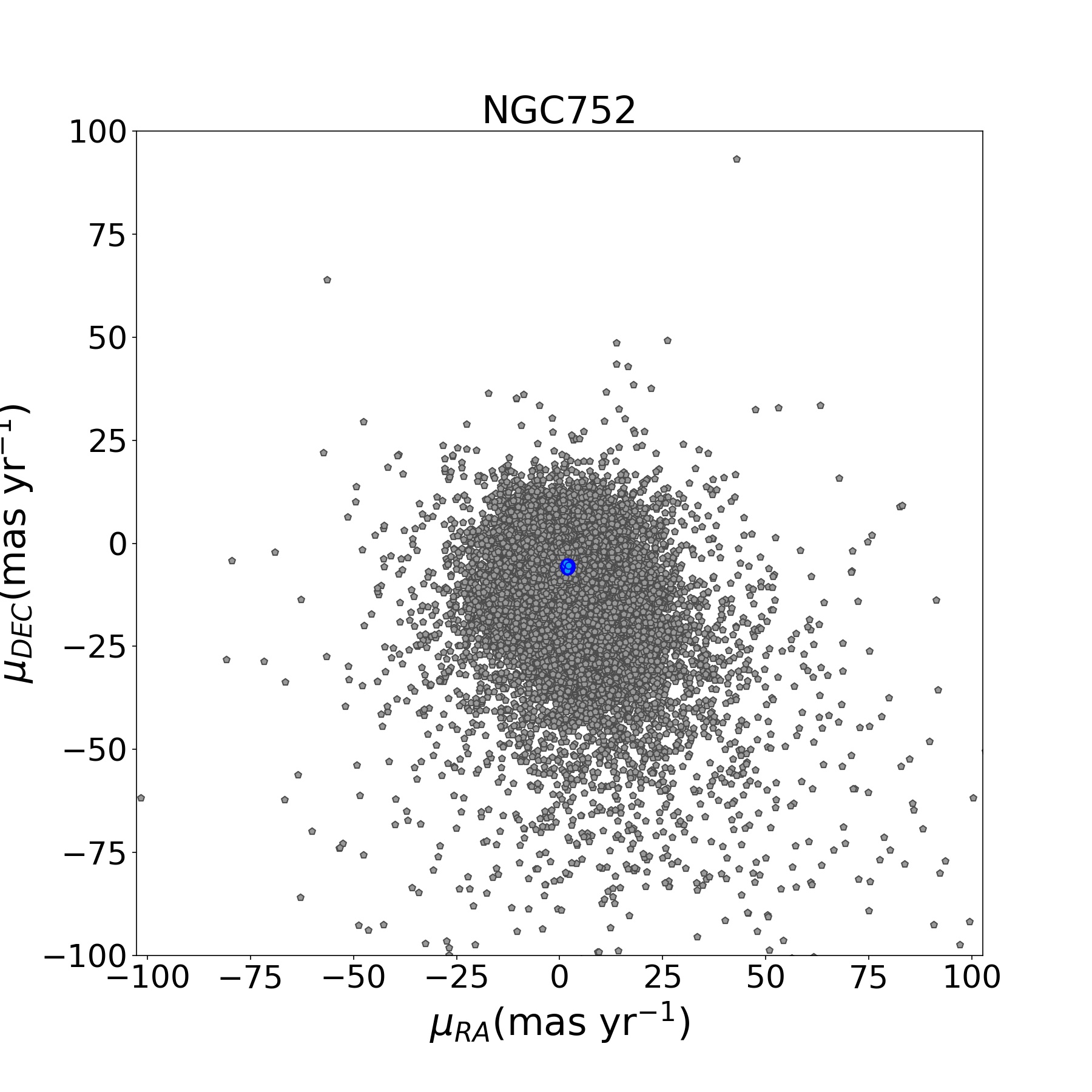}

        \end{subfigure}
        \begin{subfigure}{0.44\textwidth}
                \centering

                \includegraphics[width=\textwidth]{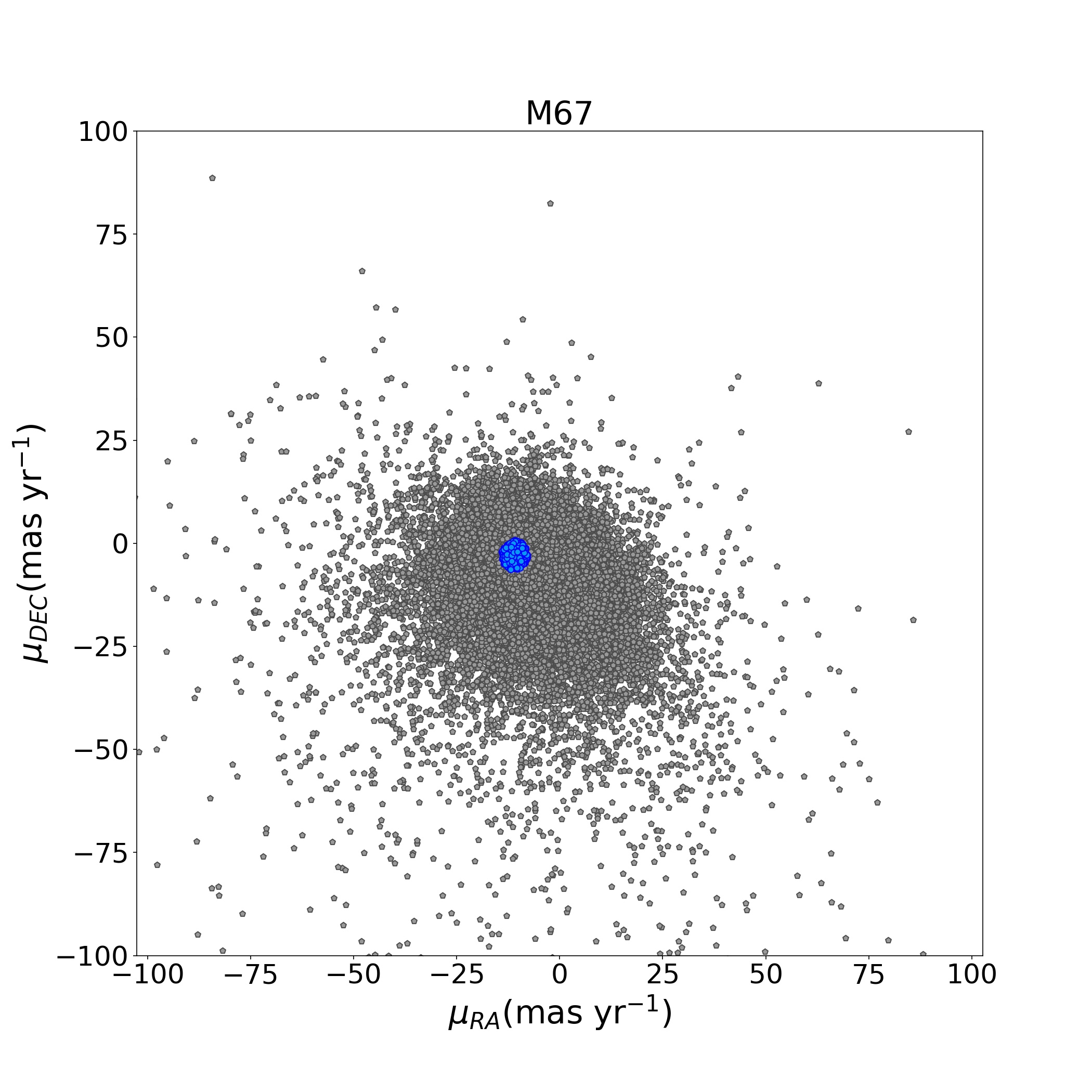}

        \end{subfigure}
        \begin{subfigure}{0.44\textwidth}
                \centering

                \includegraphics[width=\textwidth]{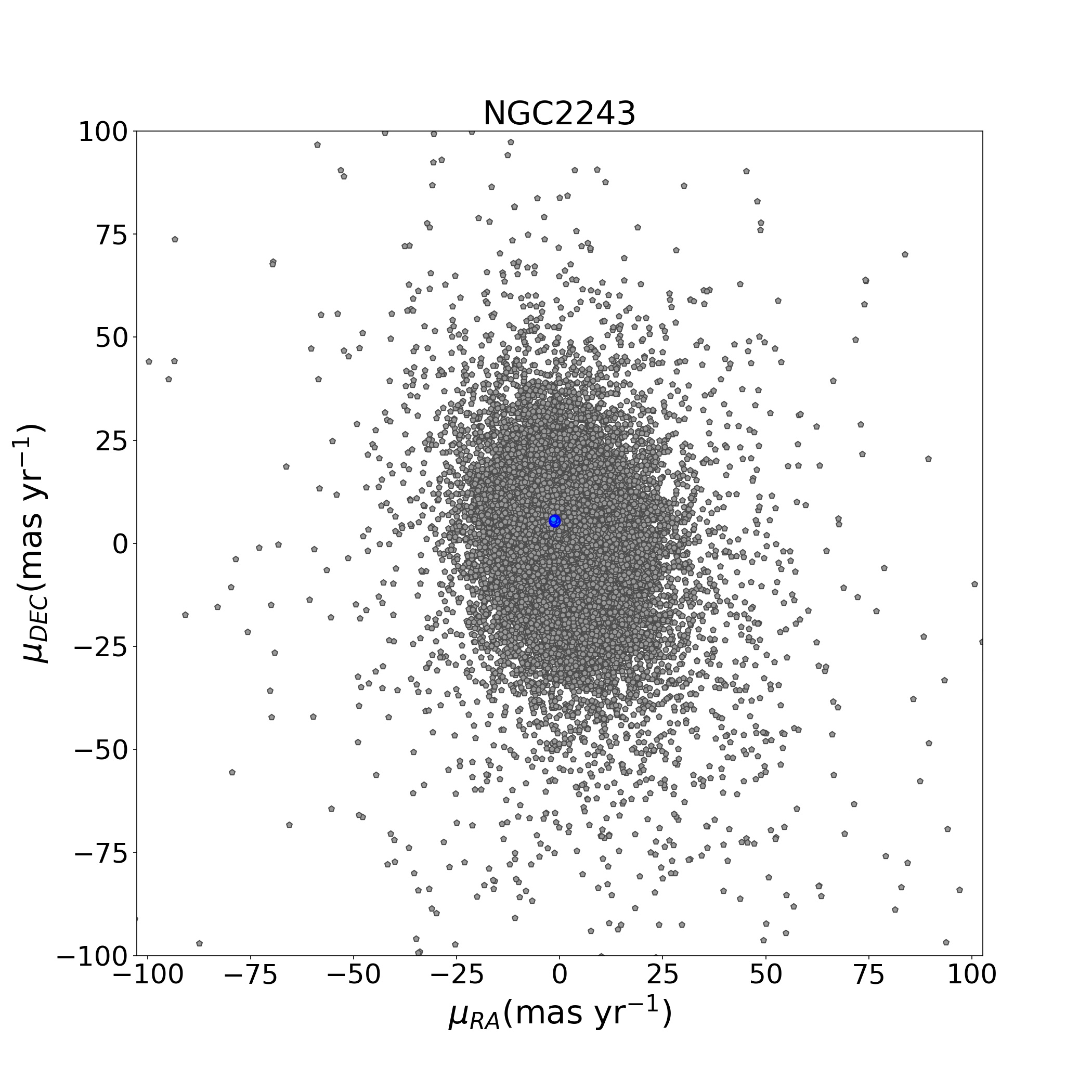}

        \end{subfigure}

  \caption{Proper motion of the cluster member star candidates from the field stars by DBSCAN, vertical axis is $pmRA$ and the horizontal axis is $pmDEC$, grey dots show the field stars and blue dots show stars that were selected by DBSCAN}
  \label{proper motion of dbscan.fig}
\end{figure}

\begin{figure}
  \centering
  \captionsetup[subfigure]{labelformat=empty}
        \begin{subfigure}{0.43\textwidth}
        \centering

                \includegraphics[width=\textwidth]{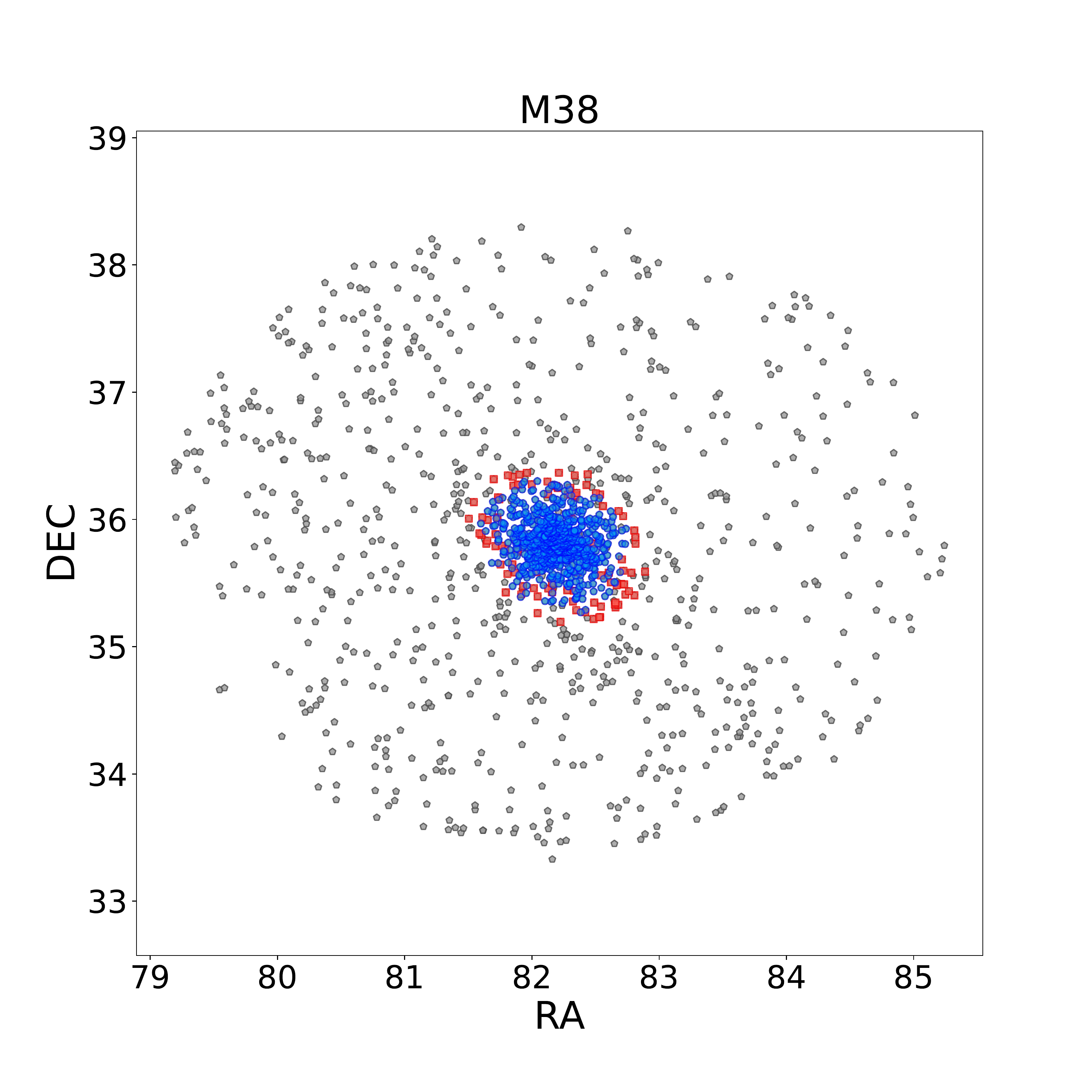}
        \end{subfigure}
        \begin{subfigure}{0.43\textwidth}

                \centering
                \includegraphics[width=\textwidth]{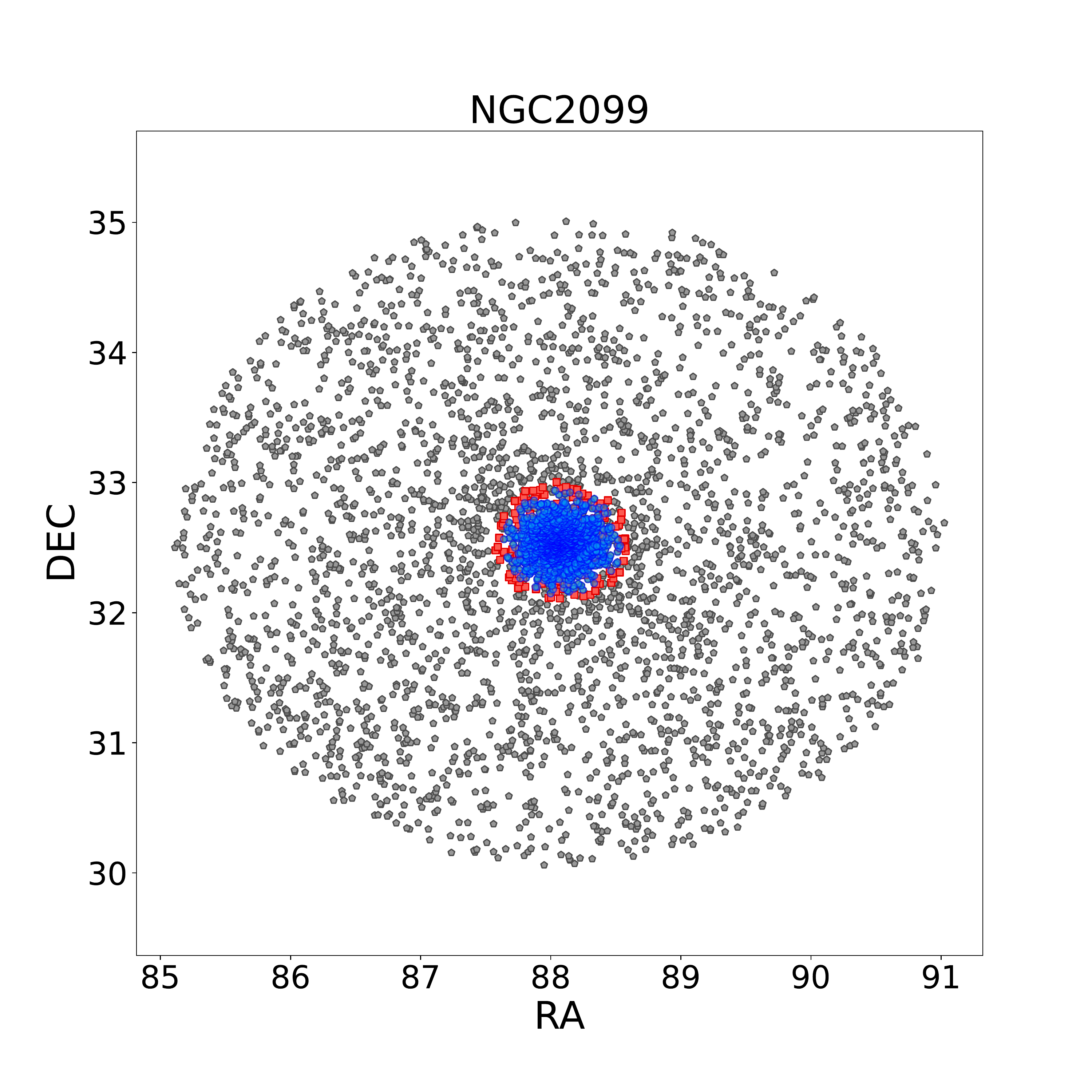}

        \end{subfigure}

  \begin{subfigure}{0.43\textwidth}
                \centering

                \includegraphics[width=\textwidth]{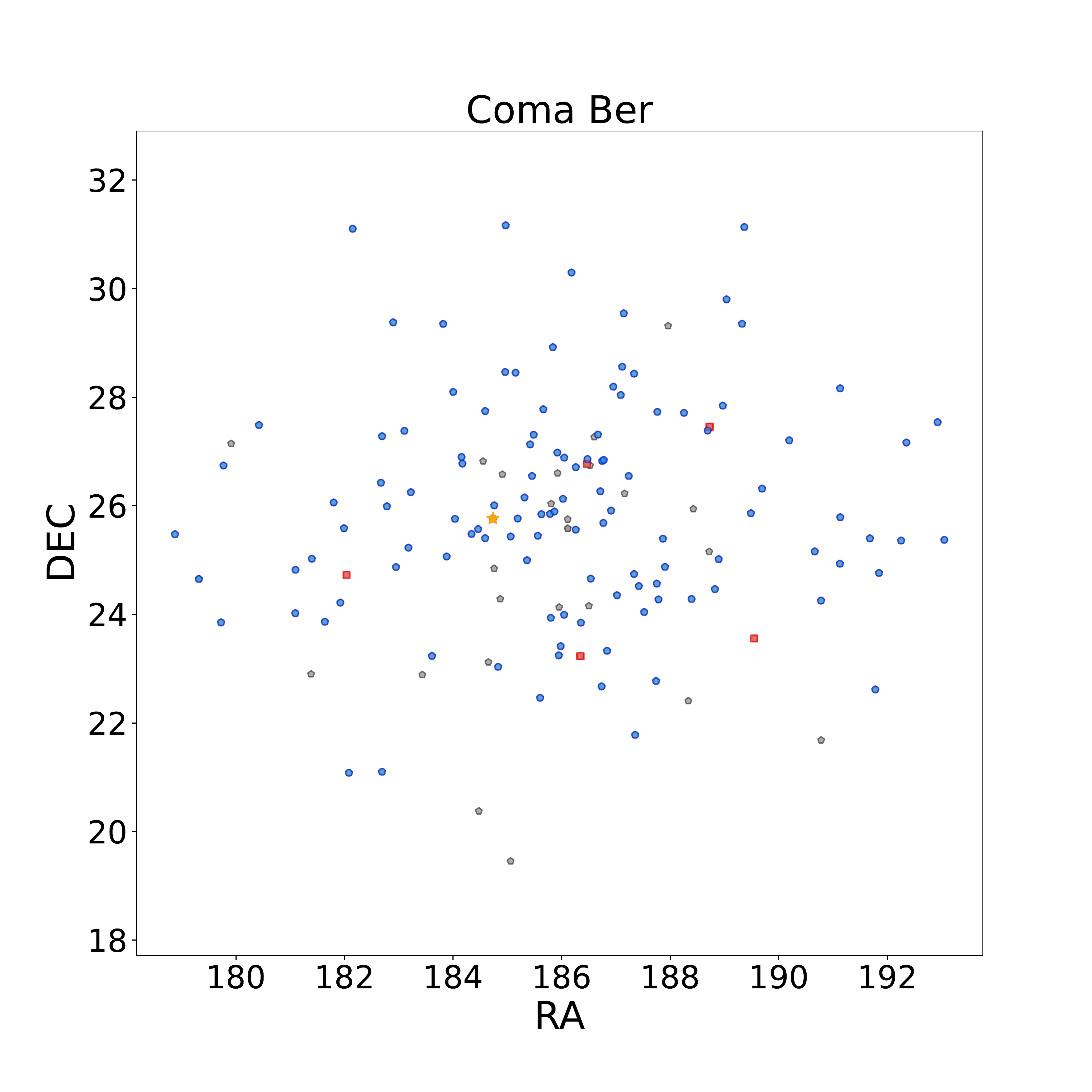}
        \end{subfigure}
        \begin{subfigure}{0.43\textwidth}
                \centering

                \includegraphics[width=\textwidth]{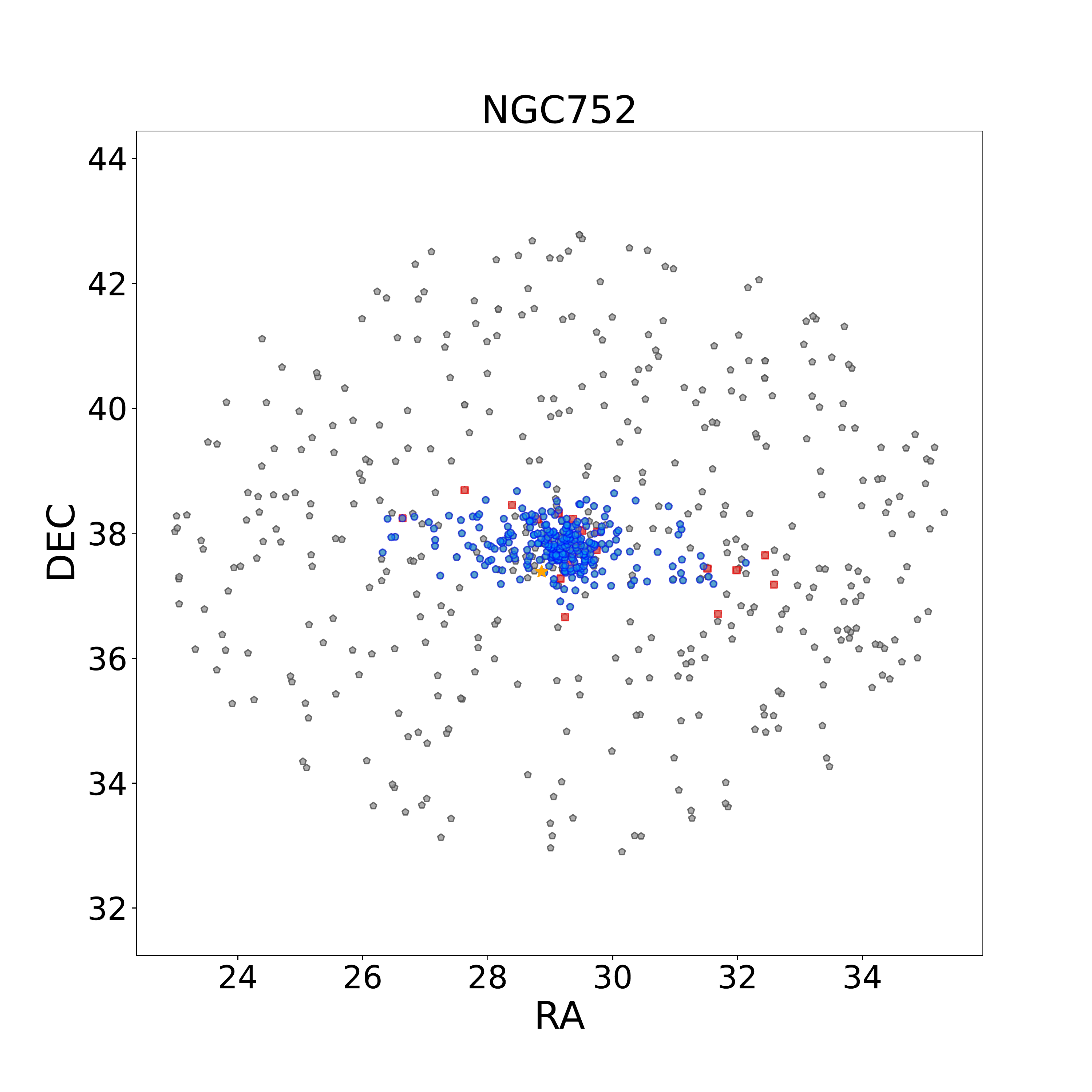}

        \end{subfigure}
        \begin{subfigure}{0.43\textwidth}
                \centering

                \includegraphics[width=\textwidth]{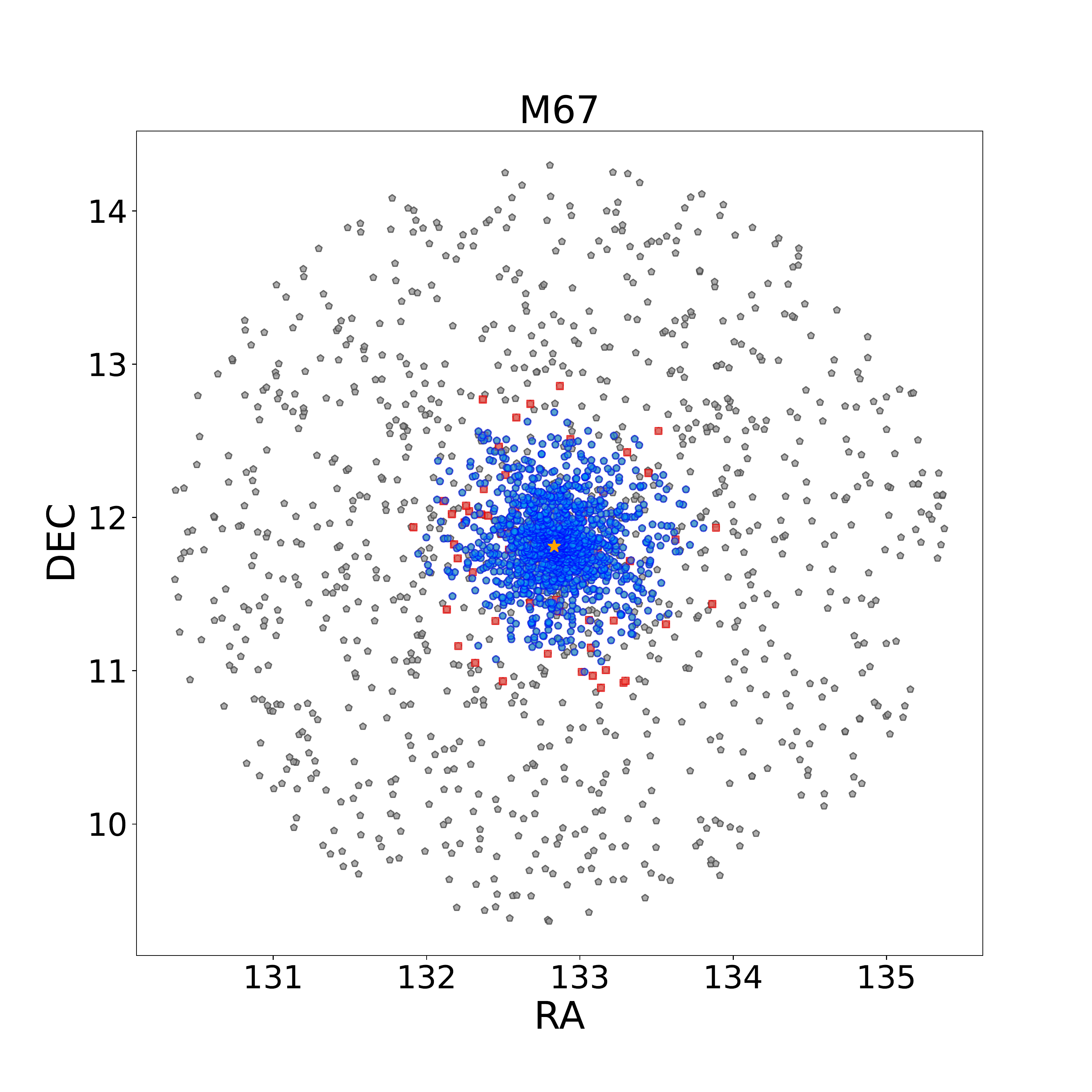}

        \end{subfigure}
        \begin{subfigure}{0.43\textwidth}
                \centering

                \includegraphics[width=\textwidth]{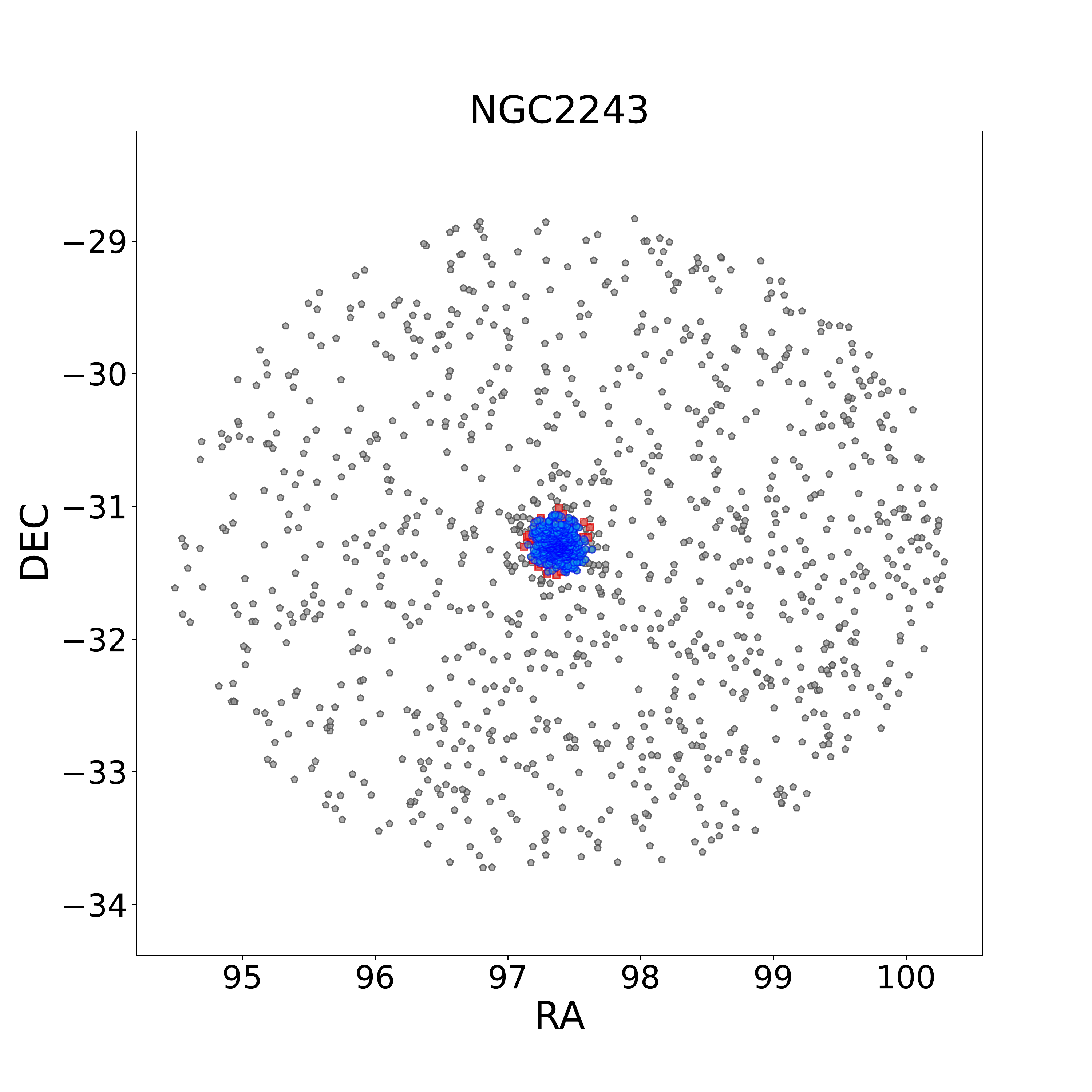}

        \end{subfigure}
  \caption{Position of stars selected by DBSCAN and GMM, grey dots show selected stars by DBSCAN but are not selected by GMM algorithm. Red dots show stars that were selected by GMM algorithm and have a probability of membership between $>0.5$ and $<0.8$ and blue dots show stars that were selected by GMM algorithm and have a probability $>0.8$. White dwarfs are shown by orange stars.}
  \label{position of dbscan and GMM.fig}
\end{figure}

\begin{figure}
  \centering
  \captionsetup[subfigure]{labelformat=empty}
        \begin{subfigure}{0.43\textwidth}
        \centering

                \includegraphics[width=\textwidth]{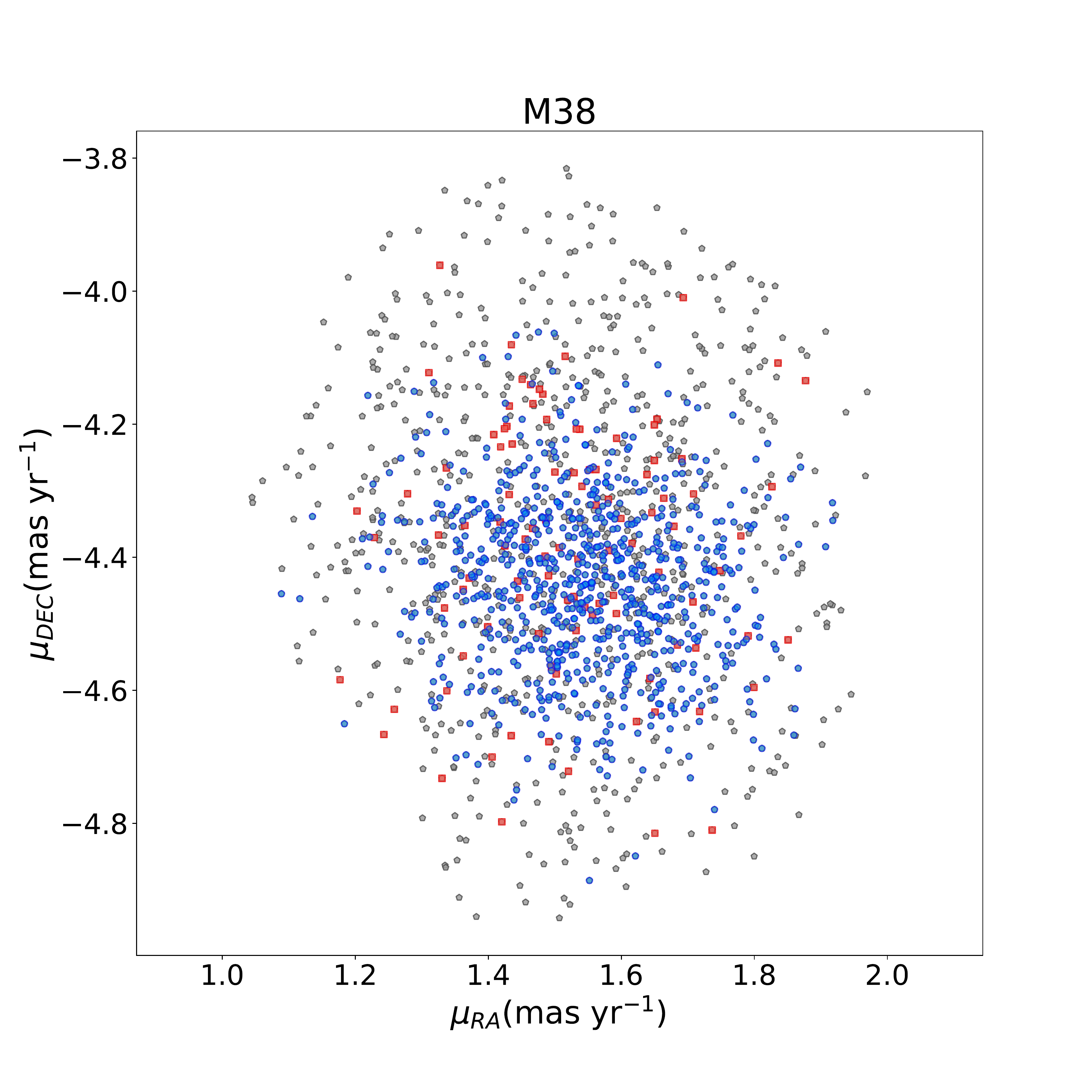}
        \end{subfigure}
        \begin{subfigure}{0.43\textwidth}

                \centering
                \includegraphics[width=\textwidth]{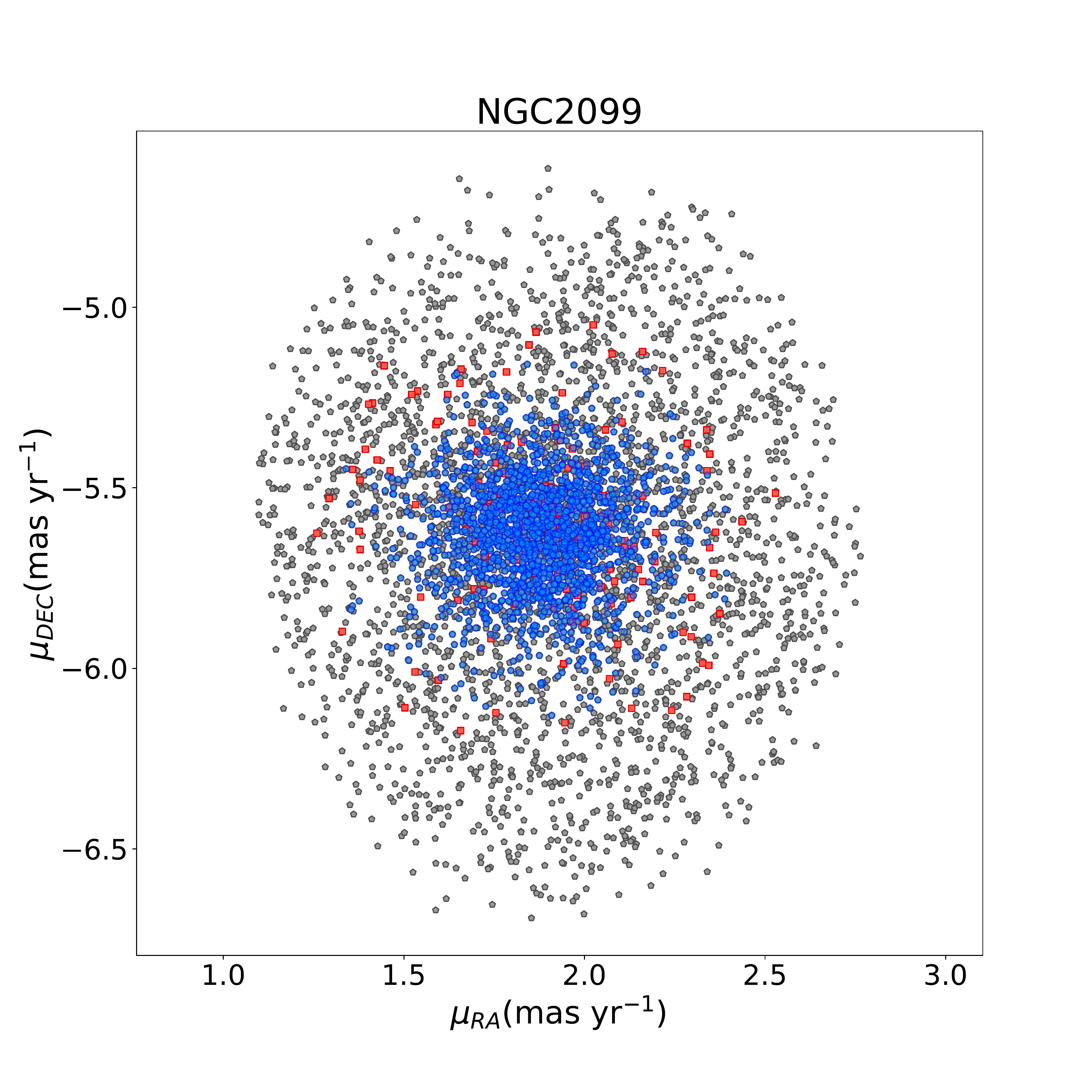}

        \end{subfigure}
  \begin{subfigure}{0.43\textwidth}
                \centering

                \includegraphics[width=\textwidth]{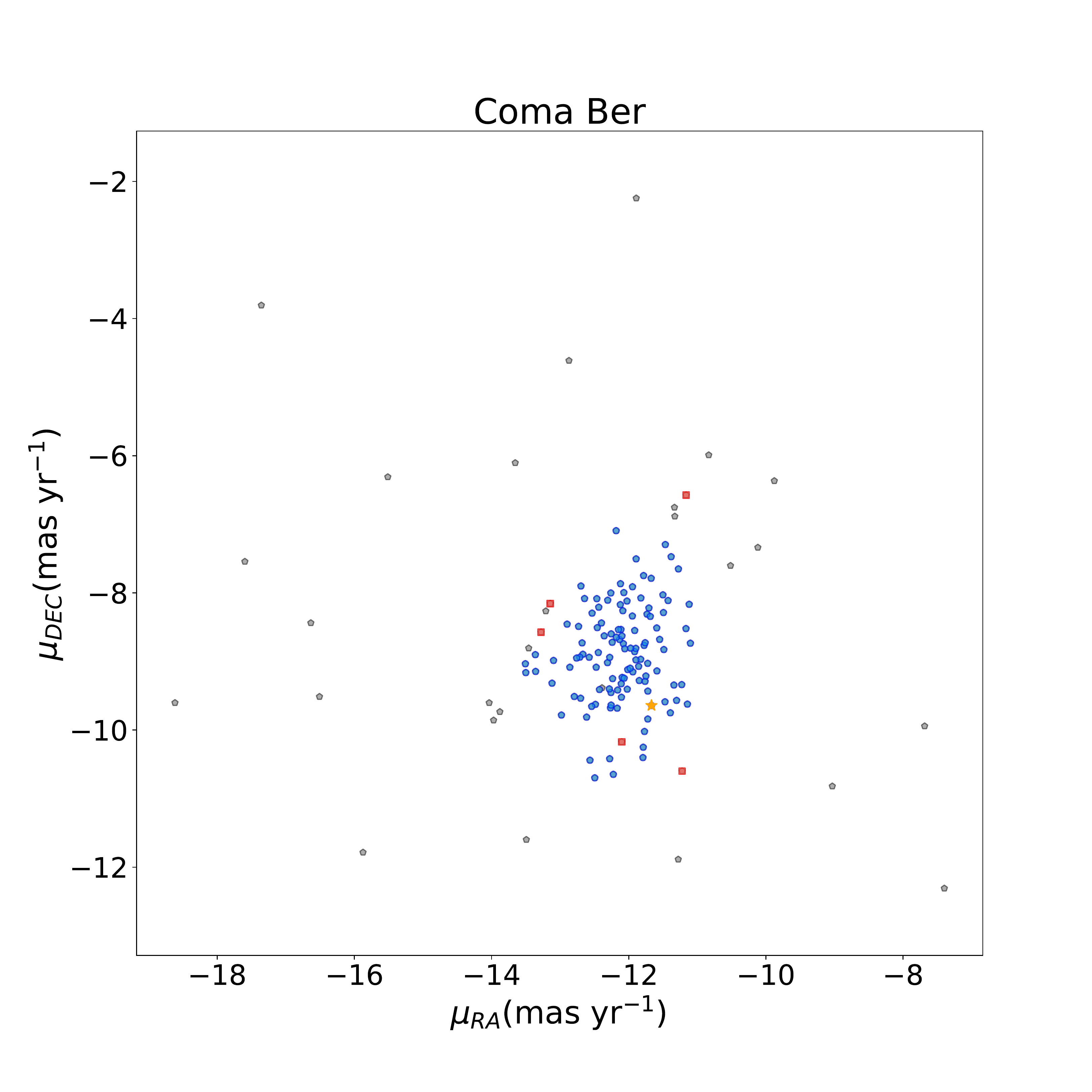}
        \end{subfigure}
        \begin{subfigure}{0.43\textwidth}
                \centering

                \includegraphics[width=\textwidth]{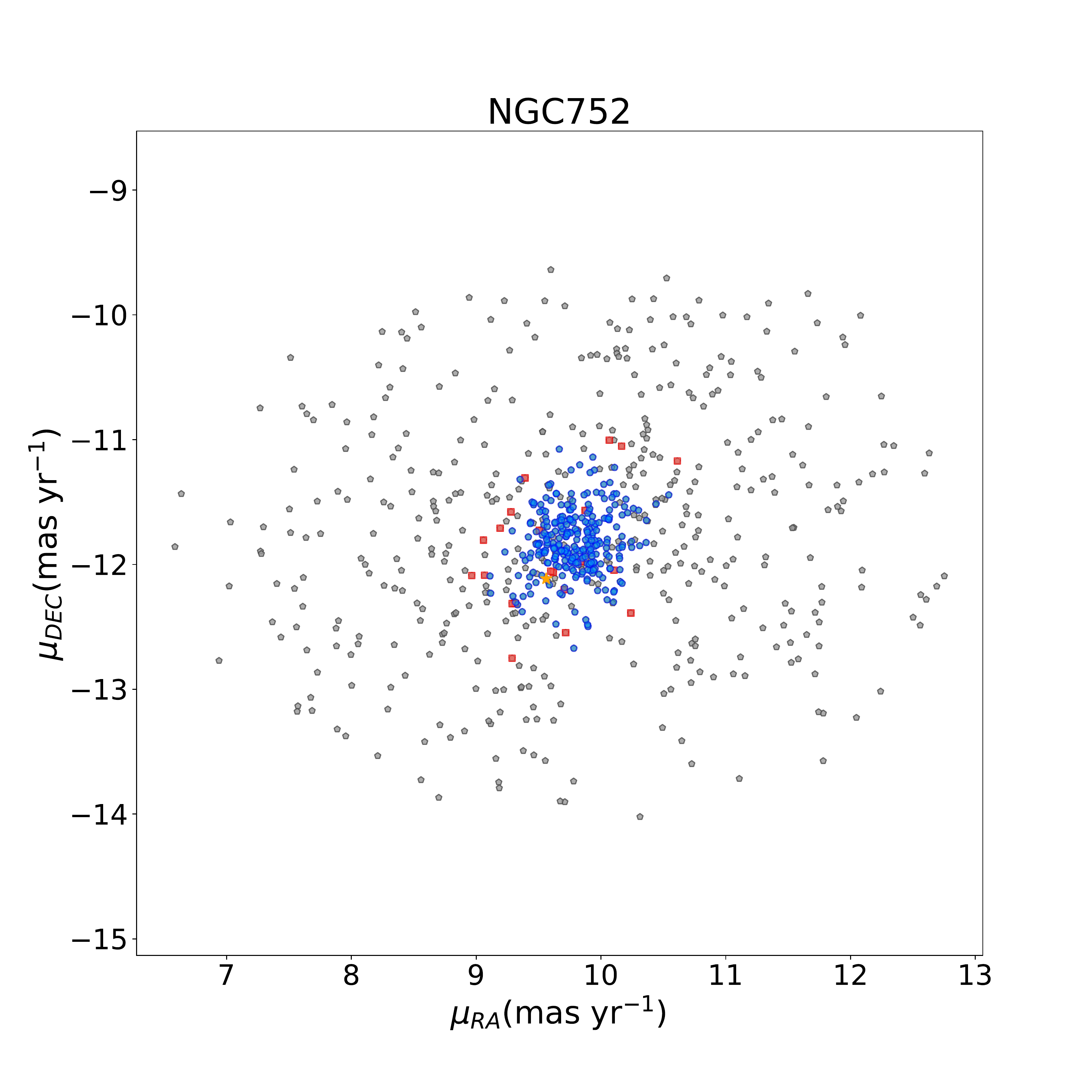}

        \end{subfigure}
        \begin{subfigure}{0.43\textwidth}
                \centering

                \includegraphics[width=\textwidth]{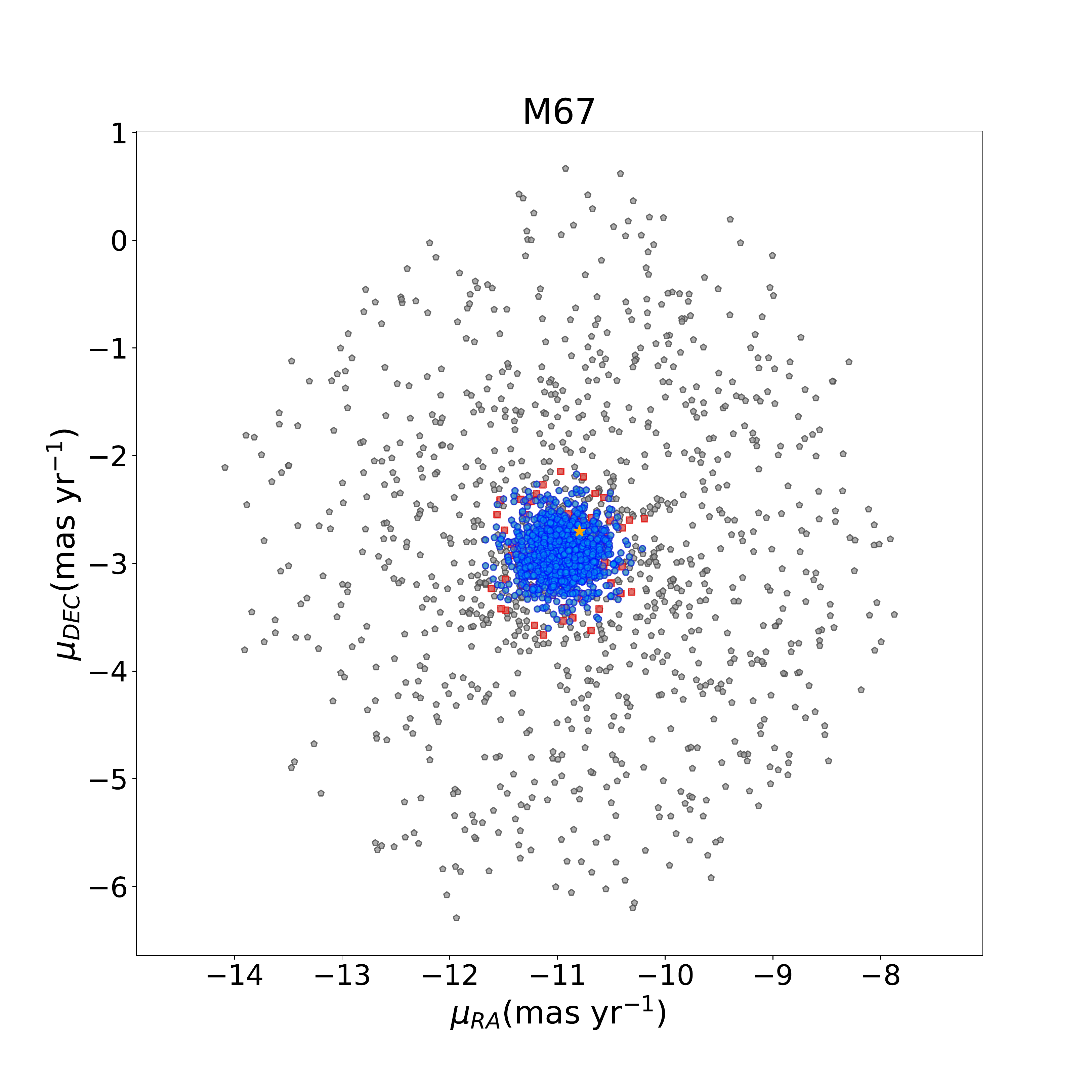}

        \end{subfigure}
        \begin{subfigure}{0.43\textwidth}
                \centering

                \includegraphics[width=\textwidth]{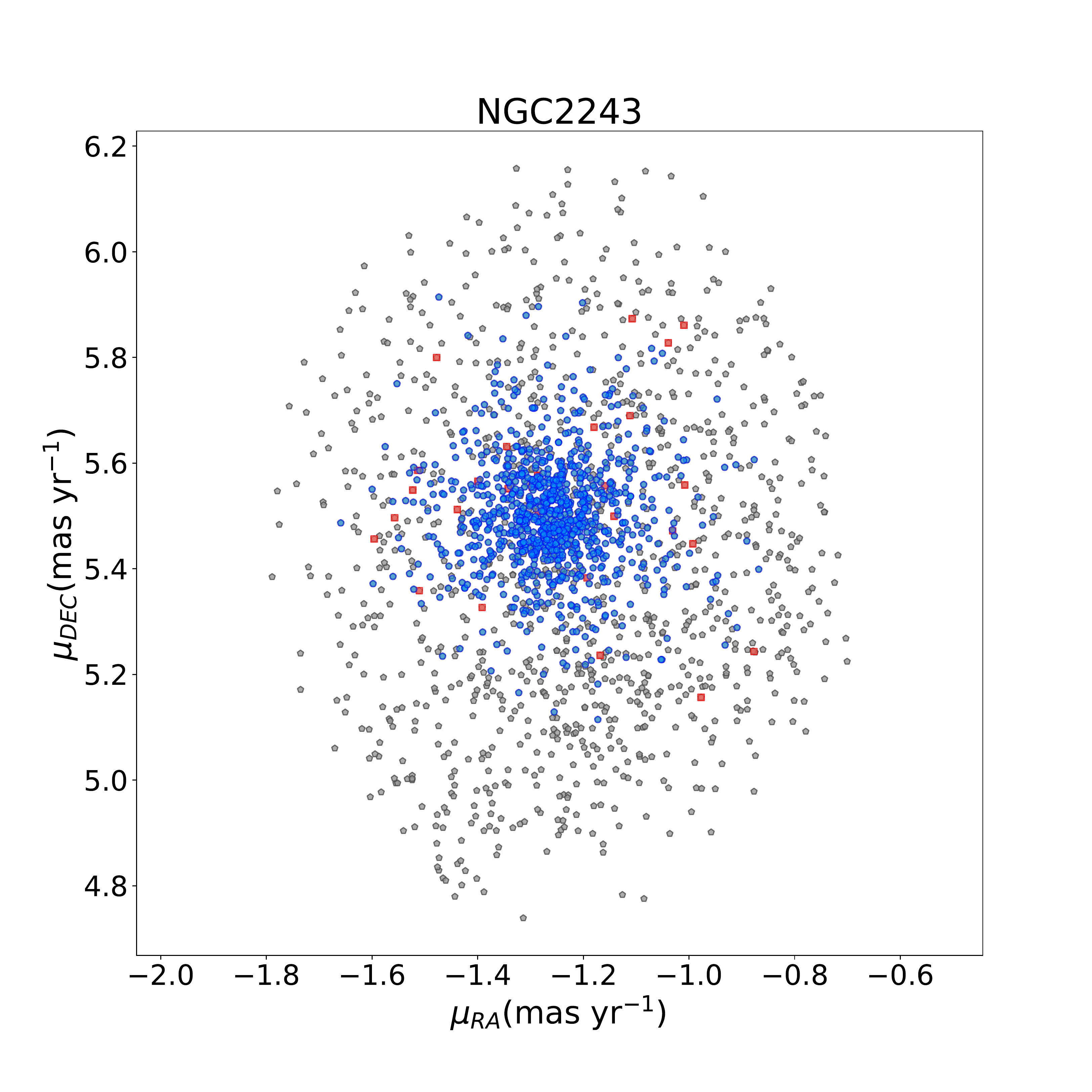}

        \end{subfigure}
  \caption{Proper motion of stars selected by DBSCAN and GMM, grey dots show selected stars by DBSCAN but are not selected by GMM algorithm. Red dots show stars that were selected by GMM algorithm and have probability of membership between $>0.5$ and $<0.8$ and blue dots show show stars that were selected by GMM algorithm and have probability $>0.8$. White dwarfs are shown by orange stars.}
  \label{proper motion of dbscan and GMM.fig}
\end{figure}

\begin{figure}
  \centering
  \captionsetup[subfigure]{labelformat=empty}
        \begin{subfigure}{0.43\textwidth}
        \centering

                \includegraphics[width=\textwidth]{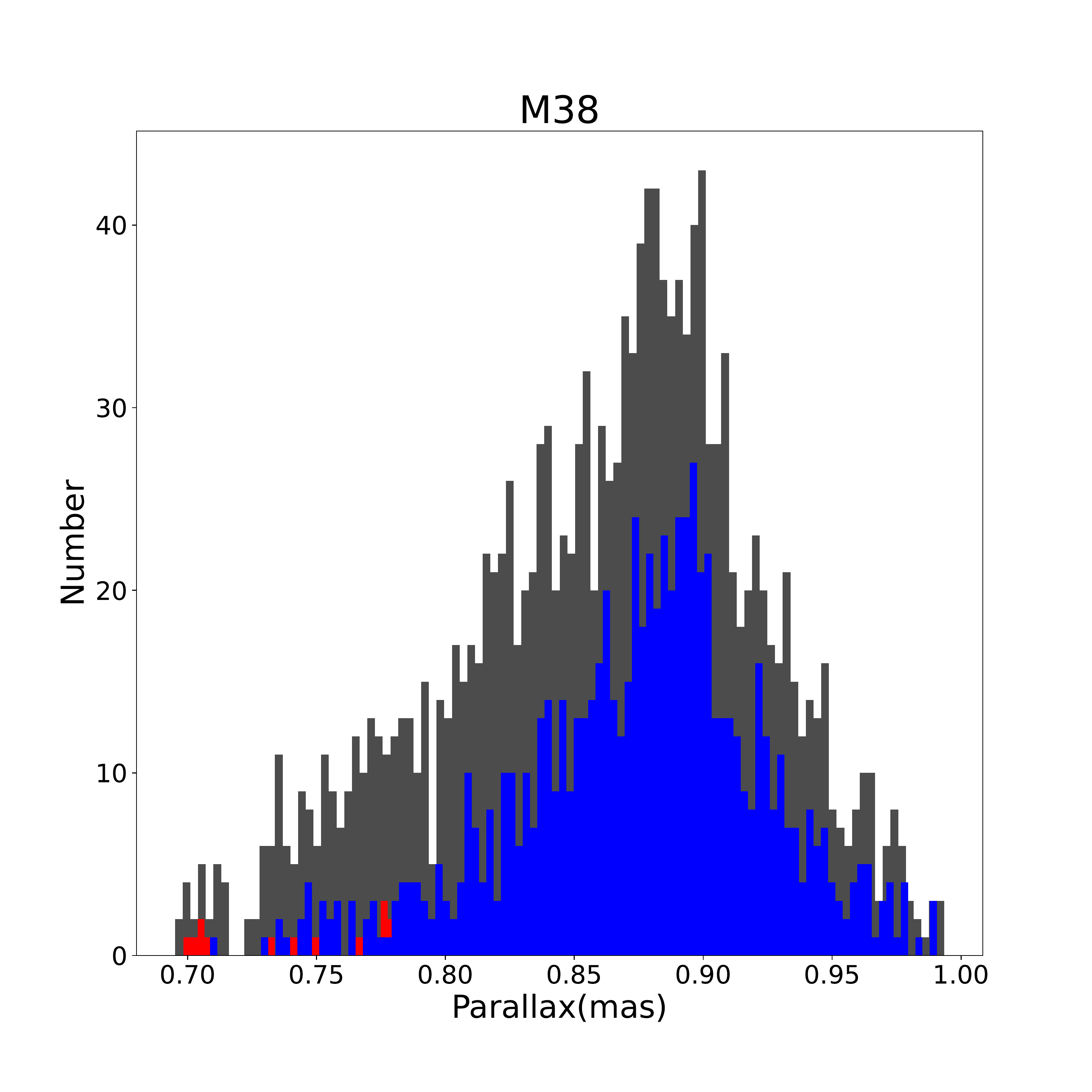}
        \end{subfigure}
        \begin{subfigure}{0.43\textwidth}

                \centering
                \includegraphics[width=\textwidth]{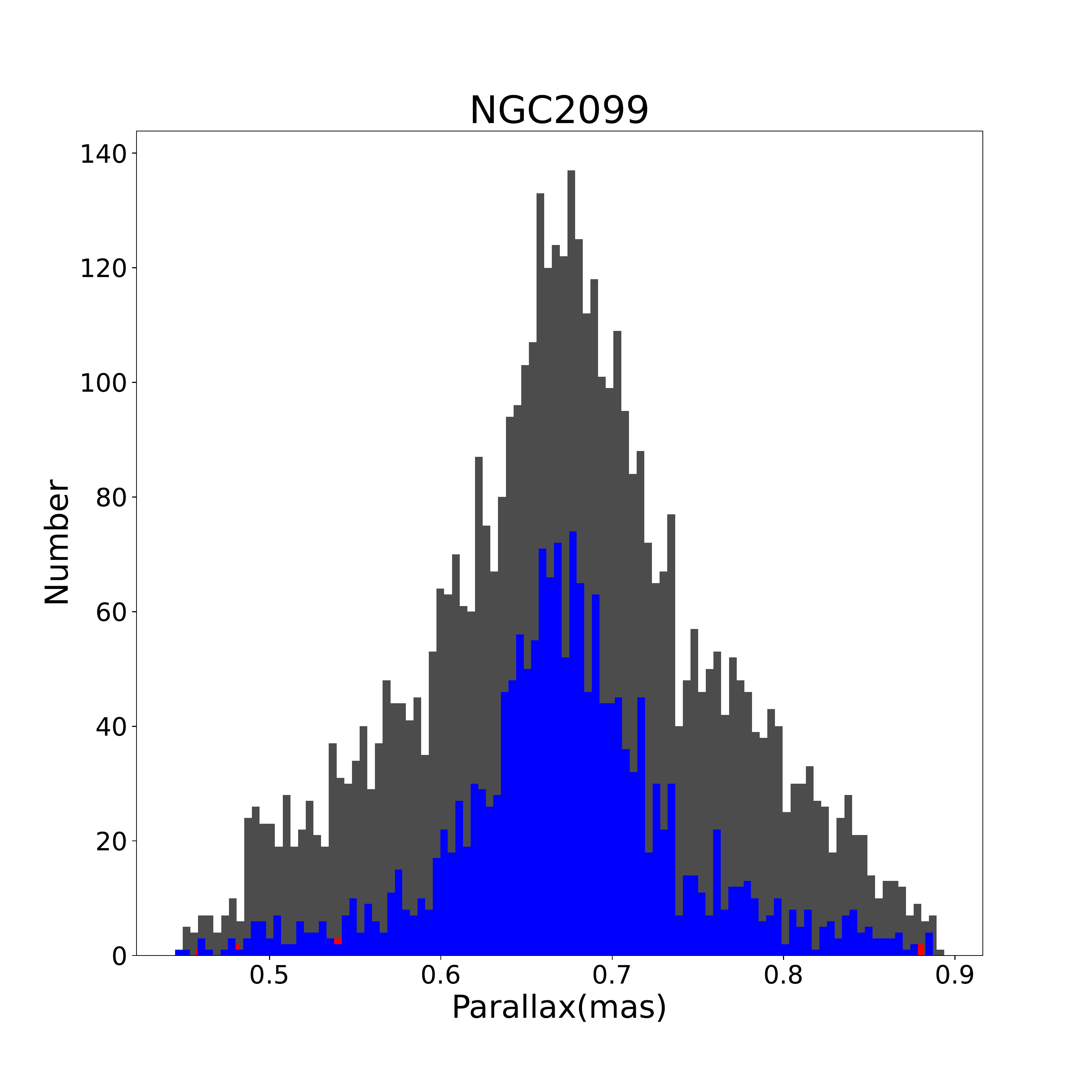}

        \end{subfigure}
        \begin{subfigure}{0.43\textwidth}
                \centering

                \includegraphics[width=\textwidth]{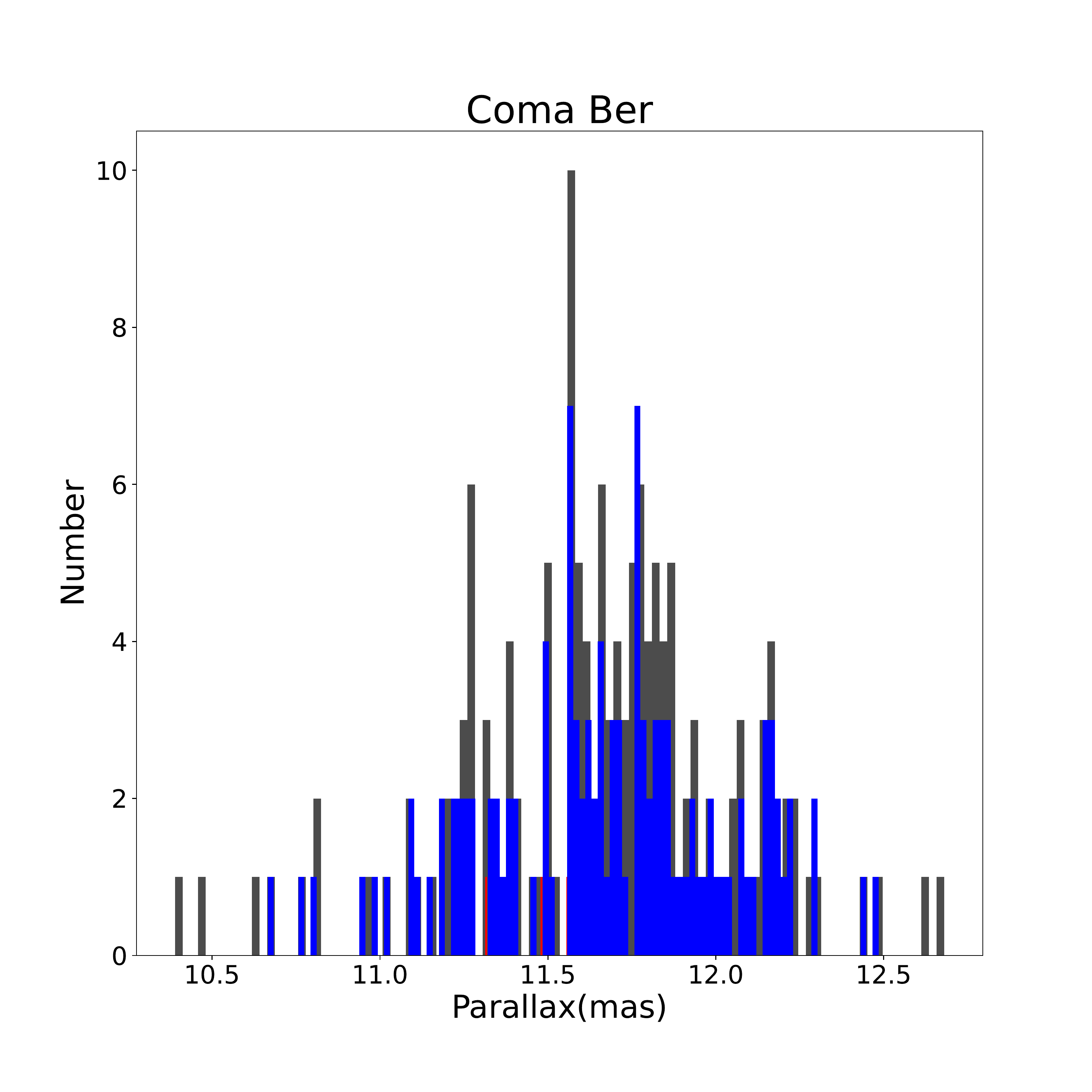}
        \end{subfigure}
        \begin{subfigure}{0.43\textwidth}
                \centering

                \includegraphics[width=\textwidth]{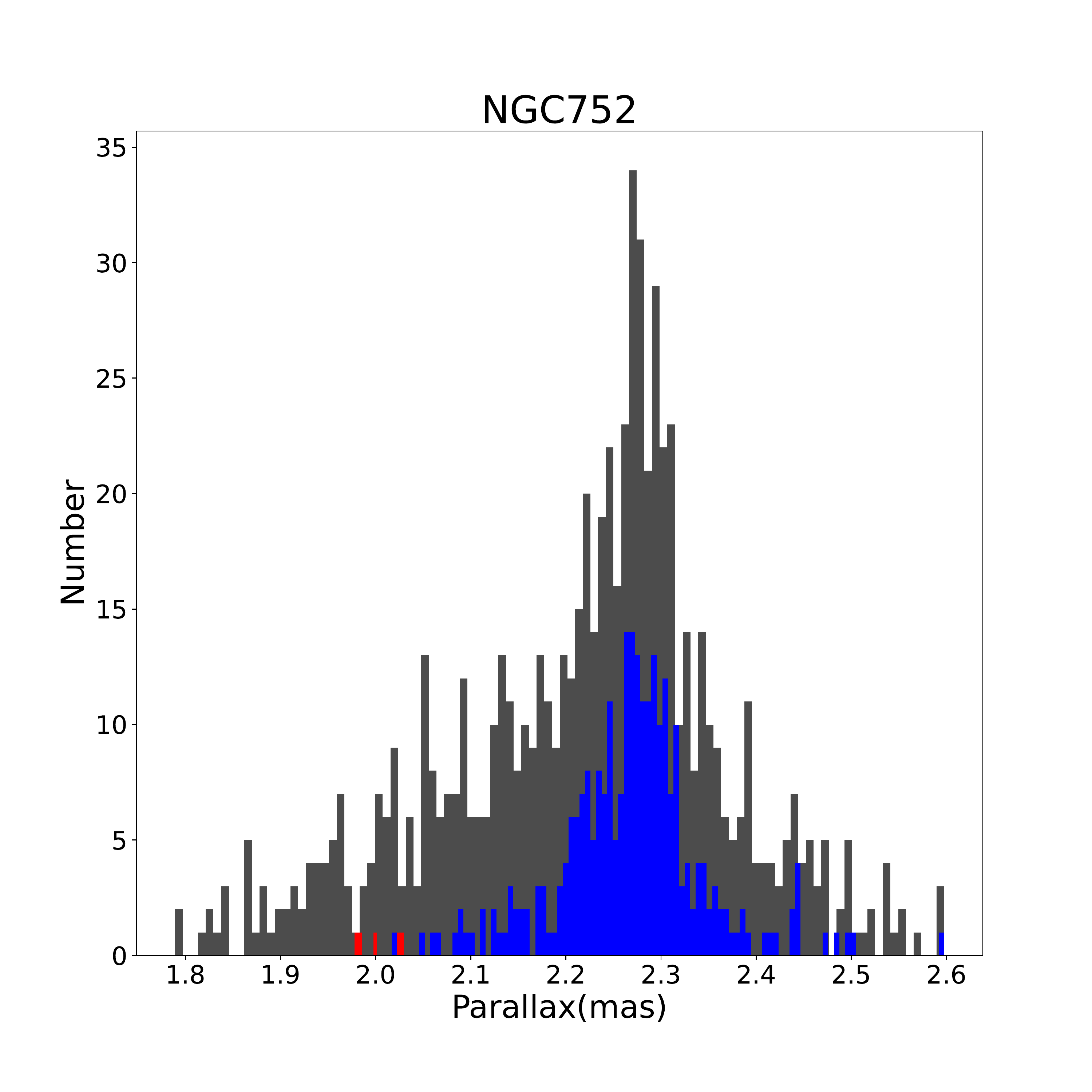}

        \end{subfigure}
        \begin{subfigure}{0.43\textwidth}
                \centering

                \includegraphics[width=\textwidth]{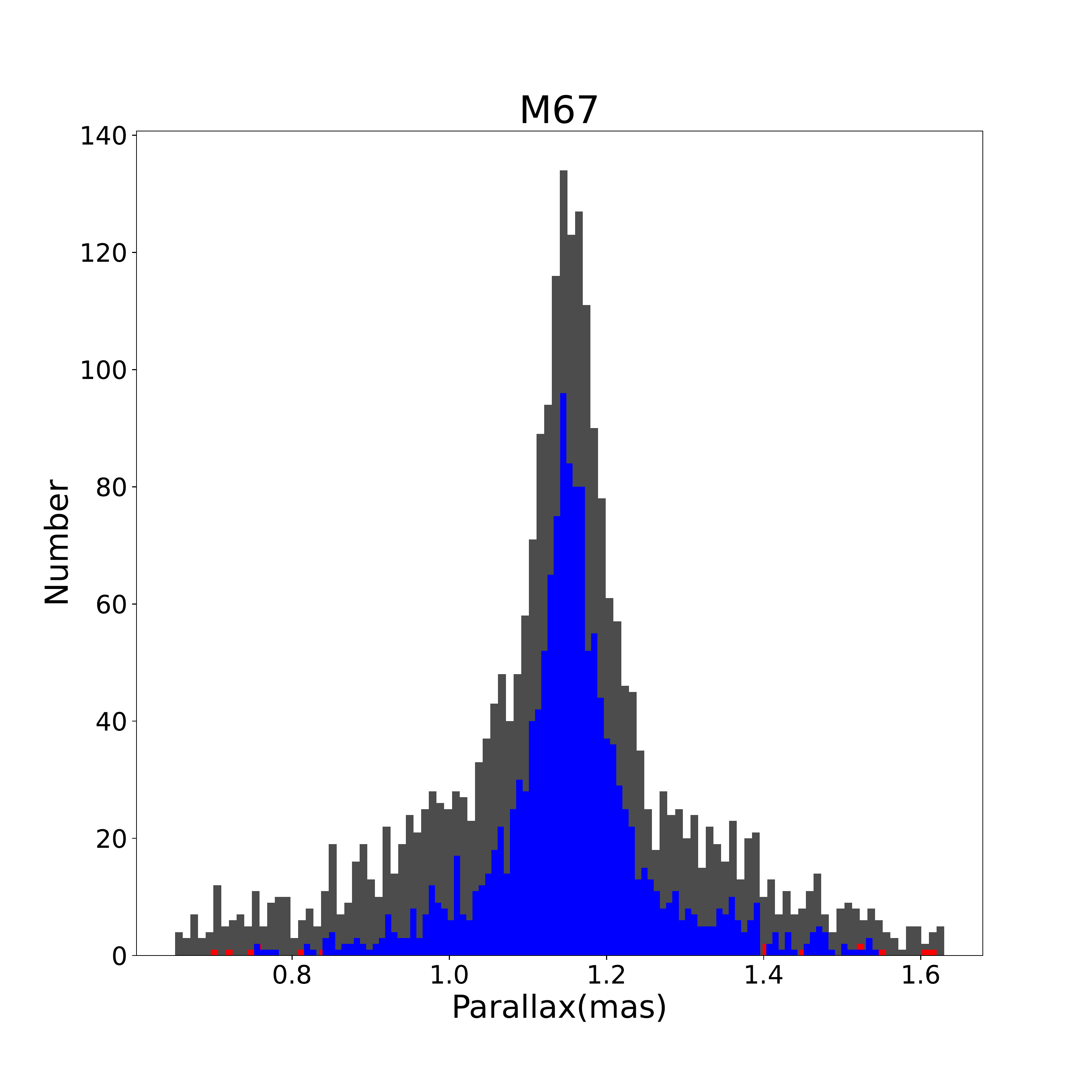}

        \end{subfigure}
        \begin{subfigure}{0.43\textwidth}
                \centering

                \includegraphics[width=\textwidth]{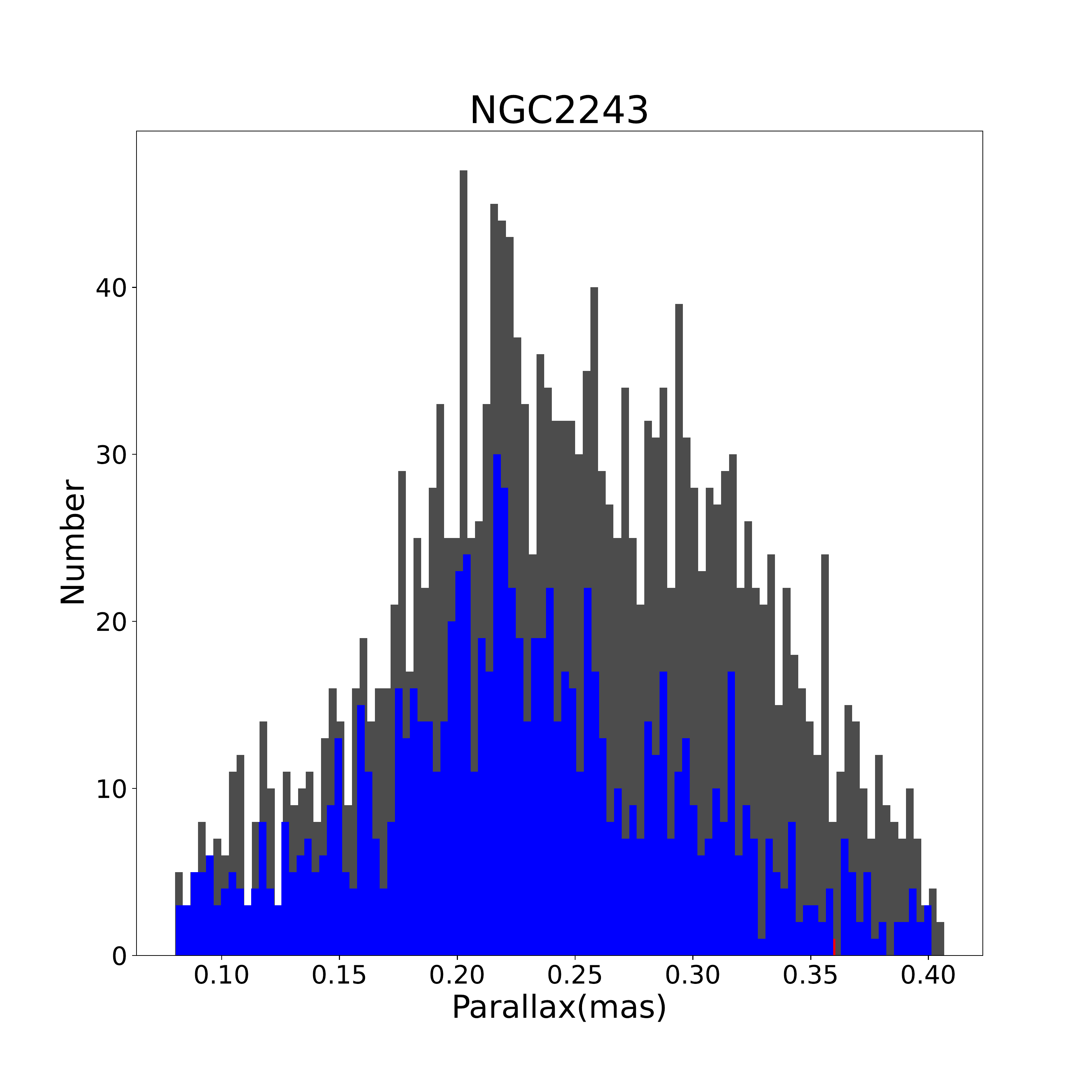}

        \end{subfigure}
  \caption{Parallax of cluster member star candidates by DBSCAN and GMM. Grey lines show selected stars by DBSCAN but were not selected by GMM algorithm. Red lines show stars that were selected by GMM algorithm and have probability of membership between $>0.5$ and $<0.8$ and blue lines show stars that were selected by GMM algorithm and have probability $>0.8$}
  \label{parallax of dbscan and GMM.fig}
\end{figure}

\begin{figure}
  \centering
  \captionsetup[subfigure]{labelformat=empty}
        \begin{subfigure}{0.43\textwidth}
        \centering

                \includegraphics[width=\textwidth]{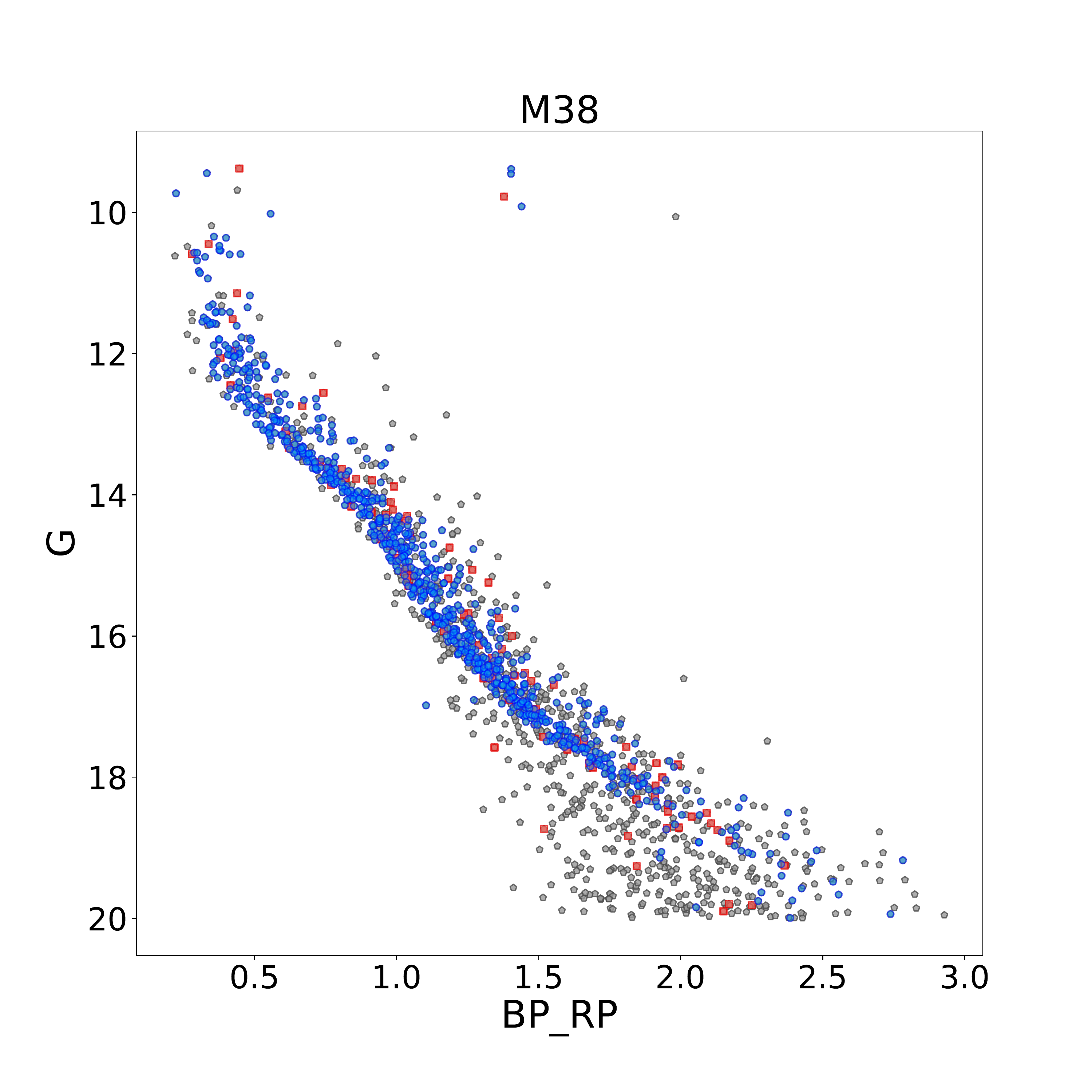}
        \end{subfigure}
        \begin{subfigure}{0.43\textwidth}

                \centering
                \includegraphics[width=\textwidth]{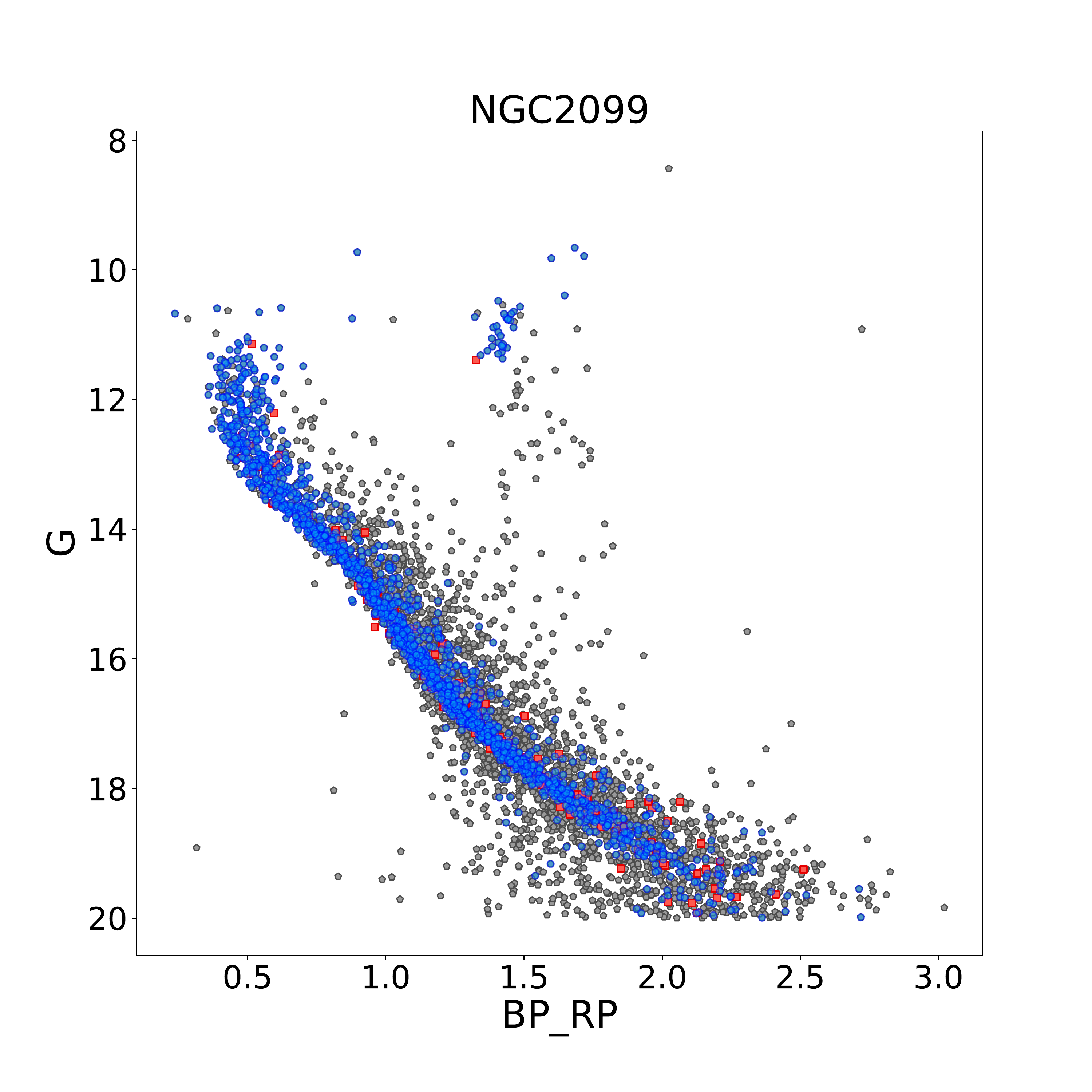}

        \end{subfigure}
        \begin{subfigure}{0.43\textwidth}
                \centering

                \includegraphics[width=\textwidth]{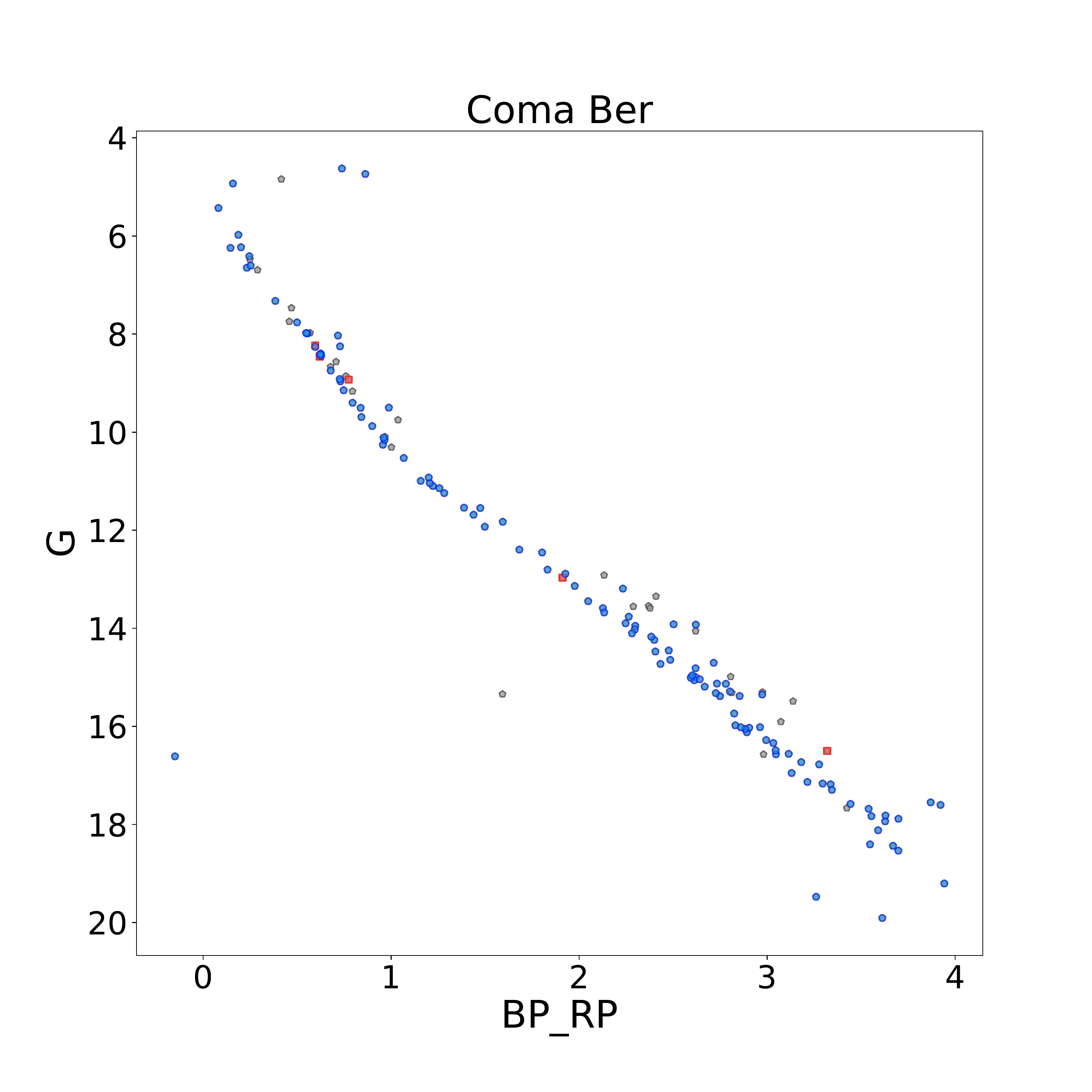}
        \end{subfigure}
        \begin{subfigure}{0.43\textwidth}
                \centering

                \includegraphics[width=\textwidth]{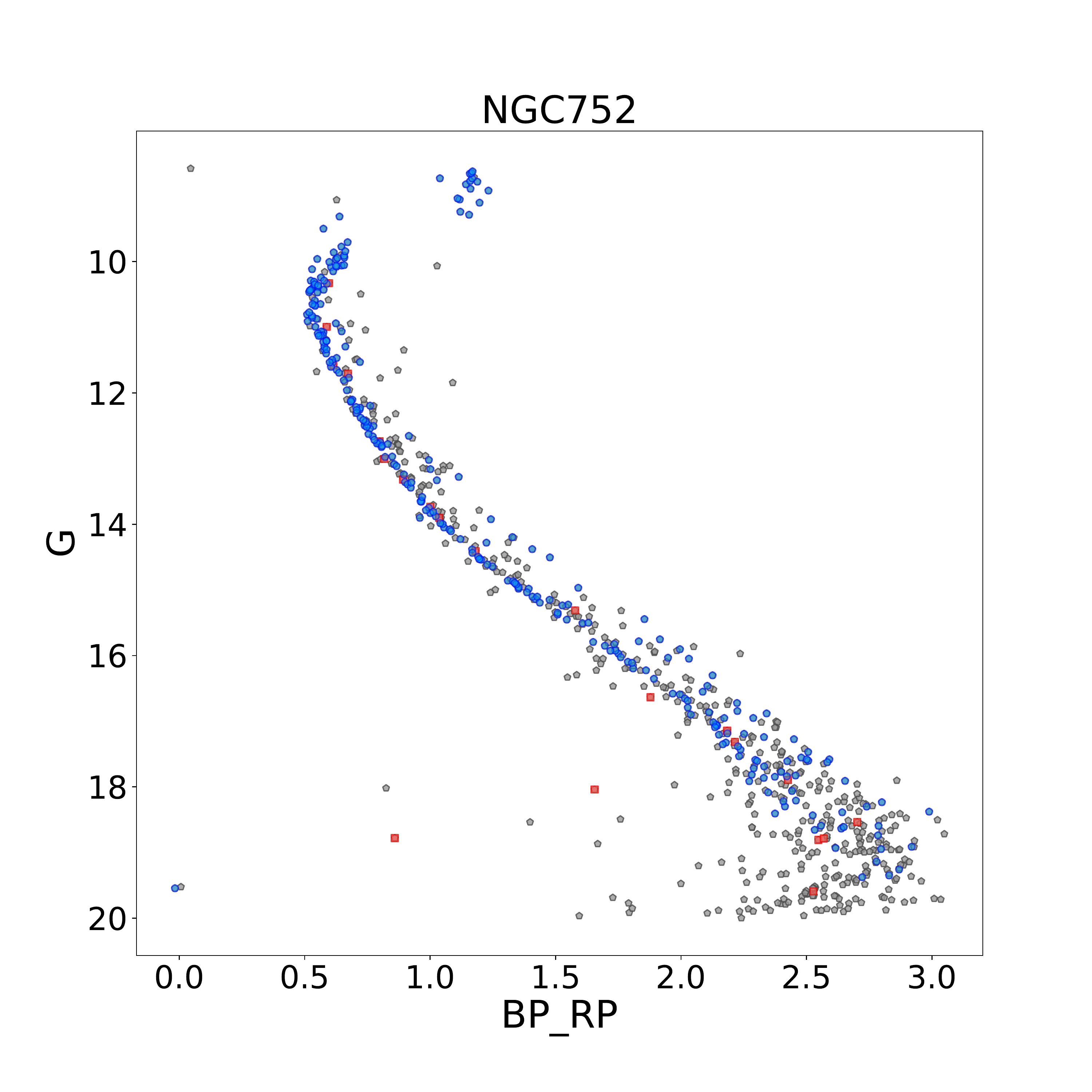}

        \end{subfigure}
        \begin{subfigure}{0.43\textwidth}
                \centering

                \includegraphics[width=\textwidth]{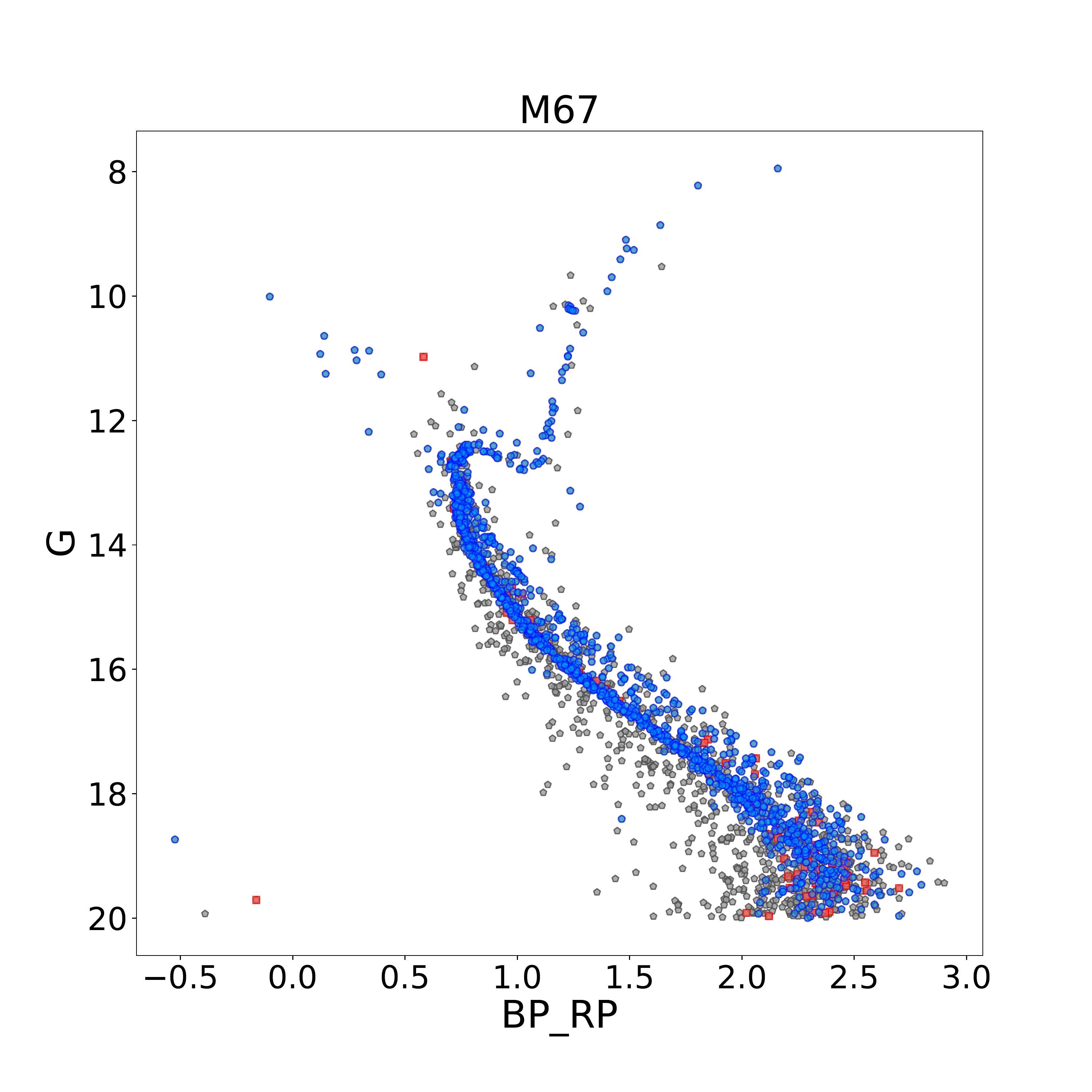}

        \end{subfigure}
        \begin{subfigure}{0.43\textwidth}
                \centering

                \includegraphics[width=\textwidth]{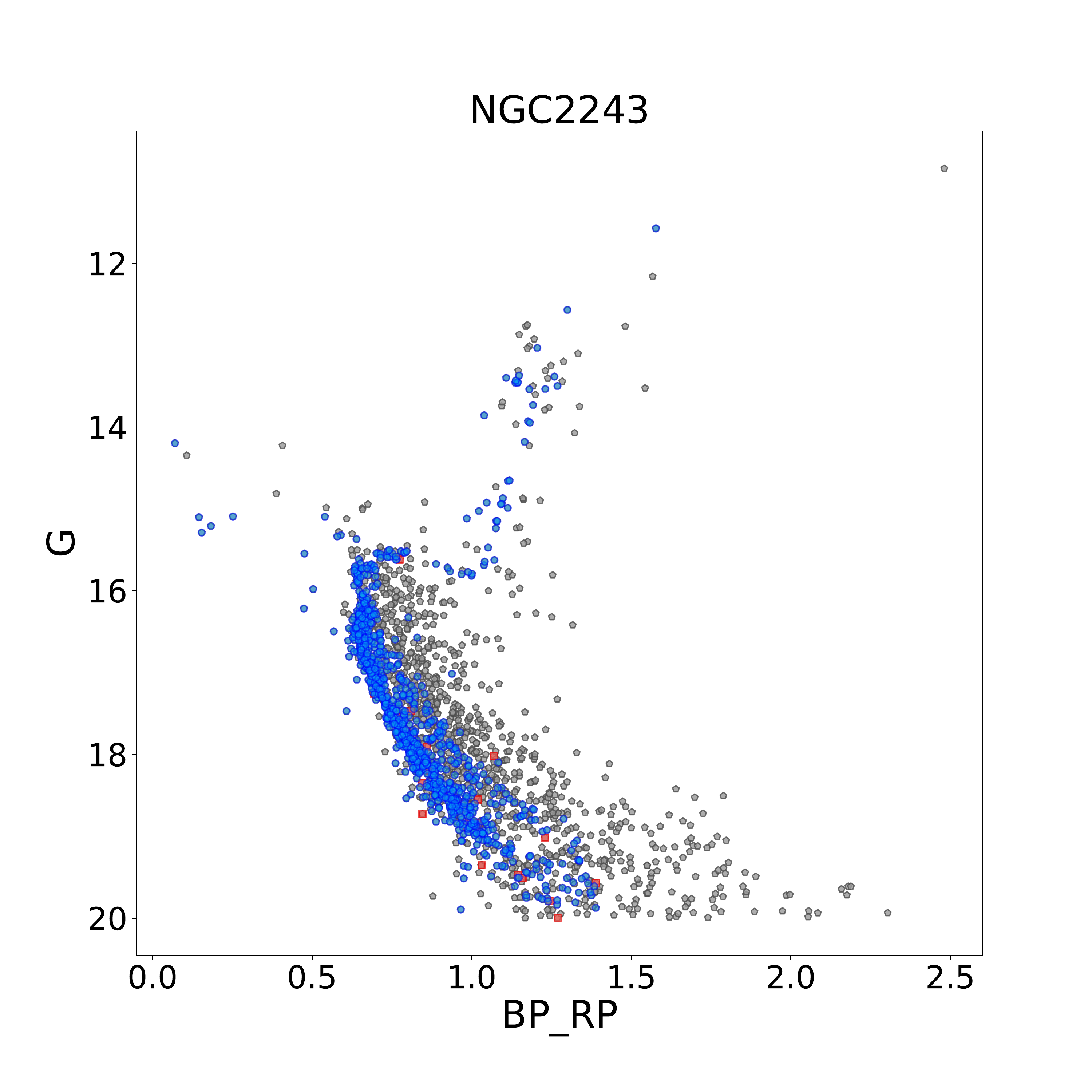}

        \end{subfigure}
  \caption{CMD of cluster member stars candidate from field stars by DBSCAN and GMM, grey dots show selected stars by DBSCAN that were not selected by GMM algorithm, red dots show stars that were selected by GMM algorithm and have a probability of membership $<0.8$ and blue dots show cluster members, the vertical axis is G magnitude, and the horizontal axis is $B-R$ that shows colors indicate.}
  \label{CMD of dbscan and GMM.fig}
\end{figure}

\begin{figure}
  \centering
  \captionsetup[subfigure]{labelformat=empty}
  
        \begin{subfigure}{0.49\textwidth}
        \centering

                \includegraphics[width=\textwidth]{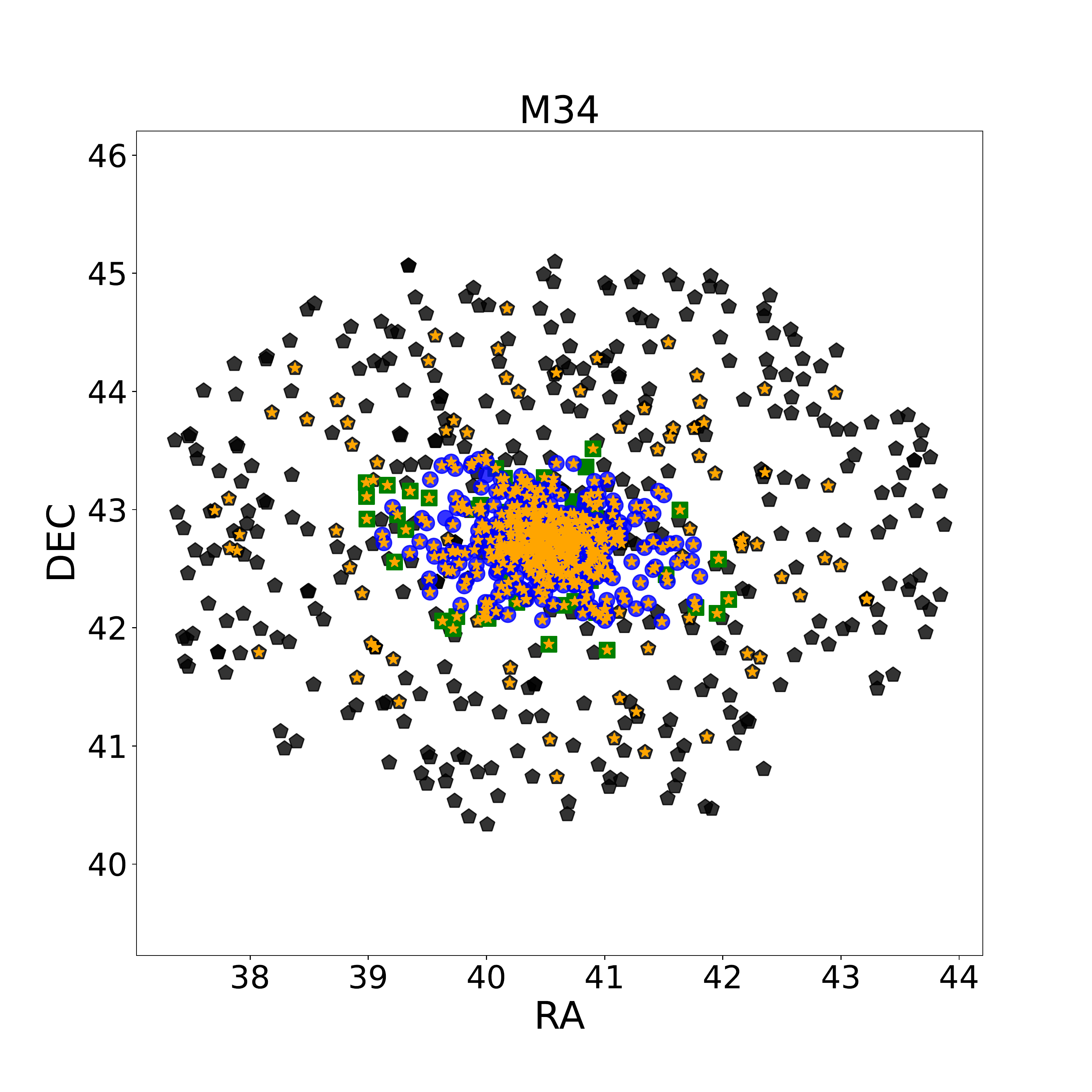}
        \end{subfigure}
        \begin{subfigure}{0.49\textwidth}

                \centering
                \includegraphics[width=\textwidth]{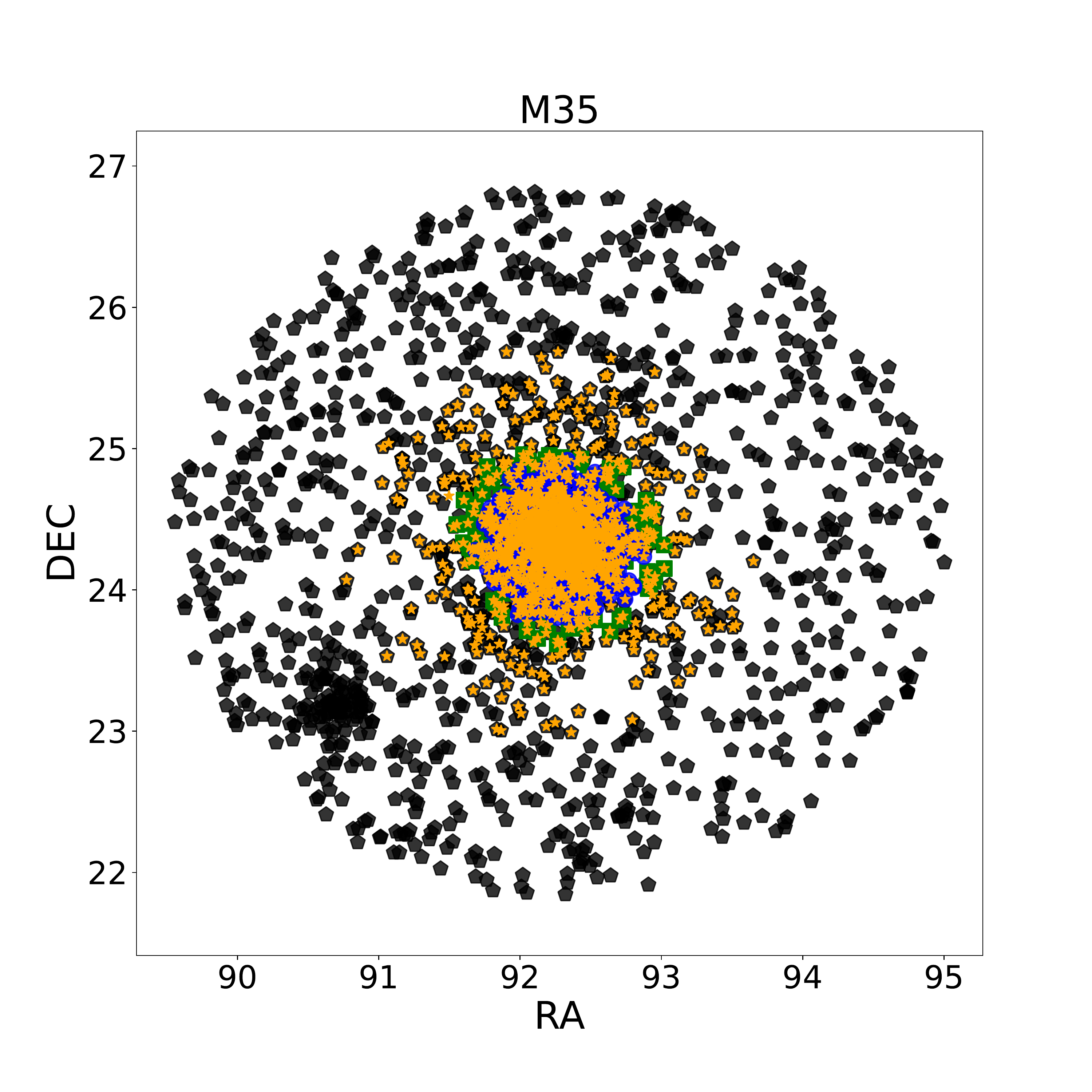}

        \end{subfigure}
        \begin{subfigure}{0.49\textwidth}

                \centering
                \includegraphics[width=\textwidth]{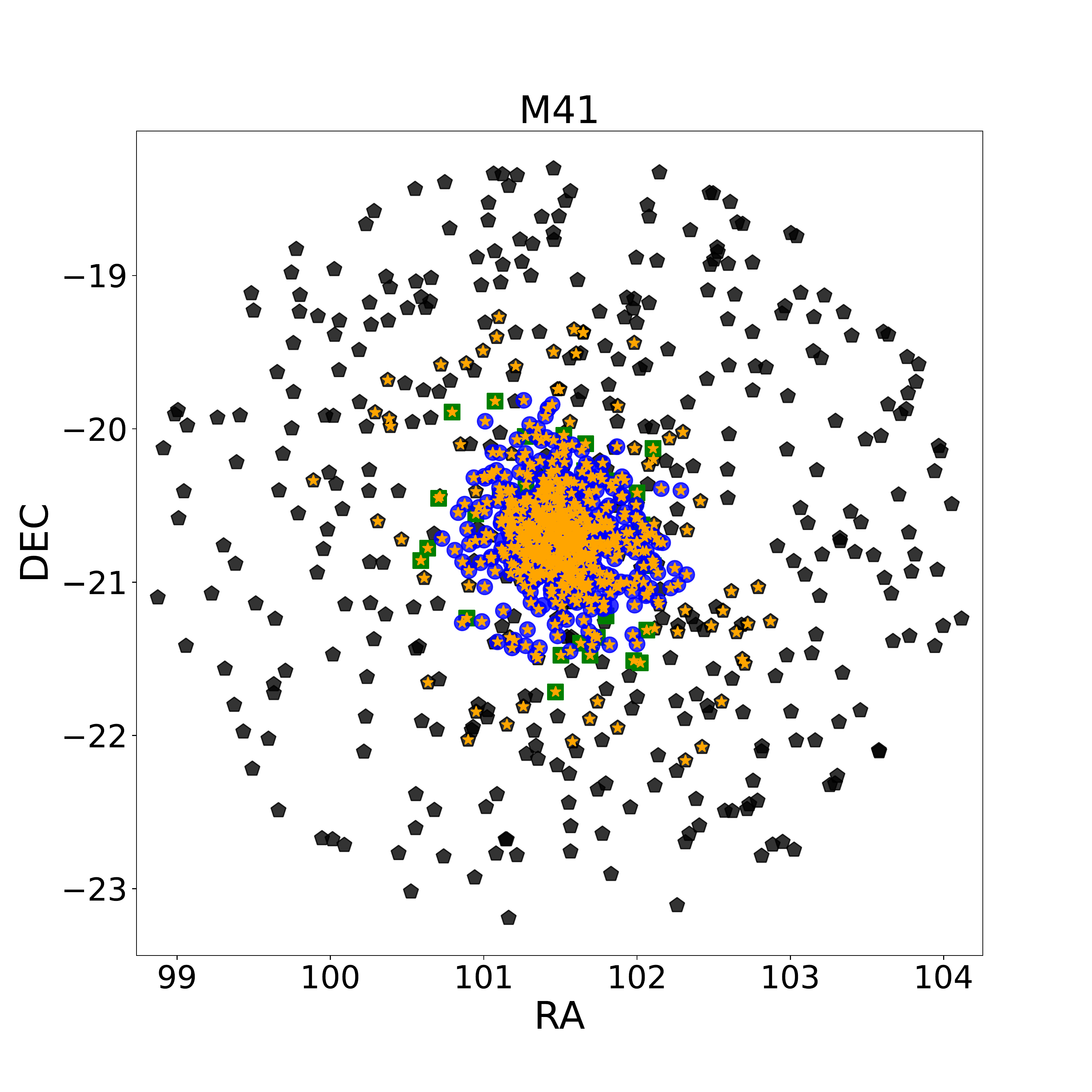}

        \end{subfigure}
        \begin{subfigure}{0.49\textwidth}

                \centering
                \includegraphics[width=\textwidth]{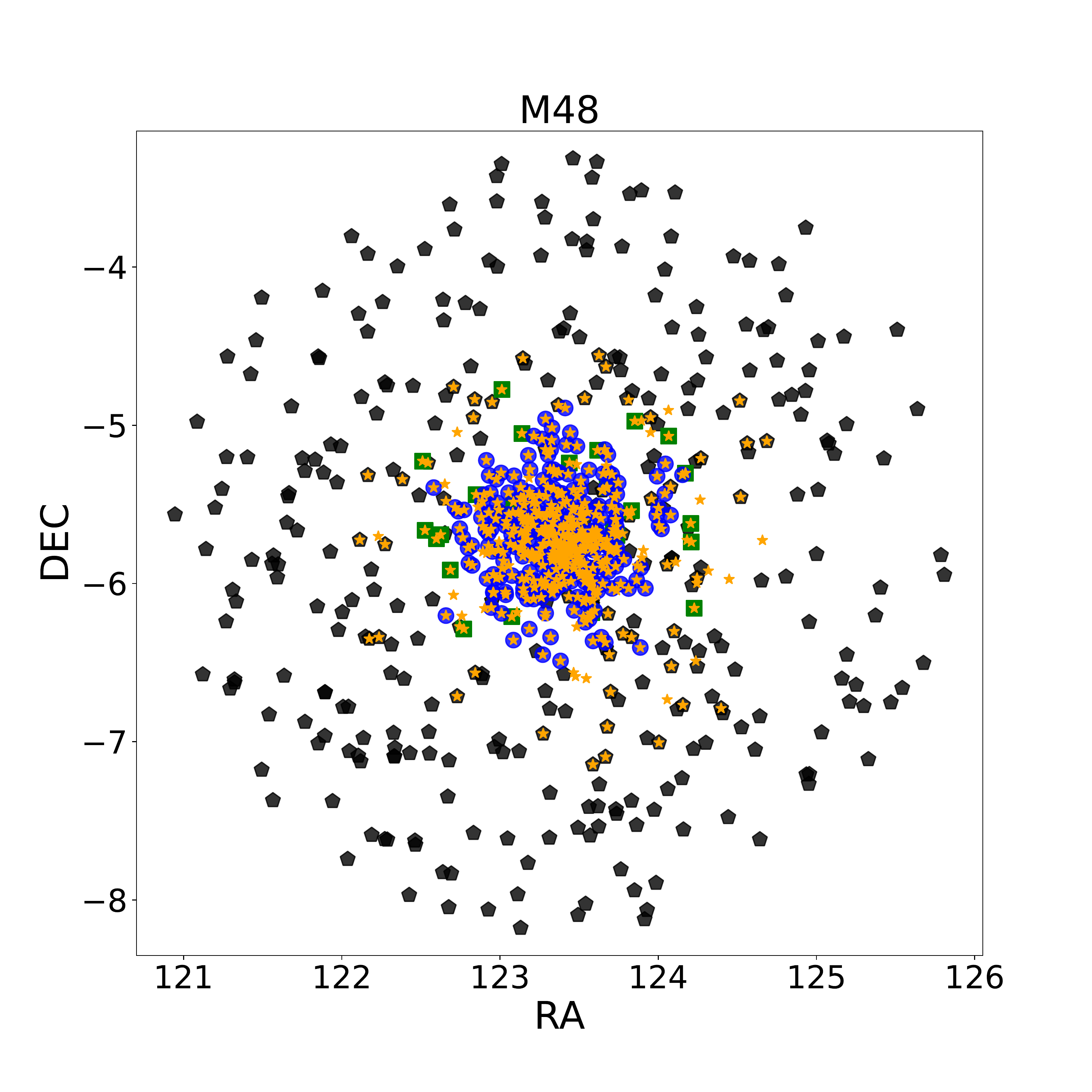}

        \end{subfigure}
  
  \caption{Position of stars selected by DBSCAN and GMM in 150 arcmin based on Gaia DR2 and $G<18$, black dots show selected stars by DBSCAN but are not selected by GMM algorithm. Green dots show stars that were selected by GMM algorithm and have a probability of membership between $>0.5$ and $<0.8$ and blue dots show stars that were selected by GMM algorithm and have a probability $>0.8$. Stars that detected by Cantat-Gaudin2018 with probability $>0.5$ are shown by orange stars.}
  \label{mgc.fig}
\end{figure}

\begin{figure}
  \centering
  \captionsetup[subfigure]{labelformat=empty}
        \begin{subfigure}{0.43\textwidth}
        \centering

                \includegraphics[width=\textwidth]{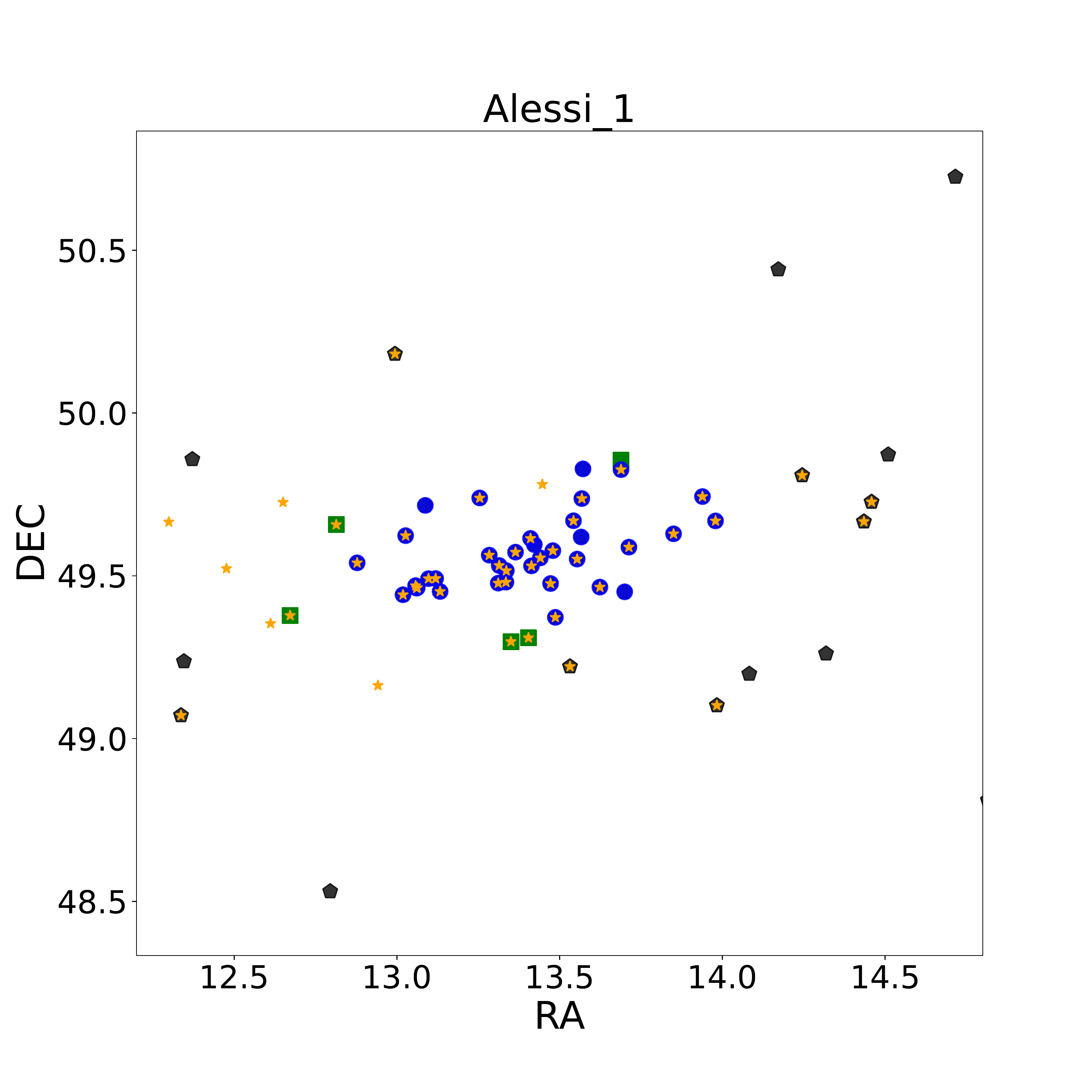}
        \end{subfigure}
        \begin{subfigure}{0.43\textwidth}
        \centering

                \includegraphics[width=\textwidth]{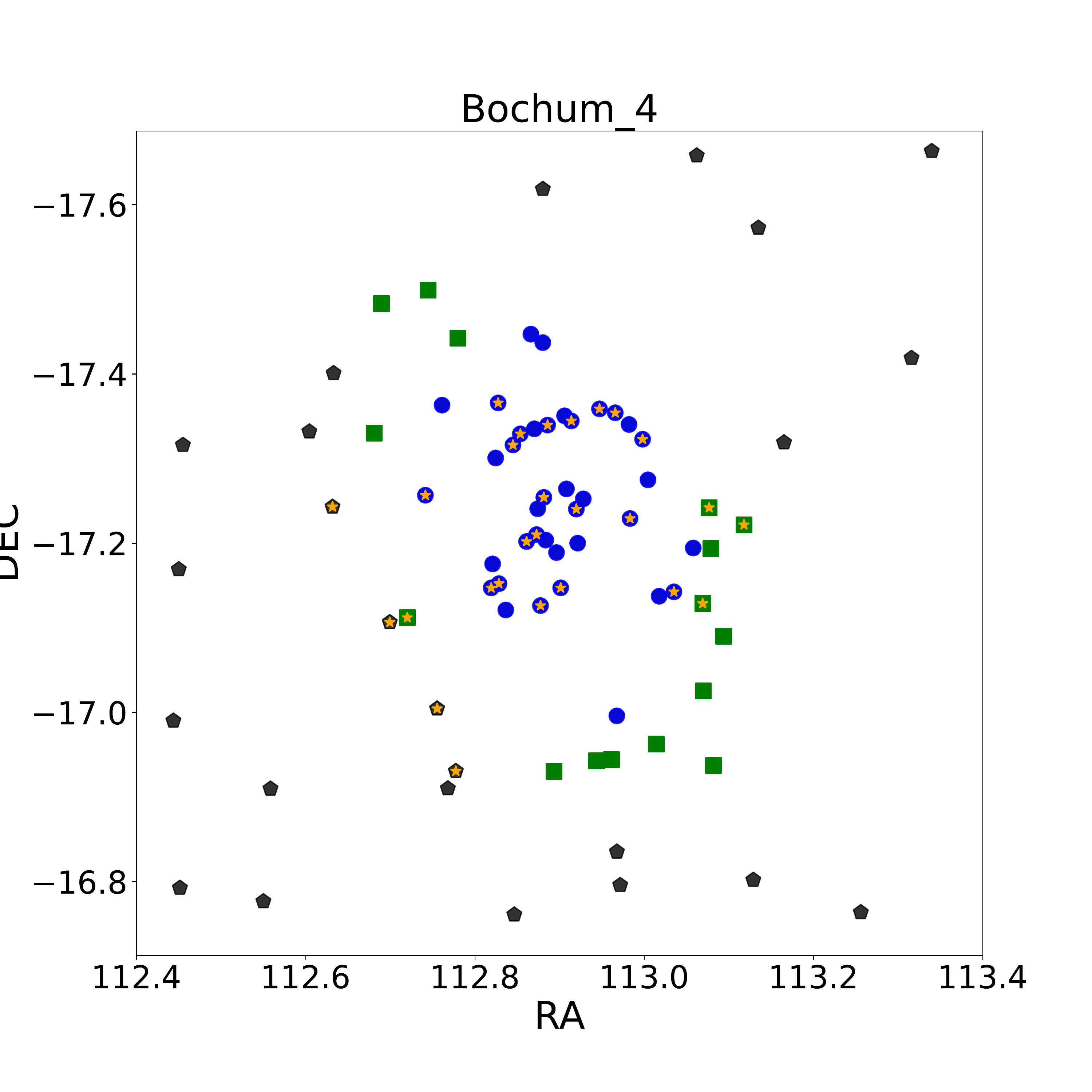}
        \end{subfigure}
        \begin{subfigure}{0.43\textwidth}

                \centering
                \includegraphics[width=\textwidth]{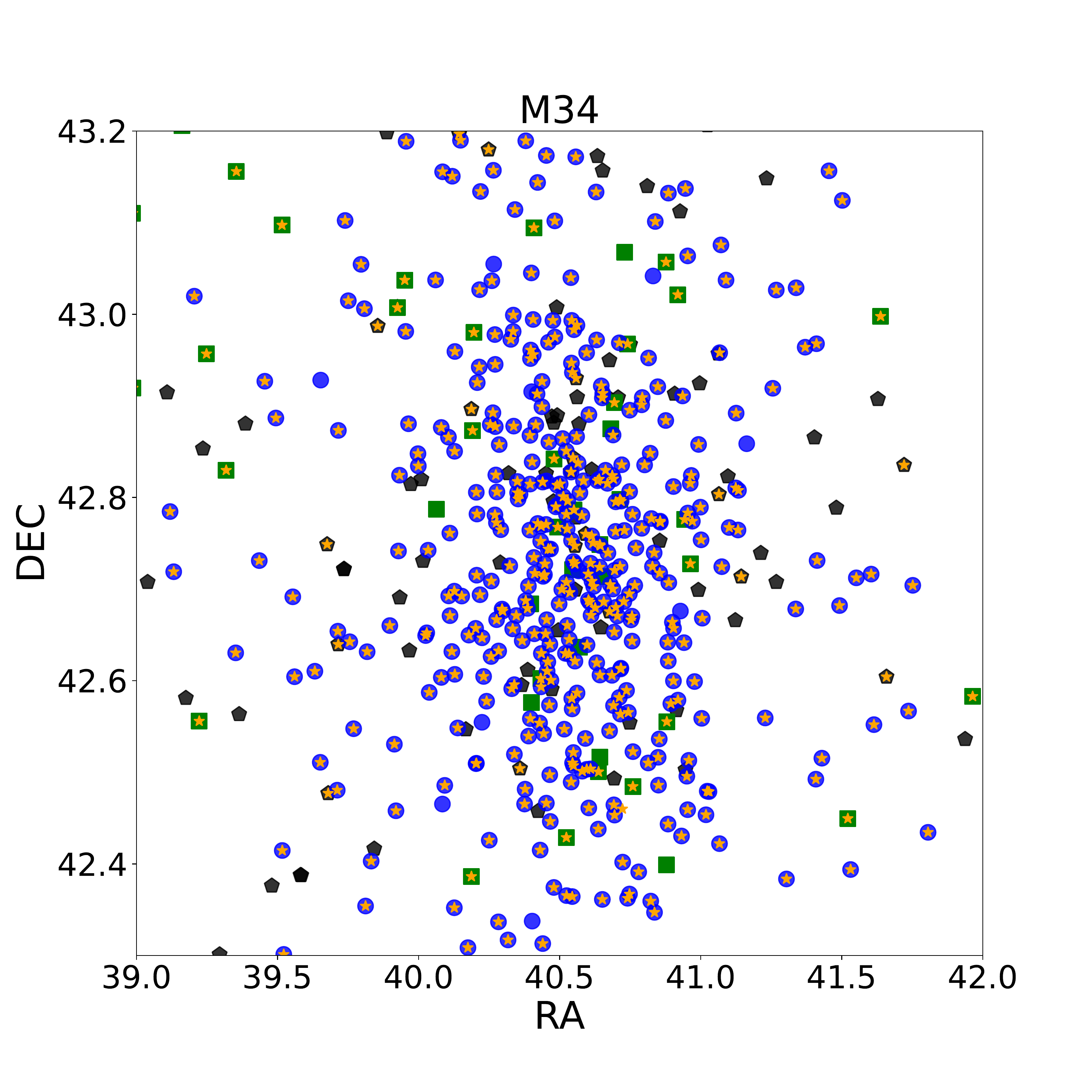}

        \end{subfigure}
        \begin{subfigure}{0.43\textwidth}

                \centering
                \includegraphics[width=\textwidth]{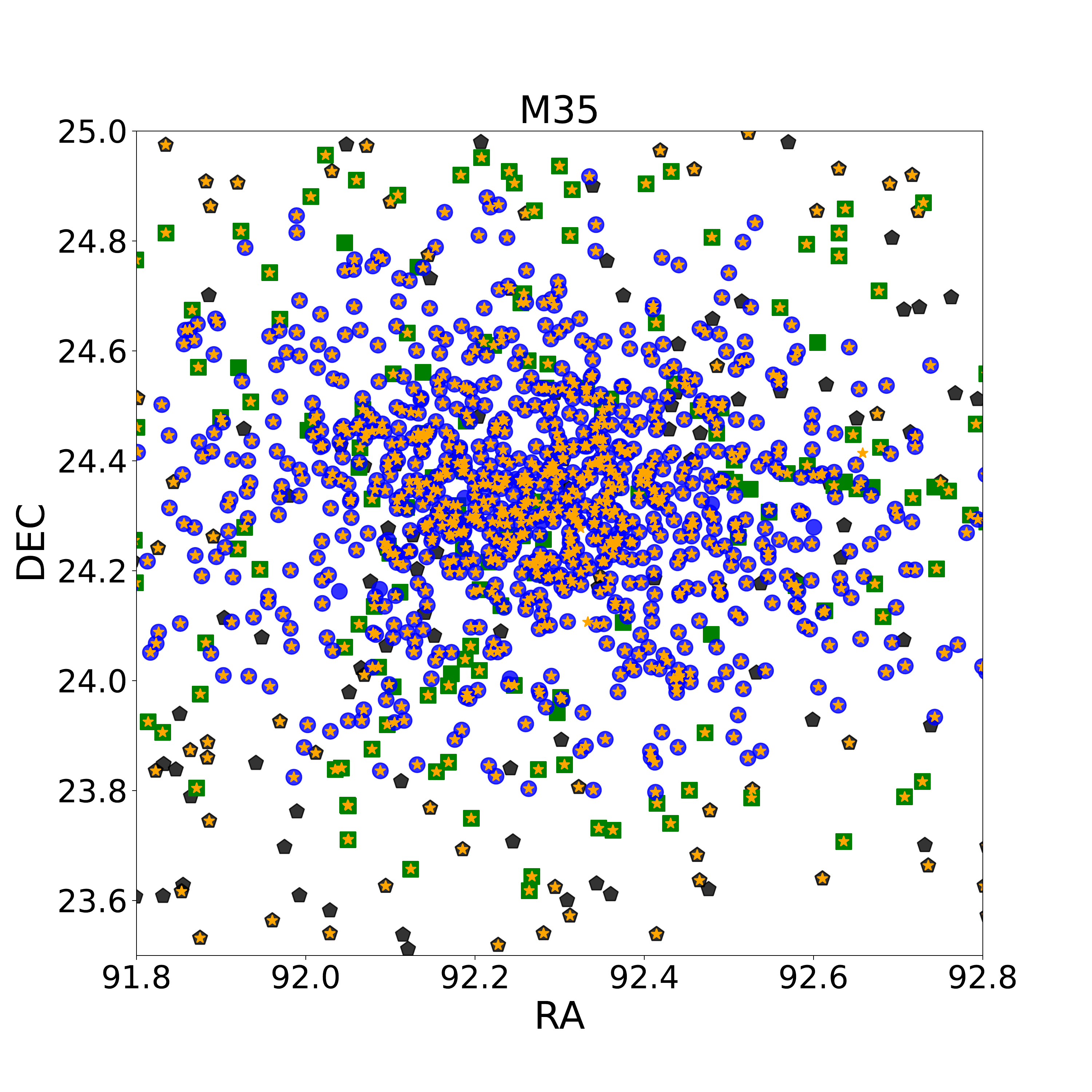}

        \end{subfigure}
        \begin{subfigure}{0.43\textwidth}

                \centering
                \includegraphics[width=\textwidth]{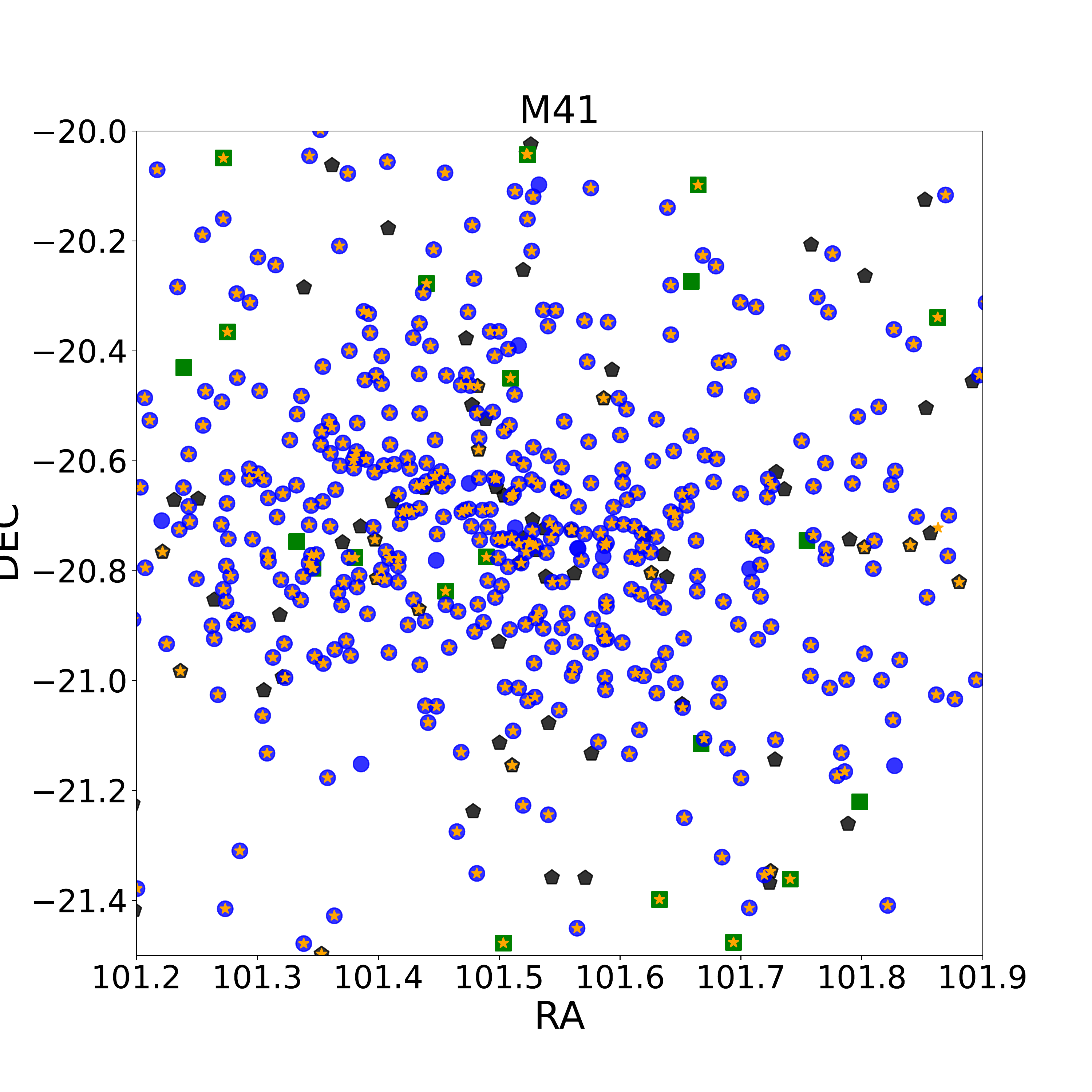}

        \end{subfigure}
        \begin{subfigure}{0.43\textwidth}

                \centering
                \includegraphics[width=\textwidth]{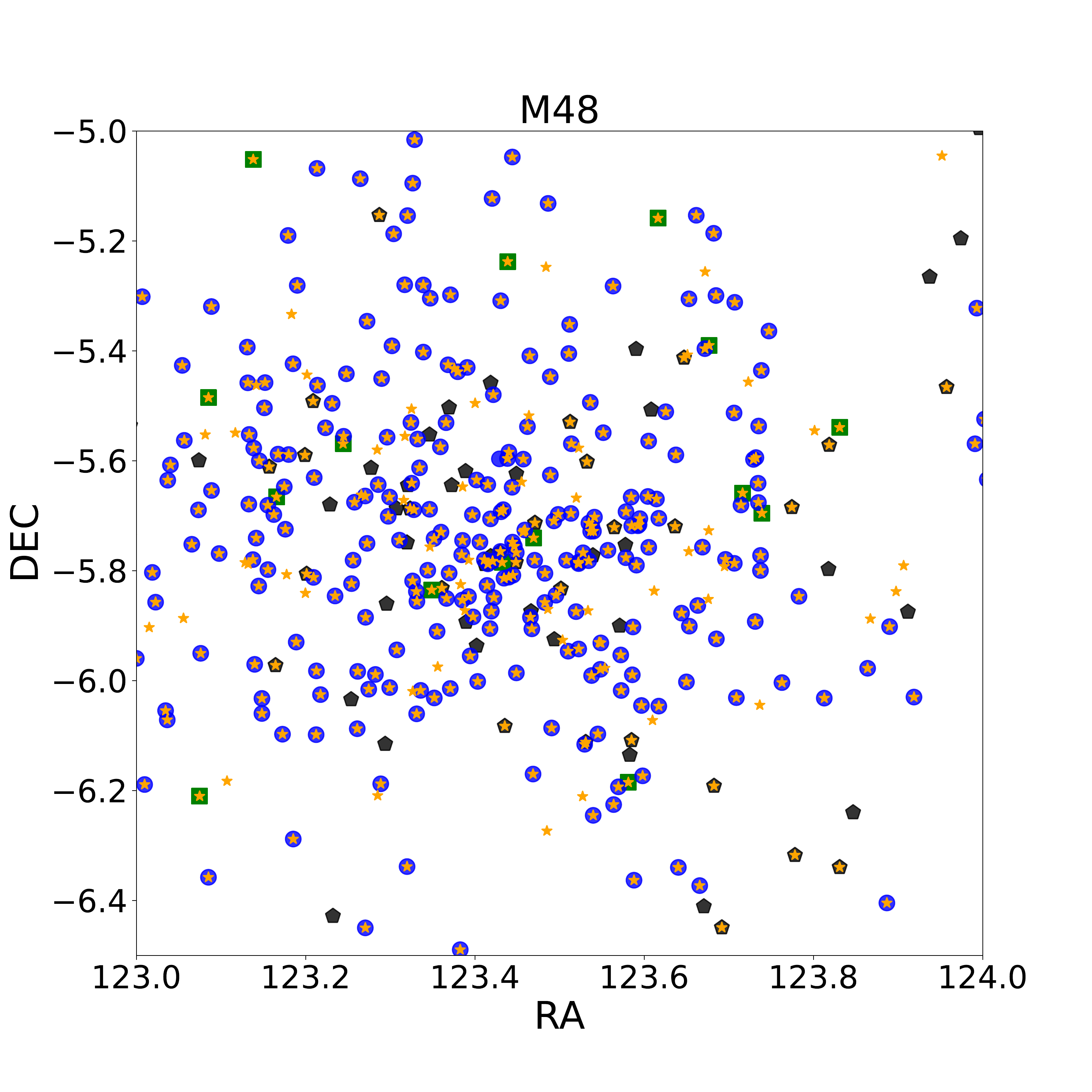}

        \end{subfigure}
  \caption{Position of stars selected by DBSCAN and GMM in cluster region based on Gaia DR2 and $G<18$, black dots show selected stars by DBSCAN but are not selected by GMM algorithm. Green dots show stars that were selected by GMM algorithm and have a probability of membership between $>0.5$ and $<0.8$ and blue dots show stars that were selected by GMM algorithm and have a probability $>0.8$. Stars that detected by Cantat-Gaudin2018 with probability $>0.5$ are shown by orange stars.}
  \label{gaudin.fig}
\end{figure}

In the first step, the cluster member candidates are selected by the DBSCAN  algorithm in three dimensions including  pmRA,  pmDEC, and Parallax.
Fig~\ref{proper motion of dbscan.fig} shows the proper-motion-selected cluster members, which are denoted by the blue dots among the stars in the field shown with the grey dots. For other six clusters see supplementary\ref{a.proper motion of dbscan.fig}. 
As seen in Fig~\ref{proper motion of dbscan.fig},  stars detected by DBSCAN show concentrated points in the proper motion space among stars in the field.\\
\noindent
Then the GMM algorithm is applied to clear the contamination of candidates from the field stars in five dimensions including RA, DEC, pmRA,  pmDEC, and Parallax. In the case of low-member clusters, Alessi01 and Bochum04, we applied GMM just in RA and DEC. It should be noted that MinPts and Eps are to be selected in coordination with GMM. Stars with a membership probability higher than $0.8$ are considered as cluster member candidates.\\
\noindent
Figs.~\ref{position of dbscan and GMM.fig}-\ref{CMD of dbscan and GMM.fig}, shows the result when the GMM algorithm is applied to cluster member candidates that have been selected by DBSCAN in the first step. For other six clusters see supplementary\ref{a.position of dbscan and GMM.fig.}-\ref{a.CMD of dbscan and GMM.fig}. Grey points represent non-members, red points represent the cluster members with a membership probability between $0.5$ to $0.8$, and blue points the stars with a membership probability higher than $0.8$.\\ 
\noindent
The astrometric parameters for each cluster are shown in Table~\ref{astro.tab}. The cluster distances are taken from \cite{Bailer_Jones_2018}. Table~\ref{results.tab} shows the main results of this work in each step. We consider all stars that have a membership probability higher than $0.8$ as cluster members.  \\
As can be seen, \cite{Cantat-Gaudin2018} detected 48, 27, 474, 1710, 578, 1325, 645, 479 and 691 members for Alessi01, Bochum04, M38, NGC2099, M34, M35, M41, M48, and M67 respectively, by using the K-means algorithm in three astrometric data (two proper motions and one parallax) and stars brighter than 18 mag based on UPMASK~\cite{upmask}. \\ 
Also,~\cite{Gao_spectral_clustering} detected 40 members, for Coma Ber by using the Spectral clustering algorithm with stars brighter than 18 mag. Moreover, \cite{ML/mnras/stab118} found 1194, 1640, and 583 members for M67, NGC2099, and NGC2243 respectively by using the ML-MCO method, which is formed KNN and GMM algorithm, with probability high than $0.6$ and stars brighter than 20 mag. In addition, stars that have a probability between 0.2 to 0.6 provided that have parallax in the range of stars with a probability higher than 0.8.\\
\noindent
\cite{Gao_m67_gmm_rf} detected 1361 members with a probability high than 0.8 for M67 by using combination of GMM and Random Forest methods. Moreover,\cite{ngc752ti} found 282 members for NGC752 with a probability high than 0.6 and in addition stars that have a probability between 0.2 to 0.6 provided that have parallax in the range of stars with a probability higher than 0.8 by using ML-MCO method with stars brighter than 20 mag.\\ 
It is important to point out that to carry out an accurate comparison, our method was initially implemented on the data of Gaia DR2 (with the exception of NGC752, because \cite{ngc752ti} used Gaia DR3) taking into consideration any conditions that were applied in each article. The results are shown in Table~\ref{gdr2.tab}.\\
\noindent
It is worth mentioning that in the case of  M67, \cite{Gao_m67_gmm_rf} did not find any white dwarf and \cite{ML/mnras/stab118} considered three white dwarfs as the sample sources, while here in our study, two white dwarfs were detected as cluster members with a probability higher than 0.5 and one of them has a probability higher than 0.8.\\
As shown in Fig~\ref{position of dbscan and GMM.fig} the stars with probability less than 0.8 are mostly distributed in the outer part of the clusters (except for Coma Ber and NGC752) because we used GMM in RA and DEC. This shows that GMM in RA and DEC has a good efficiency in the outskirts of clusters. In Fig~\ref{CMD of dbscan and GMM.fig}, it can be seen that NGC2243 displays two separate sequences that one of them appears to be a binary sequence.\\
Fig~\ref{mgc.fig} presents a comparison between this work and \cite{Cantat-Gaudin2018} regarding M34, M35, M41 and, M48 based on the Gaia DR2 and stars with G<18 mag in 150 arcmin. As illustrated in the Fig~\ref{mgc.fig}, the stars that remained undetected by this work are located outside of the cluster dense region. This is due to GMM's selection of stars in five parameters (RA , DEC , Parallax , pmRA , pmDEC).\\
To show the details of \cite{Cantat-Gaudin2018} and this study, we show cluster region in Fig\ref{gaudin.fig} regarding Alessi01, Bochum04, M34, M35, M41 and, M48 based on the Gaia DR2 and stars with G<18. As seen in Fig\ref{gaudin.fig}, some members in the cluster region were detected in this work but were not detected by \cite{Cantat-Gaudin2018}. To show the dense region of clusters with details, the axes are not equal and for this reason, the shape of the cluster does not show itself.
\section{COMPARISON WITH ANOTHER TWO COMBINATION METHODS}\label{section.compare}
According to \cite{Gao_m67_gmm_rf},  which is a combination of Random Forest and GMM methods, at the first step, data is filtered based on the distance for GMM condition. Despite this filtering, the GMM could not detect reliable members for M67 around 150 arcmin; hence, they considered 60 arcmin for member detection. In the next step, data that GMM obtained, trained Random Forest in photometric and astrometric parameters. Here, in this paper,  we did not apply any filters on data except for the two conditions of positive parallax and $G<20$. We looked for members within 150 arcmin by combining two methods of  DBSCAN and GMM and found reliable members for M67 just in astrometric data. Moreover, for data selection, we considered other ranges for some cases like Coma Ber (400 arcmin)
\onecolumn

\begin{table}
\centering
\caption{Physical parameters of clusters that are analyzed in this work.}
\begin{tabular}{c c c c c}
    \hline
    \hline
    Name & Parallax~(mas) & pmRA~(mas\,yr$^{-1}$) & pmDEC~(mas\,yr$^{-1}$) & D~(pc) \\
    \hline
    M38 & $0.87\pm0.06$ & $1.54\pm0.06$ & $-4.43\pm0.05$ & $1160.62$  \\
    NGC2099 & $0.67\pm0.07$ & $1.87\pm0.08$ & $-5.62\pm0.05$ & $1571.41$ \\
    Coma Ber & $11.67\pm0.06$ & $-12.1\pm0.06$ & $-8.89\pm0.06$ & $85.47$ \\
    NGC752 & $2.26\pm0.06$ & $9.79\pm0.06$ & $-11.82\pm0.07$ & $437.33$ \\
    M67 & $1.15\pm0.09$ & $-10.96\pm0.09$ & $-2.90\pm0.07$ & $864.30$ \\
    NGC2243 & $0.23\pm0.08$ & $-1.26\pm0.09$ & $5.50\pm0.09$ & $3660.65$ \\
    Alessi01 & $1.42\pm0.05$ & $6.50\pm0.04$ & $-6.41\pm0.04$ & $693.10$ \\
    Bochum04 & $0.75\pm0.04$ & $-3.00\pm0.04$ & $1.91\pm0.04$ & $1292.49$ \\
    M34 & $2.01\pm0.13$ & $0.644\pm0.13$ & $-5.78\pm0.13$ & $503.84$ \\
    M35 & $1.16\pm0.08$ & $2.26\pm0.09$ & $-2.88\pm0.06$ & $901.04$ \\
    M41 & $1.36\pm0.08$ & $-4.37\pm0.05$ & $-1.35\pm0.07$ & $752.10$ \\
    M48 & $1.29\pm0.07$ & $-1.28\pm0.07$ & $1.03\pm0.06$ & $769.71$ \\
    \hline
  \end{tabular}\label{astro.tab}
  \end{table}

\begin{table}
\caption{Results in every step. (1): Clusters name. (2): The number of sample sources from each cluster when filtered by photometric and astrometric conditions in this work. (3): Stars that DBSCAN selected among sample sources. (4) and (5): Two free parameters of the DBSCAN algorithm (see text for more details). (6): Number of stars that are selected by the GMM algorithm as cluster members with membership probability higher than 0.5. (7): Cluster members with membership probability higher than 0.6. (8): Cluster members with membership probability higher than 0.8. (9): Cluster age. [1]: }\citet{mnras/m38/age}. [2]: \citet{ngc2099age:400}. [3]: \citet{Hartman_ngc2099_550}. [4]: \citet{Tang_2018_comaber}. [5]: \citet{ngc752_age_Carrera}. [6]: \citet{Schiavon_3.6_m67}. [7]: \citet{Onehag_m67_4.5}. [8]: \citet{Anthony_Twarog_2005}. [9]: \citet{Jacobson_ngc2243_age}. [10]: \citet{peliades_age}. [11]: \citet{m34_age200}. [12]: \citet{m34_age220}. [13]: \citet{m34_age250}. [14]: \citet{m41av}.
 \centering
  \begin{tabular}{ccccccccc}
    \hline
    \hline
    (1) & (2) & (3) & (4) & (5) &(6) &(7) &(8)&(9) \\
    Name & Sa & DB & Eps & MinPts & GMM & CM$>0.6$ & CM$>0.8$ & Age[Gyr] \\
    \hline
    M38 & 442955 & 1588 & 0.05 & 63 & 853 & 833 & 752 & 0.29~[1] \\
    NGC2099 & 526029 & 4742 & 0.1 & 225 & 1877 & 1827 & 1725 & 0.4~[2], 0.55~[3] \\
    Coma Ber & 189859 & 148 & 0.3 & 20 & 121 & 119 & 116 & 0.8~[4] \\
    NGC752 & 474464 & 735 & 0.15 & 80 & 290 & 286 & 269 & 2~[5] \\
    M67 & 64395 & 2572 & 0.15 & 280 & 1494 & 1481 & 1422 & 3.6~[6], 4.5~[7] \\
    NGC2243 & 218841 & 2014 & 0.05 & 158 & 964 & 957 & 936 & 3.8~[8], 4.7~[9] \\
    Alessi01 & 290461 & 152 & 0.1 & 65 & 52 & 52 & 43 & $-$ \\
    Bochum04 & 165839 & 200 & 0.06 & 55 & 55 & 52 & 38 & $-$ \\
    M34 & 219422 & 1693 & 0.18 & 223 & 852 & 824 & 743 & 0.2~[11], 0.22~[12], 0.25~[13] \\
    M35 & 513262 & 3635 & 0.11 & 316 & 1360 & 1304 & 1114 & 0.15~[13] \\
    M41 & 455740 & 1273 & 0.12 & 50 & 873 & 846 & 783 & 0.243~[14] \\
    M48 & 214036 & 1721 & 0.15 & 480 & 515 & 499 & 452 & 0.36~[14] \\
    \hline
  \end{tabular}
  \label{results.tab}
\end{table}

\twocolumn
\noindent
and  NGC752 (5 degree).\\
In another study, \cite{ML/mnras/stab118} used a combination of KNN and GMM methods. However, the KNN method  is a supervised algorithm that find the same data based on the nearest distance and needs suitable data train,  they used this algorithm as unsupervised.\\
They applied KNN on proper motion and prepared data for GMM. In the next step, they applied GMM on data in three astrometric parameters, i.e., one parallax and two proper motions. In the present study, we used DBSCAN as a density-based unsupervised  algorithm. 
Since two free parameters of MinPts and Eps are used in DBSCAN method, there is more control over the input data for GMM and more acceptable results are obtained. We applied  GMM on five astrometric parameters (i.e., one parallax, two positions, and two proper motions) and obtained better results even for the case of Coma Ber.
\section{DISCUSSION}\label{section.discus}

The high-precision photometric and astrometric data from Gaia-DR3 offer a unique opportunity to identify possible members of open clusters. The main goal of this study is to find cluster members and derive astrometric parameters and physical characteristics from Gaia DR3 data alone using the new machine-learning-based method. Here, we discuss the evolution pattern of open clusters based on the color-magnitude diagram (CMD), the spatial distribution of stars, and evidence of mass segregation in studied clusters.\\
We need information about the extinction values of the clusters to determine the absolute magnitude. The previous works were applied to achieve this aim and these values are shown in Table~\ref{tab_discussion}.\\

\subsection{Distribution of stars in the ten clusters}\label{section.discus.distr}

\begin{figure}
  \centering
  \captionsetup[subfigure]{labelformat=empty}
        \begin{subfigure}{0.44\textwidth}
        \centering

                \includegraphics[width=\textwidth]{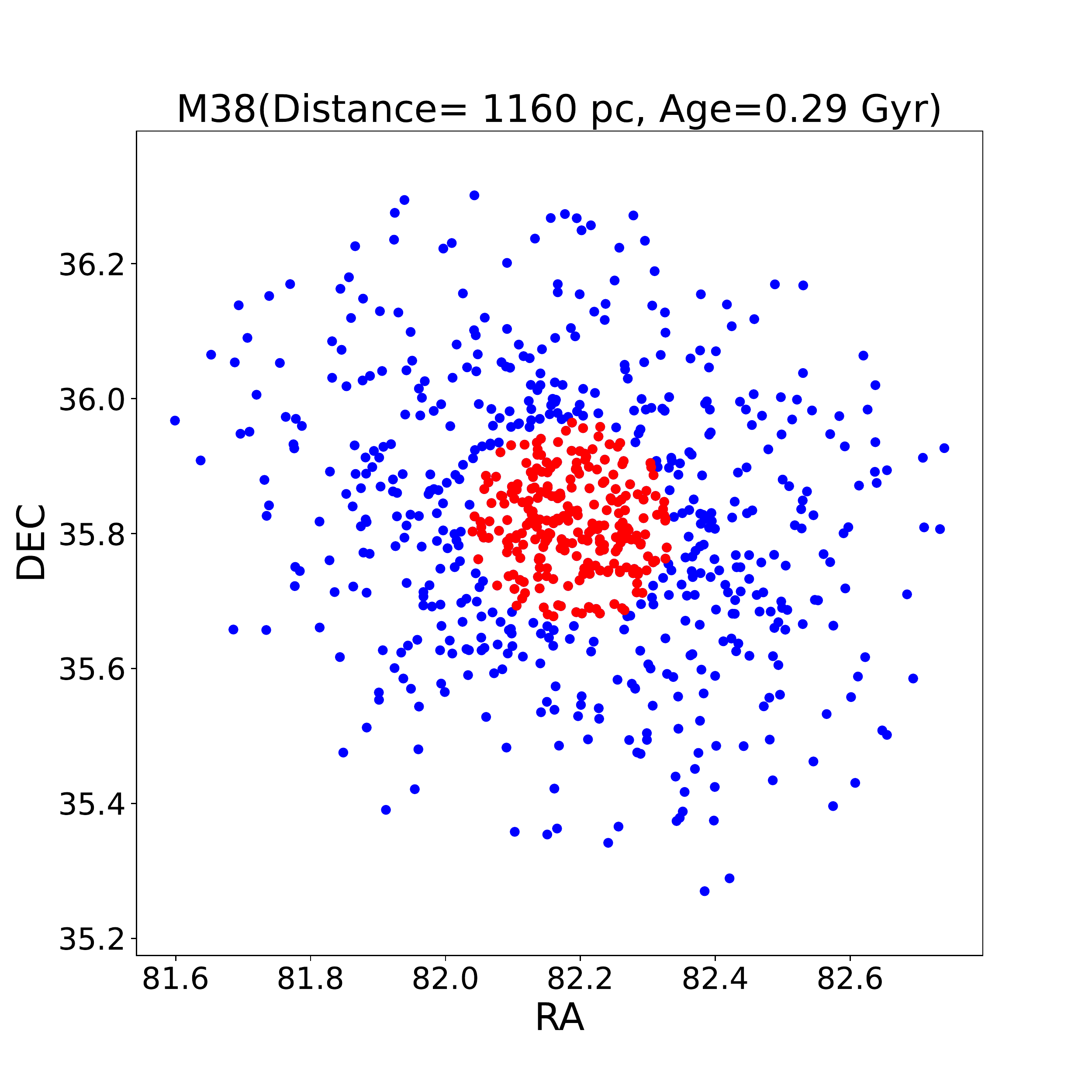}
        \end{subfigure}
        \begin{subfigure}{0.44\textwidth}

                \centering
                \includegraphics[width=\textwidth]{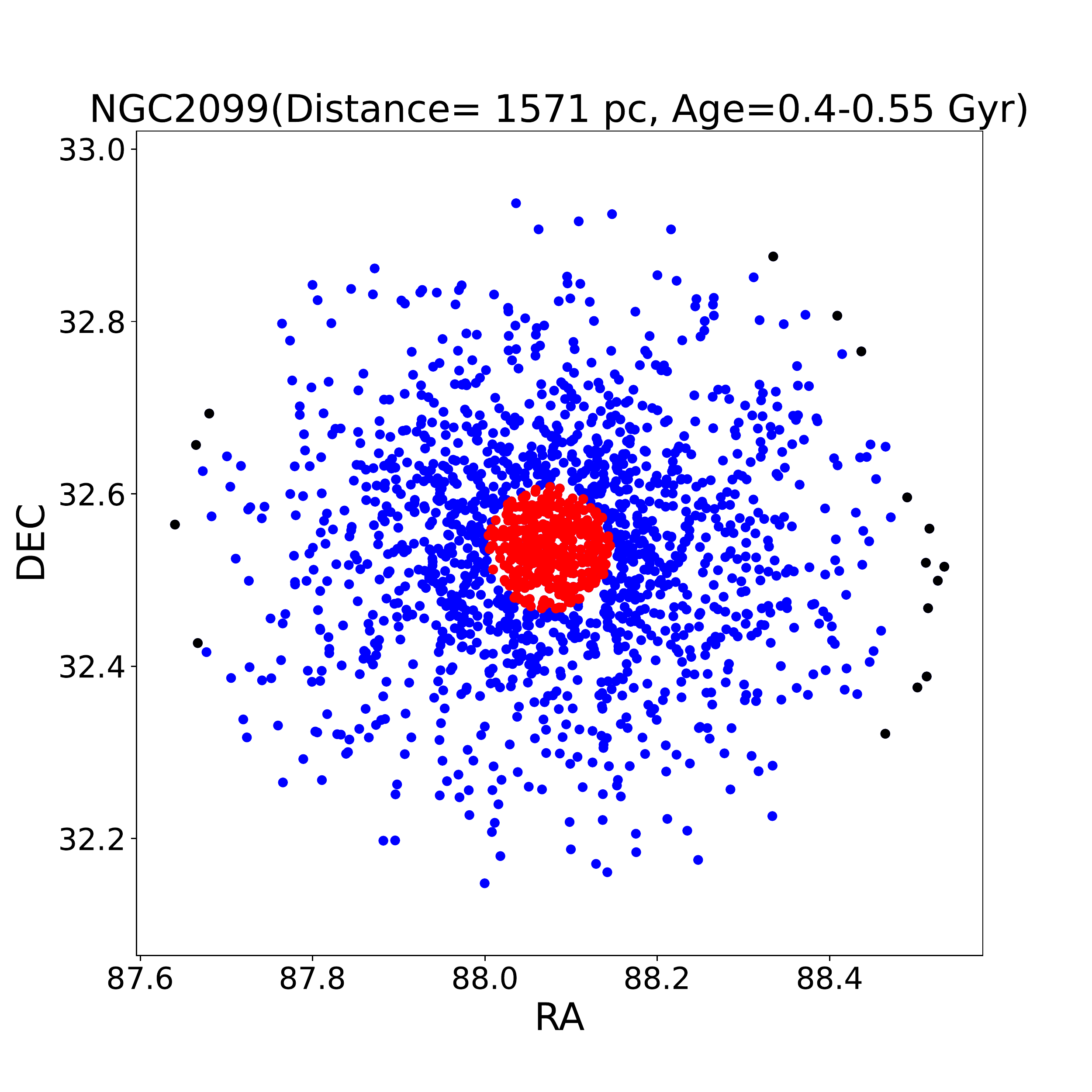}

        \end{subfigure}
        \begin{subfigure}{0.44\textwidth}
                \centering

                \includegraphics[width=\textwidth]{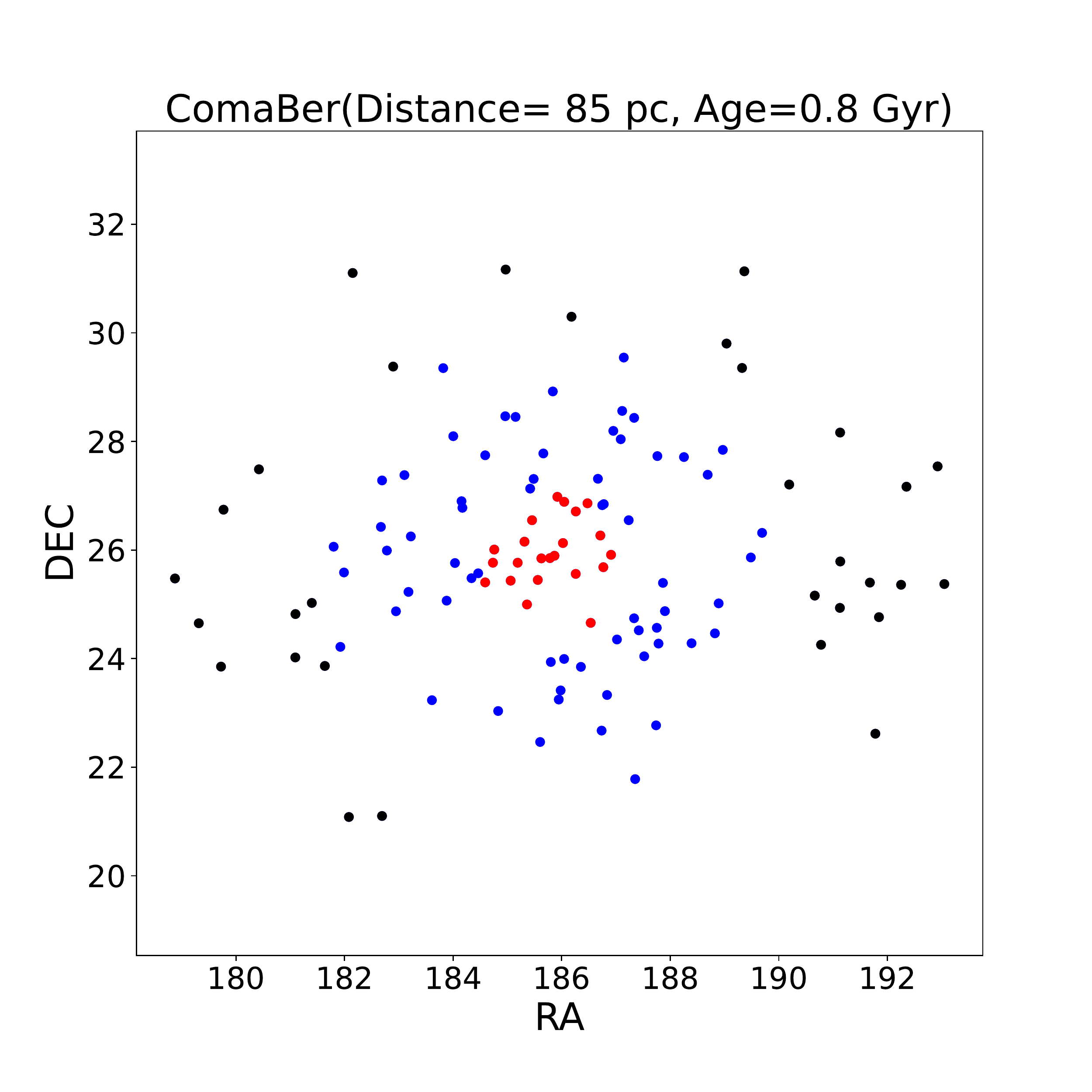}
        \end{subfigure}
        \begin{subfigure}{0.44\textwidth}
                \centering

                \includegraphics[width=\textwidth]{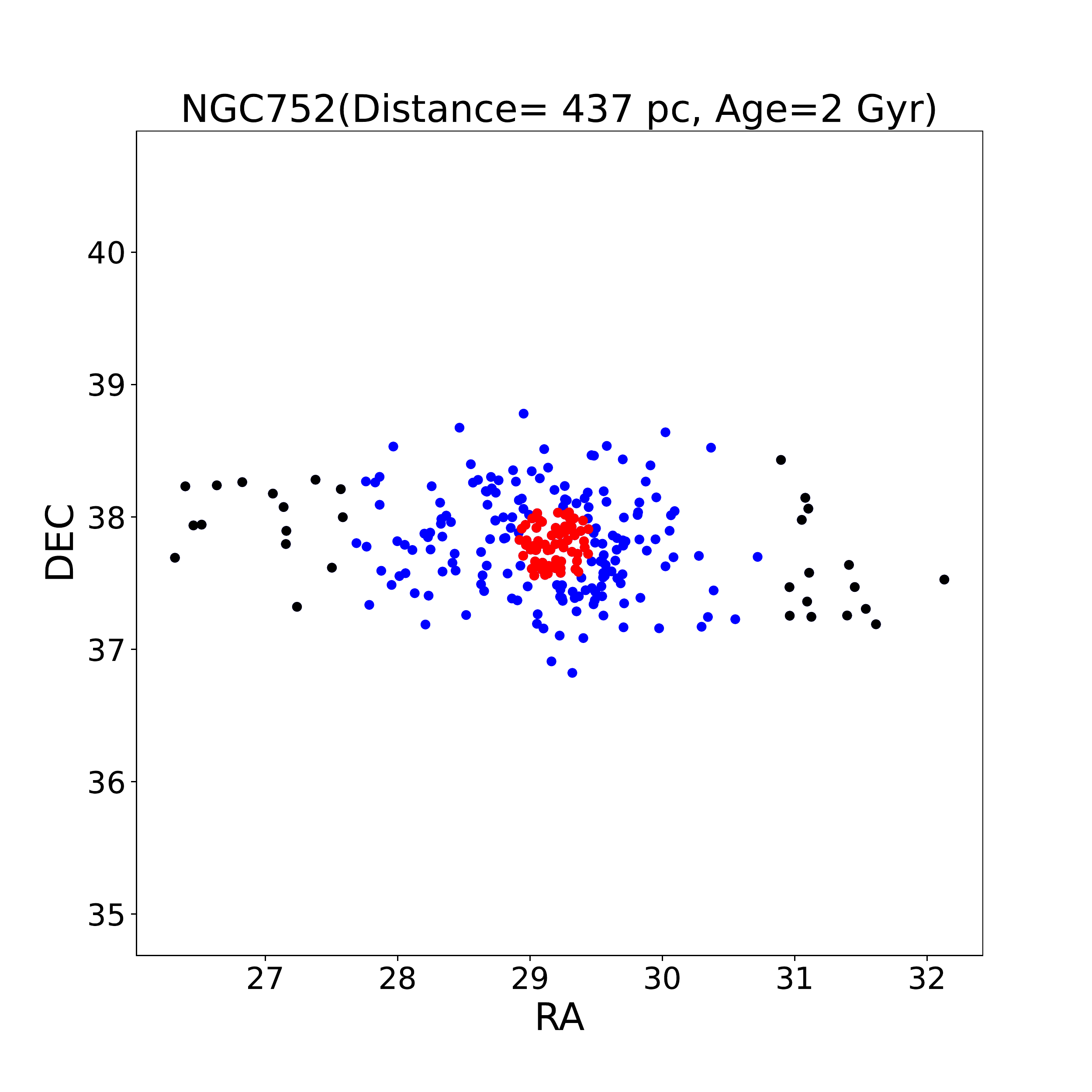}

        \end{subfigure}
        \begin{subfigure}{0.44\textwidth}
                \centering

                \includegraphics[width=\textwidth]{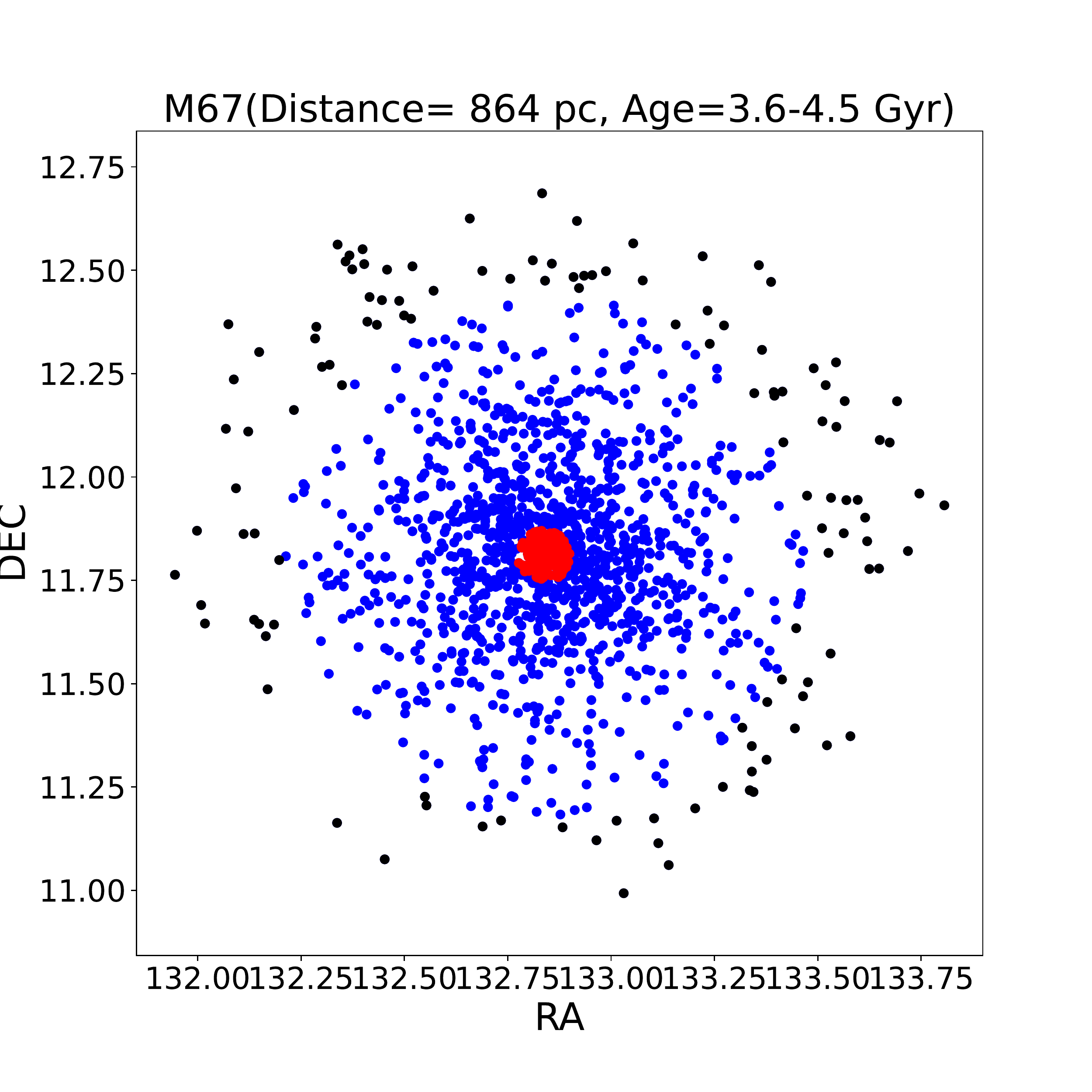}

        \end{subfigure}
        \begin{subfigure}{0.44\textwidth}
                \centering

                \includegraphics[width=\textwidth]{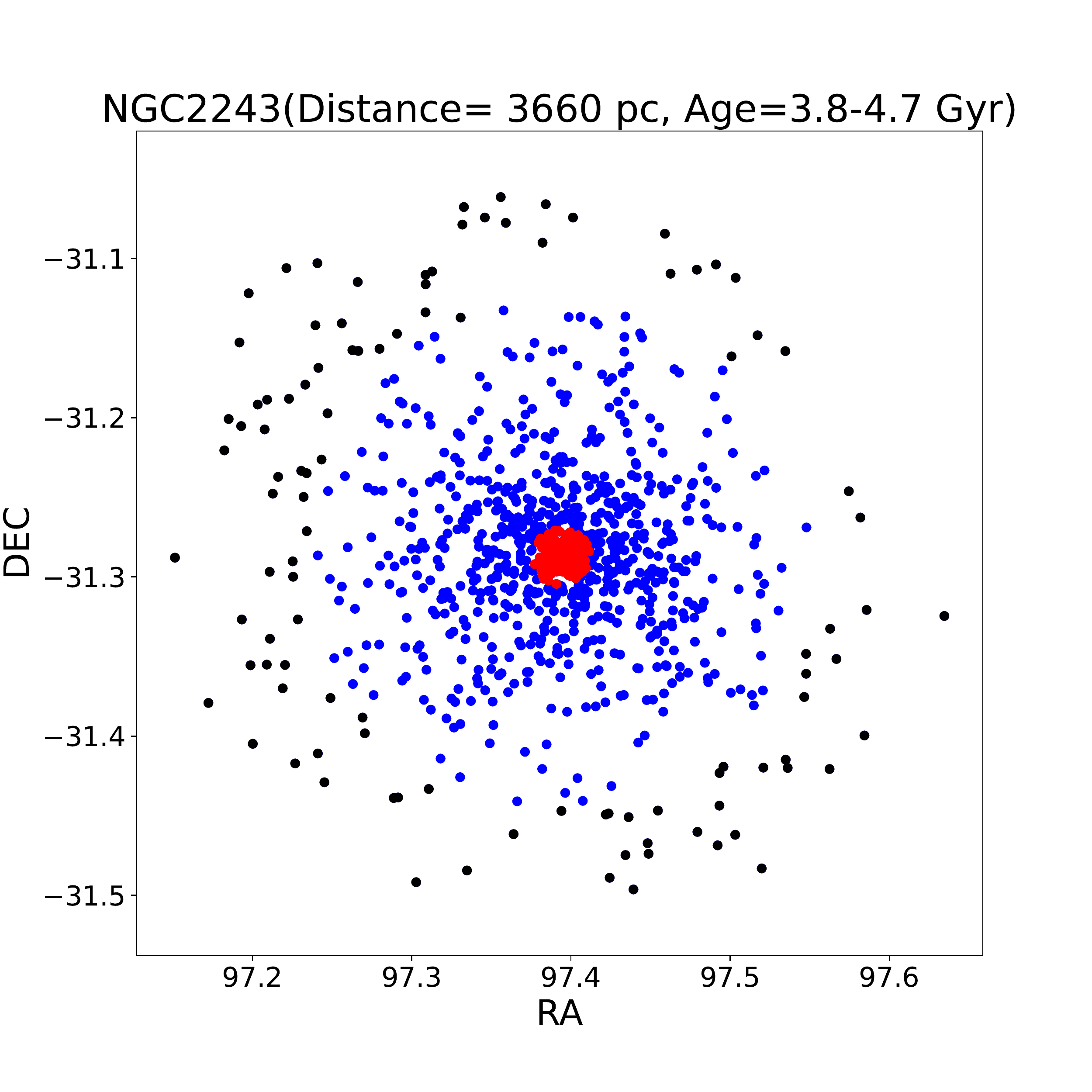}

        \end{subfigure}
  \caption{Position of cluster stars. Red dots show stars within the cluster core, and black dots show stars that are distributed outside the tidal radius.}
  \label{position from fields.fig}
\end{figure}

The tidal radius for the ten under study clusters( M38, NGC2099, Coma Ber, NGC752, M67, NGC2243, M34, M35, M41 and M48 is calculated by dividing the clusters into a number of rings, to find the star density in each ring. To find the tidal and core radius the King surface density profile $f(r)$  is fitted \citep{king1962} as follows\\
\begin{equation}\label{king.eq}
  f(r)=f_b+\frac{f_0}{1+(\frac{r}{R_C})^{2}},
\end{equation}
where $f_b$ is the background surface density, $f_0$ is the central peak
surface density and $R_C$ is the cluster's core radius. The tidal radius, $R_t$, is calculated as follows \citep{tidal}:
\begin{equation}\label{tidal.eq}
  R_t=R_C\sqrt{\frac{f_0}{3\sigma_b}-1},
\end{equation}
where $\sigma_b$ is surface density error.
Fitting of the King profile to the observed number density profile is achieved by adjusting the core and tidal radius of each cluster as three free parameters in our analysis. For comparison, the best-fitting model for each cluster is shown in Fig. \ref{stars distribotion.fig}. For other four clusters see supplementary\ref{a.stars distribotion.fig}. Our overall results and the values of core and tidal radius for ten open clusters are summarized in Table \ref{tab_discussion}.\\
Fig\ref{position from fields.fig} shows the 2D distribution of stars in each cluster. For other four clusters see supplementary\ref{a.position from fields.fig}. Stars outside the tidal radius (black points) and in cluster core(red points). For the case of Coma Ber, the obtained tidal radius of $6.61$ pc is  in agreement with \citet{Tang_2019_comaber} and \citet{Kraus_2007} who found $6.9$ pc  and $6.6$ pc, respectively. In addition, the tidal radius of 46 arcmin  that is obtained for the case of M38 is also in good agreement with what \citet{m38tidal} found for the density of contaminating field stars by CCD observation on stars that have magnitude $<18$ (i.e., $45$ arcmin). They did not find cluster core radius while we found 2.36 pc for cluster core radius. Regarding the case of NGC2099, we found $R_t=11.44$ and $R_c=1.96$ pc  that meet with $25.04$ and $4.29$ arcmin, for tidal and core radius respectively which is comparable with previous results, i.e., $R_t=18$ and $R_c=5.3$ arcmin obtained by  \citet{ngc2099tidal}. For the case of NGC752, the  tidal radius of $R_t=11.92$ pc in this work is larger than $9.52$ pc obtained by \citet{ngc752ti}. For NGC2243 9.24 pc and 1.20 pc are found for tidal and core radius respectively. 

\subsection{Evidence for mass segregation}\label{section.discus.MASS}

Star clusters can undergo significant changes not only at birth but also during the course of their dynamical evolutions. It is therefore essential to specify to what extent the present-day properties of a star cluster, e.g. their degree of mass segregation, are imprinted by early evolution and the formation processes, and to what extent they are the outcome of long-term dynamical evolution.\\
In a star cluster with different stellar masses, the more massive stars sink towards the central regions, while the lighter stars on average move further away from the center via relaxation processes. This so-called dynamical mass segregation is a natural outcome of energy equipartition driven by two-body encounters and happens on a time scale, which is of the order of several two-body relaxation times \citep{massegeration}.\\
However, a number of observational studies \citep{Hillenbrand1998, deGrijs2002, Stolte2006, Sheikhi2016} have found evidence of mass segregation in clusters with ages shorter than the time needed to produce the observed segregation via two-body relaxation.\\
It has been suggested that the observed mass segregation in young clusters could be primordial and possibly imprinted by the early star formation process \citep{Bonnell2002, Bonnell1998, Bonnell2006}.\\
Such mass segregation could be due to the higher accretion rate of protostars in high-density regions of molecular clouds fragmenting into clumps.\\
Moreover, there are some exceptional diffuse outer-halo globular clusters (e.g. Palomar 4, Palomar 14) with present-day half-mass relaxation times exceeding the Hubble time. Therefore, no dynamical mass segregation is expected in these clusters. \citet{Frank2012, Frank2014} have found clear evidence for mass segregation of main-sequence stars in Pal 4 and Pal 14. Because of the large two-body relaxation time scales of these clusters, this could be interpreted as evidence of PMS. \citet{Zonoozi2011, Zonoozi2014, Zonoozi2017} have presented a comprehensive set of N-body computations of Pal 14 and Pal 4 over a Hubble time, and compared the results to the observed parameters from \citet{Jordi2009} and \citet{Frank2012} and  showed that  a very high degree of PMS would be necessary to explain the observed characteristics.\\
To estimate the degree of mass segregation for each open cluster in our sample, we divided the luminosity of cluster stars into three groups ($L/L_{\odot}>1$, $1>L/L_{\odot}>0.1$, and $L/L_{\odot}<0.02$) and counted the number of stars in each group.\\
The radial cumulative density function calculated from the cluster center is shown  in Fig~\ref{Mass segregation.fig}. For other four clusters see supplementary\ref{a.Mass segregation.fig}. Since the faint stars with a magnitude higher than 20 are removed in our analysis, the low-mass stars in M38, NGC2099, and NGC2243 are lost.\\ 
\onecolumn
\begin{table}
\centering
\caption{Physical parameters of clusters in GDR2 (except for the case of  NGC752) for comparison this method from another methods}
\subcaption*{This work}
\begin{tabular}{c c c c c c c}
    \hline
    \hline
    Name & Parallax~(mas) & pmRA~(mas\,yr$^{-1}$) & pmDEC~(mas\,yr$^{-1}$) & D~(pc) & N & (Eps,MinPts)\\
    \hline
    M38$(G<18)$ & $0.86\pm0.06$ & $1.60\pm0.10$ & $-4.41\pm0.08$ & $1134.58$ & 701$(P>0.5)$ & $(0.07,180)$ \\
    NGC2099$(G<20)$ & $0.66\pm0.08$ & $1.93\pm0.15$ & $-5.63\pm0.12$ & $1480.98$ & 1534$(P>0.6)$ & $(0.08,135)$ \\
    Coma Ber$(G<18)$ & $11.63\pm0.08$ & $-12.09\pm0.12$ & $-8.97\pm0.10$ & $85.80$ & 98$(P>0.5)$ & $(0.3,20)$ \\
    NGC752$(G<20)$ & $2.26\pm0.06$ & $9.79\pm0.06$ & $-11.82\pm0.07$ & $437.33$ & 286 $(P>0.6)$ & $(0.15,80)$ \\
    M67$(G<20)$ & $1.14\pm0.12$ & $-10.98\pm0.21$ & $-2.94\pm0.14$ & $857.43$ & 1614$(P>0.6)$ & $(0.12,128)$ \\
    M67 & $1.14\pm0.12$ & $-10.98\pm0.20$ & $-2.94\pm0.14$ & $857.03$ & 1590$(P>0.8)$ & $(0.12,128)$ \\
    M67$(G<18)$ & $1.13\pm0.06$ & $-10.97\pm0.10$ & $-2.95\pm0.07$ & $863.74$ & 1184$(P>0.5)$ & $(0.12,128)$ \\
    NGC2243$(G<20)$ & $0.22\pm0.09$ & $-1.27\pm0.15$ & $5.48\pm0.017$ & $3316.10$ & 717$(P>0.6)$ & $(0.05,158)$ \\
    Alessi01$(G<18)$ &$1.39\pm0.05$ & $6.49\pm0.08$ & $-6.30\pm0.10$ & $705.14$ & 41$(P>0.5)$ & $(0.09,49)$\\
    Bochum04$(G<18)$ & $0.75\pm0.05$ & $-2.98\pm0.07$ & $1.84\pm0.06$ & $1281.63$ & 54$(P>0.5)$ & $(0.06,40)$ \\
    M34$(G<18)$ & $1.95\pm0.10$ & $0.72\pm0.16$ & $-5.70\pm0.13$ & $507.43$ & 498$(P>0.5)$ & $(0.2,260)$ \\
    M35$(G<18)$ & $1.13\pm0.06$ & $2.28\pm0.11$ & $-2.90\pm0.09$ & $871.60$ & 1094$(P>0.5)$ & $(0.11,220)$ \\
    M41$(G<18)$ & $1.35\pm0.04$ & $-4.34\pm0.06$ & $-1.38\pm0.07$ & $726.19$ & 575$(P>0.5)$ & $(0.2,150)$ \\
    M48$(G<18)$ & $1.28\pm0.05$ & $-1.30\pm0.08$ & $1.01\pm0.05$ & 764.05 & 329$(P>0.5)$ & $(0.16,35)$ \\
    \hline
  \end{tabular}
\bigskip
\subcaption*{Other works. [1]:~\cite{Cantat-Gaudin2018}. [2]:~\citet{ML/mnras/stab118}. [3]:~\cite{Gao_spectral_clustering}. [4]:~\citet{ngc752ti}. [5]:~\citet{Gao_m67_gmm_rf}.}
\begin{tabular}{c c c c c c}
    \hline
    \hline
    Name & Parallax(mas) & pmRA(mas\,yr$^{-1}$) & pmDEC(mas\,yr$^{-1}$) & D~(pc) & N\\
    \hline
    M38$(G<18)$~[1] & $0.874$ & $1.580$ & $-4.424$ & 1107.4 & 474$(P>0.5)$ \\
    NGC2099$(G<20)$~[2] & $0.663\pm0.004$ & $ 1.928\pm0.004$ & $-5.636\pm0.004$ & $1466.9$ & 1640$(P>0.6)$\\
    Coma Ber$(G<18)$~[3] &$11.73\pm0.6$&$-12.2\pm0.1$&$-9.1\pm0.1$&85.2&40\\
    NGC752$(G<20)$~[4] &$2.229\pm0.005$&$9.825\pm0.018$&$-11.724\pm0.02$&$443.8$ & 282$(P>0.6)$\\
    M67$(G<20)$~[2] &$1.135\pm0.002$ & $-10.981\pm0.006$ & $-2.949\pm0.006$ & $860.7$ & 1194$(P>0.6)$\\
    M67~[5] &$1.1327\pm0.0018$ &$-10.9378\pm0.0078$ &$-2.9465\pm0.0074$& $883$ & 1361$(P>0.8)$\\
    M67$(G<18)$~[1] &$1.135$ &$-10.984$ &$-2.964$& $859.1$ & 691$(P>0.5)$\\
    NGC2243$(G<20)$~[2] &$0.213\pm0.003$&$-1.285\pm0.005$&$5.489\pm0.006$&$3606$ & 583$(P>0.6)$\\
    Alessi01$(G<18)$~[1] & $1.390$ & $6.536$ & $-6.245$ & $704.8$ & 48$(P>0.5)$ \\
    Bochum04$(G<18)$~[1] & $0.761$ & $-3.014$ & $1.848$ & $1266.4$ & 27$(P>0.5)$ \\
    M34$(G<18)$~[1] & $1.94$ & $0.72$ & $-5.68$ & 505.5 & 578$(P>0.5)$ \\
    M35$(G<18)$~[1] & $1.13$ & $2.30$ & $-2.90$ & 862.4 & 1325$(P>0.5)$ \\
    M41$(G<18)$~[1] & $1.36$ & $-4.33$ & $-1.38$ & 720.1 & 645$(P>0.5)$ \\
    M48$(G<18)$~[1] & $1.28$ & $-1.31$ & $1.02$ & 758.8 & 479$(P>0.5)$ \\
    \hline
  \end{tabular}
  \label{gdr2.tab}
\end{table}
  
\begin{table}
\caption{The distribution of stars in clusters\\
(1): cluster name . (2): The radius that we found cluster members candidate. (3): Tidal radius. (4): core radius. (5): Number of stars outside the tidal radius. (6): Number of stars distributed inside the core radius. (7): The redding values.} [1]:~\cite{m38tidal}, [2]:~\cite{ngc2099age:400}, [3]:~\cite{comaberreding} and \cite{Tang_2018_comaber}, [4]:~\cite{ngc752redding}, [5]:~\cite{Taylor_2006_m67} and \cite{Onehag_m67_4.5}, [6]:~\cite{Anthony_Twarog_2005}, [7]:~\cite{m34_age220}, [8]:~\cite{m35av}, [9]:~\cite{m41av}, [10]:~\cite{m48av}
  \centering
  \begin{tabular}{ccccccc}
    \hline
    \hline
    (1) & (2) & (3) & (4) & (5) & (6)\\
    Name & $R[pc]$ & $R_t[pc]$ & $R_c[pc]$ & $N_{t}$ &  $N_{c}$ & $E(B-V)$\\
    \hline
    M38 & $12.29\pm0.13$ & $14.83\pm0.50$ & $3.04\pm0.24$ & $0$ & $255$ & $0.25$~[1]  \\
    NGC2099 & $12.69\pm0.09$ & $11.44\pm0.76$ & $1.96\pm0.09$ & $16$ & $373$ & $0.3$~[2]\\
    Coma Ber & $10.88\pm0.07$ & $6.61\pm1.12$ & $2.03\pm0.25$ & $31$ & $22$ & $0.006$~[3]\\
    NGC752 & $22.49\pm0.18$ & $11.92\pm0.59$ & $2.21\pm0.15$ & $30$ & $72$ & $0.035$~[4]\\
    M67 & $14.73\pm0.02$ & $9.53\pm0.17$ & $0.95\pm0.02$ & $121$ & $147$ & $0.041$~[5]\\
    NGC2243 & $16.34\pm0.02$ & $10.17\pm0.49$ & $1.16\pm0.04$ & $105$ & $127$ & $0.055$~[6]\\
    M34 & $12.64\pm0.09$ & $8.14\pm0.44$ & $1.97\pm0.16$ & $50$ & $191$ & $0.07$~[7]\\
    M35 & $9.19\pm0.10$ & $8.00\pm0.44$ & $2.41\pm0.20$ & $27$ & $348$ & $0.22$~[8]\\
    M41 & $12.20\pm0.09$ & $10.91\pm0.54$ & $2.57\pm0.21$ & $7$ & $232$ & $0.03$~[9]\\
    M48 & $11.22$ & $8.32\pm0.92$ & $2.37\pm0.41$ & $40$ & $109$ & $0.08$~[10]\\
    \hline
  \end{tabular}
  \label{tab_discussion}
\end{table}
 
\onecolumn
\begin{figure}[h!]
  \centering
  \captionsetup[subfigure]{labelformat=empty}
        \begin{subfigure}{0.44\textwidth}
        \centering

                \includegraphics[width=\textwidth]{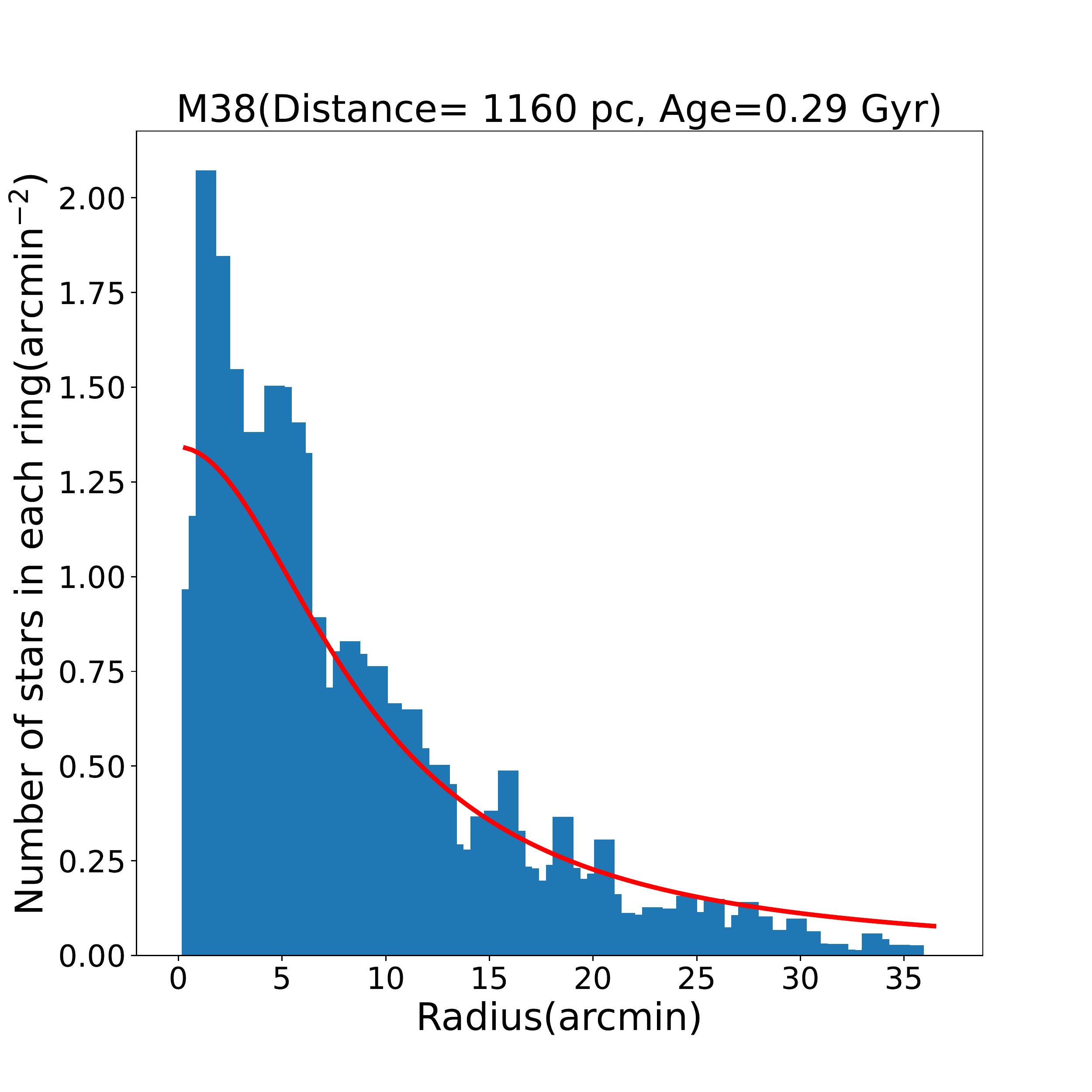}
        \end{subfigure}
        \begin{subfigure}{0.44\textwidth}

                \centering
                \includegraphics[width=\textwidth]{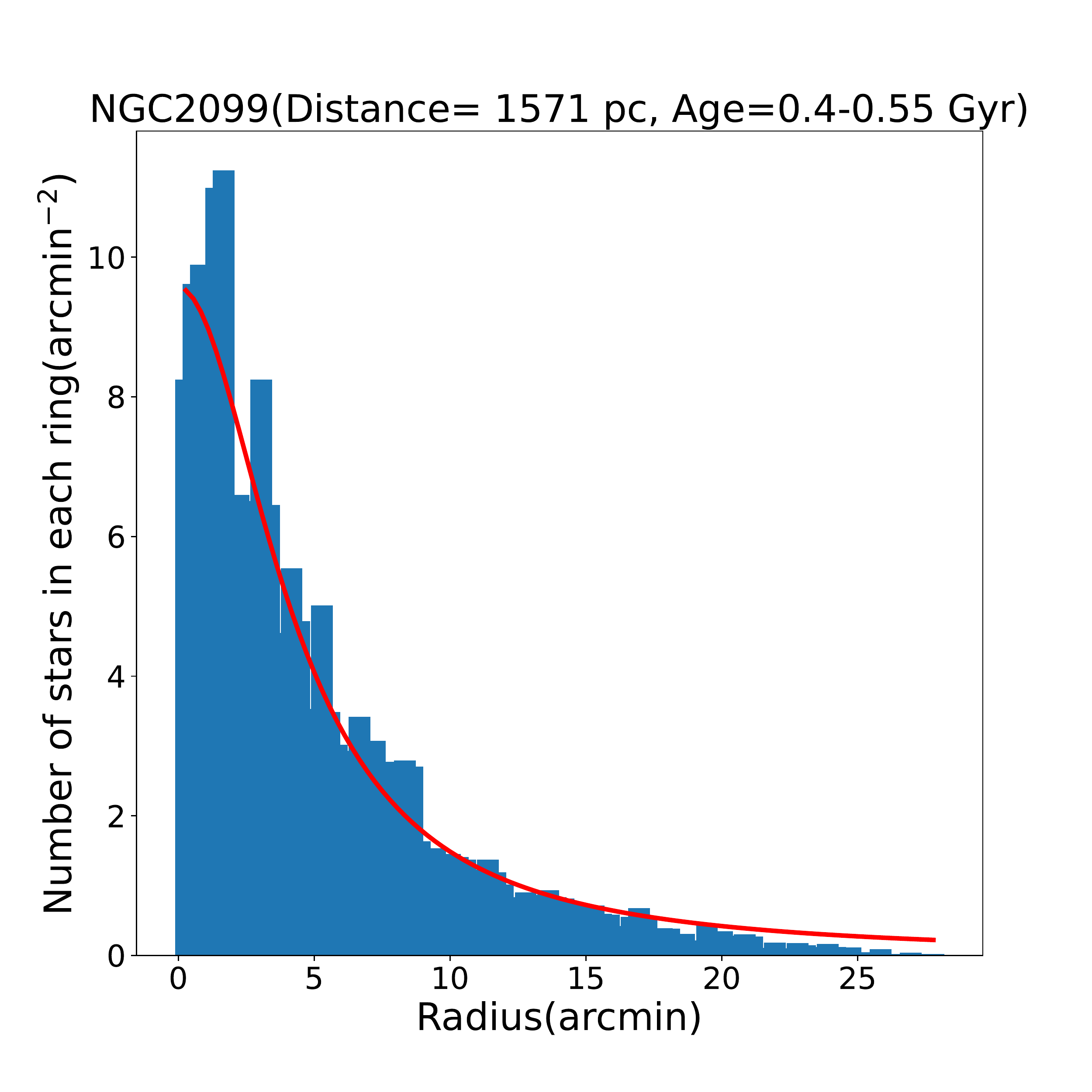}

        \end{subfigure}
        \begin{subfigure}{0.44\textwidth}
                \centering

                \includegraphics[width=\textwidth]{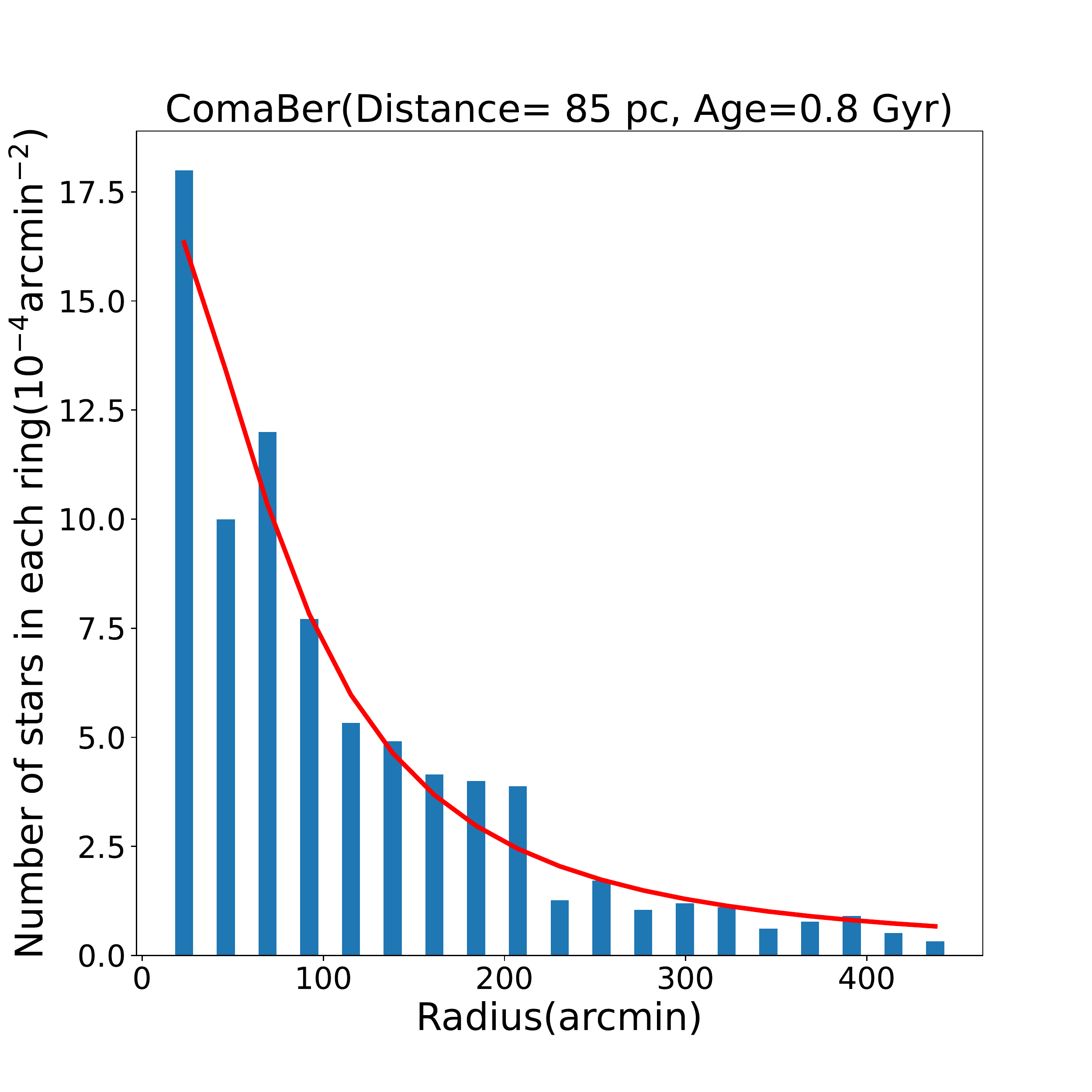}
        \end{subfigure}
        \begin{subfigure}{0.44\textwidth}
                \centering

                \includegraphics[width=\textwidth]{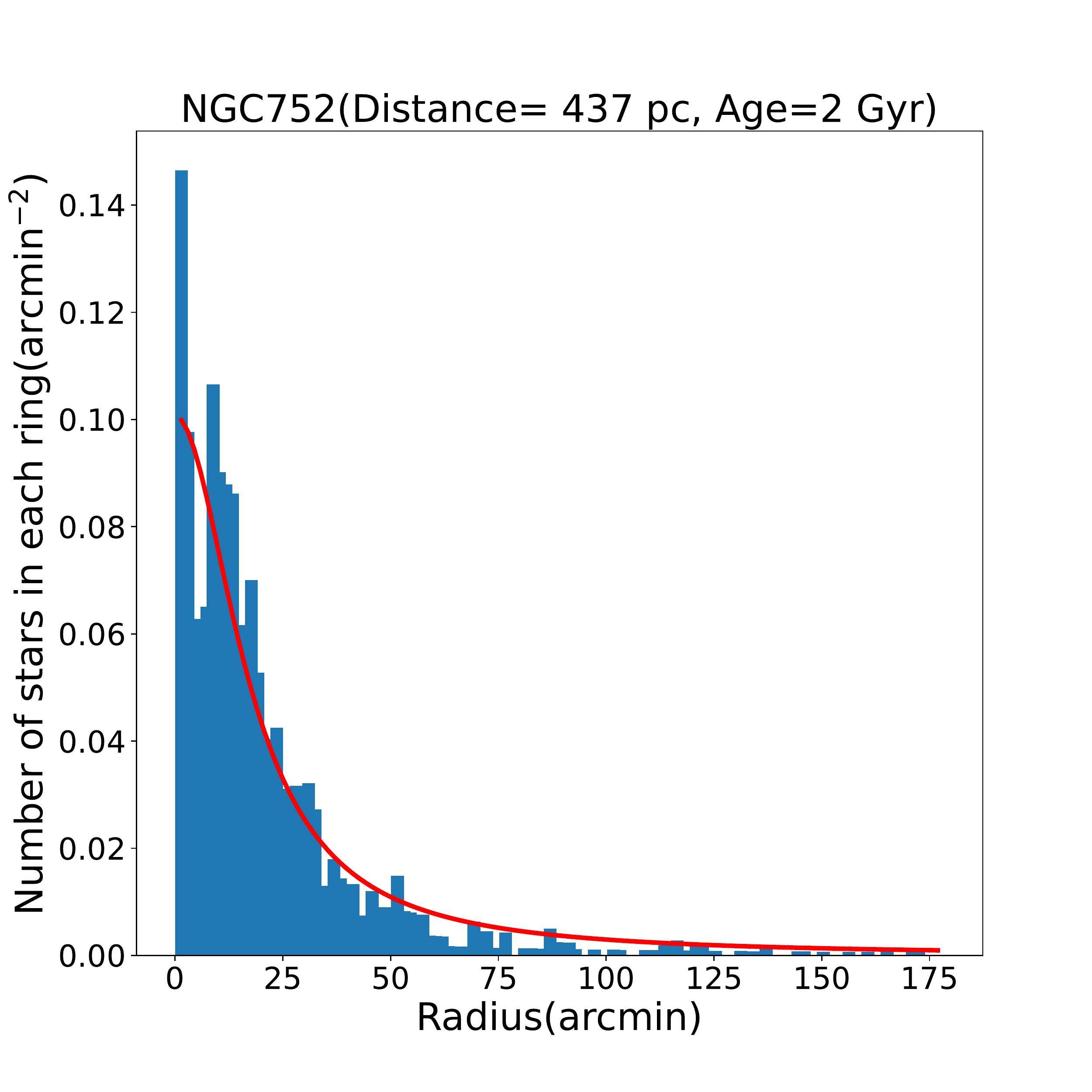}

        \end{subfigure}
        \begin{subfigure}{0.44\textwidth}
                \centering

                \includegraphics[width=\textwidth]{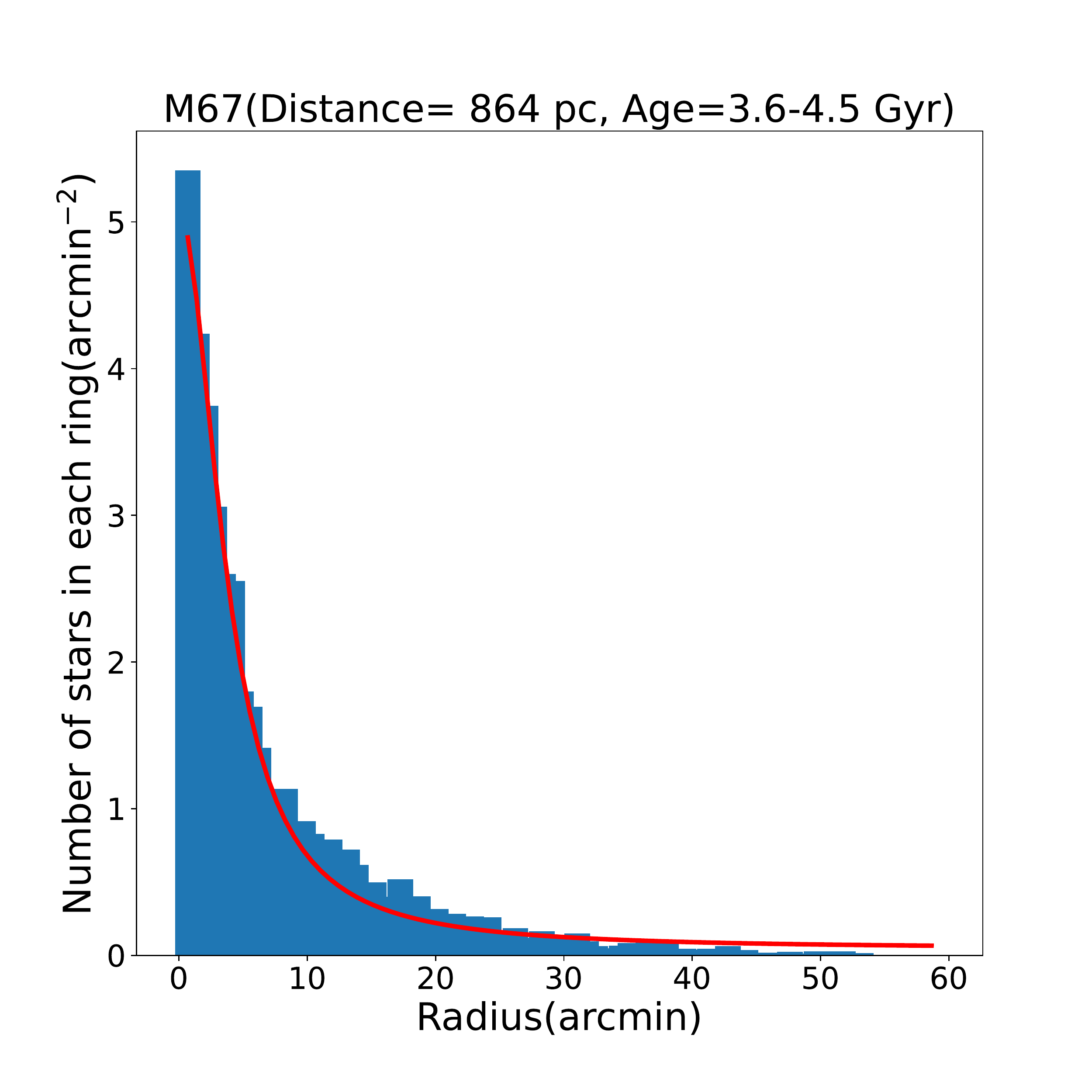}

        \end{subfigure}
        \begin{subfigure}{0.44\textwidth}
                \centering

                \includegraphics[width=\textwidth]{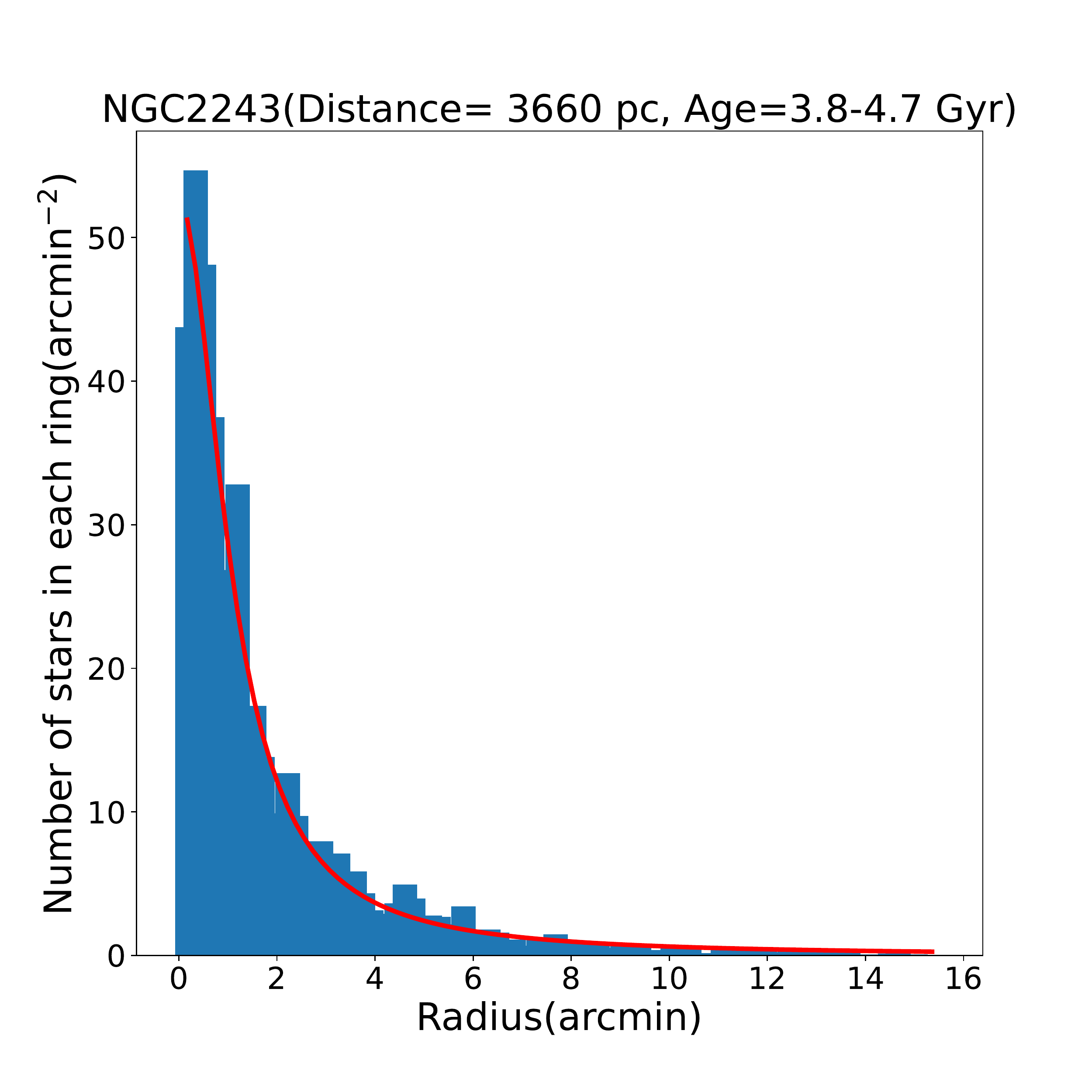}

        \end{subfigure}
  \caption{King profile fit on the distribution of stars in the cluster, the vertical axis is the density of stars number and the horizontal axis is radius.}
  \label{stars distribotion.fig}
\end{figure}
\begin{figure}[h!]
  \centering
  \captionsetup[subfigure]{labelformat=empty}
        \begin{subfigure}{0.44\textwidth}
        \centering

                \includegraphics[width=\textwidth]{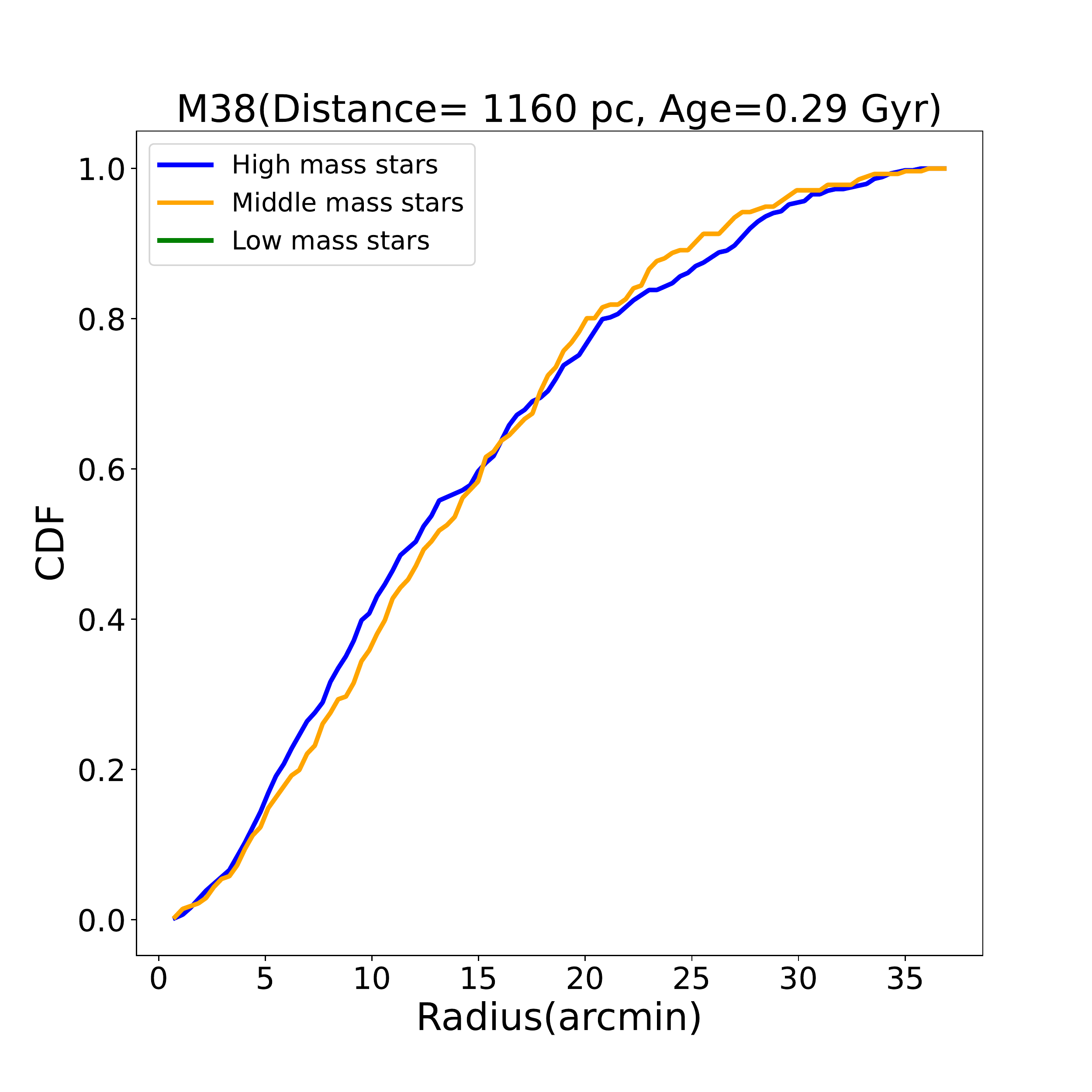}
        \end{subfigure}
        \begin{subfigure}{0.44\textwidth}

                \centering
                \includegraphics[width=\textwidth]{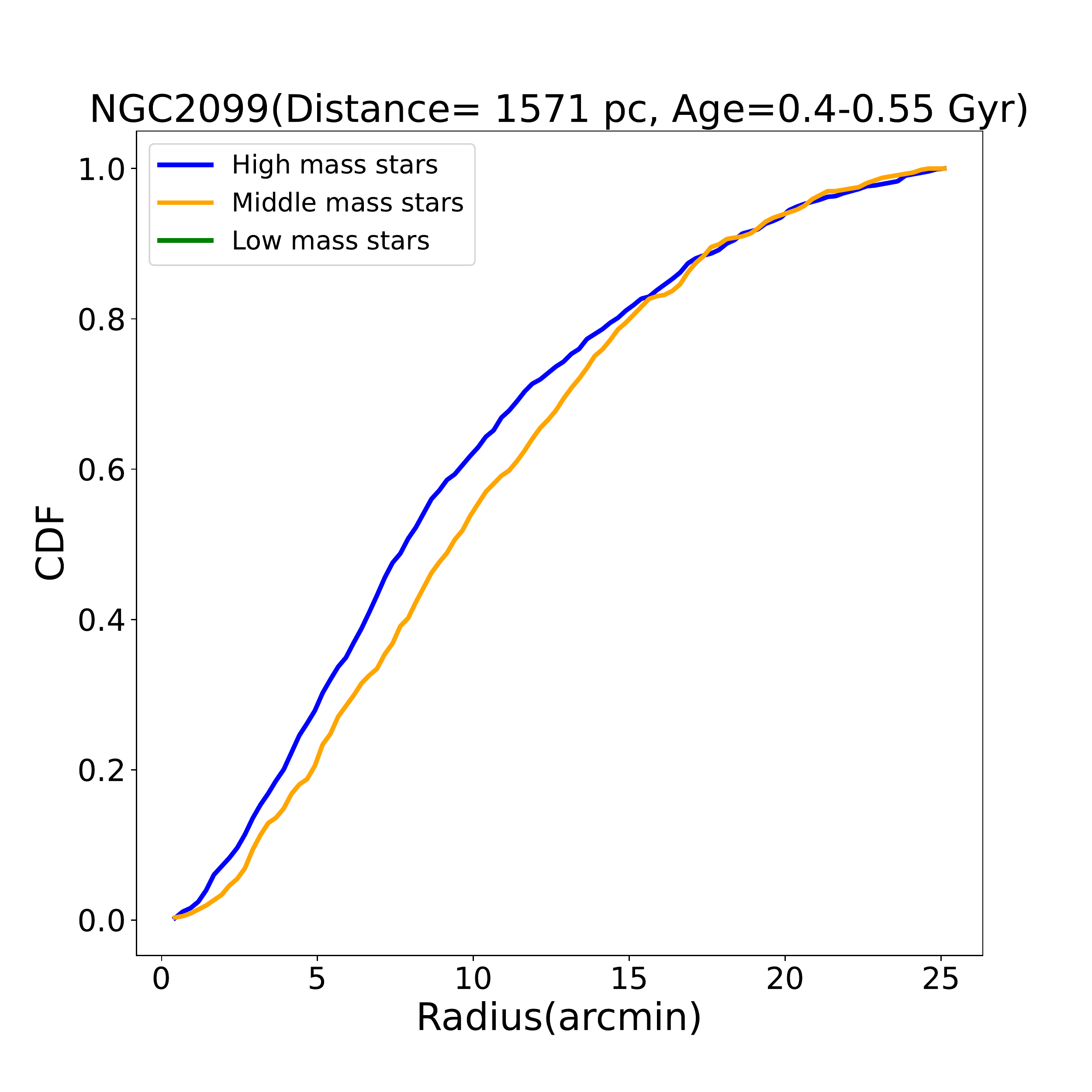}

        \end{subfigure}
        \begin{subfigure}{0.44\textwidth}
                \centering

                \includegraphics[width=\textwidth]{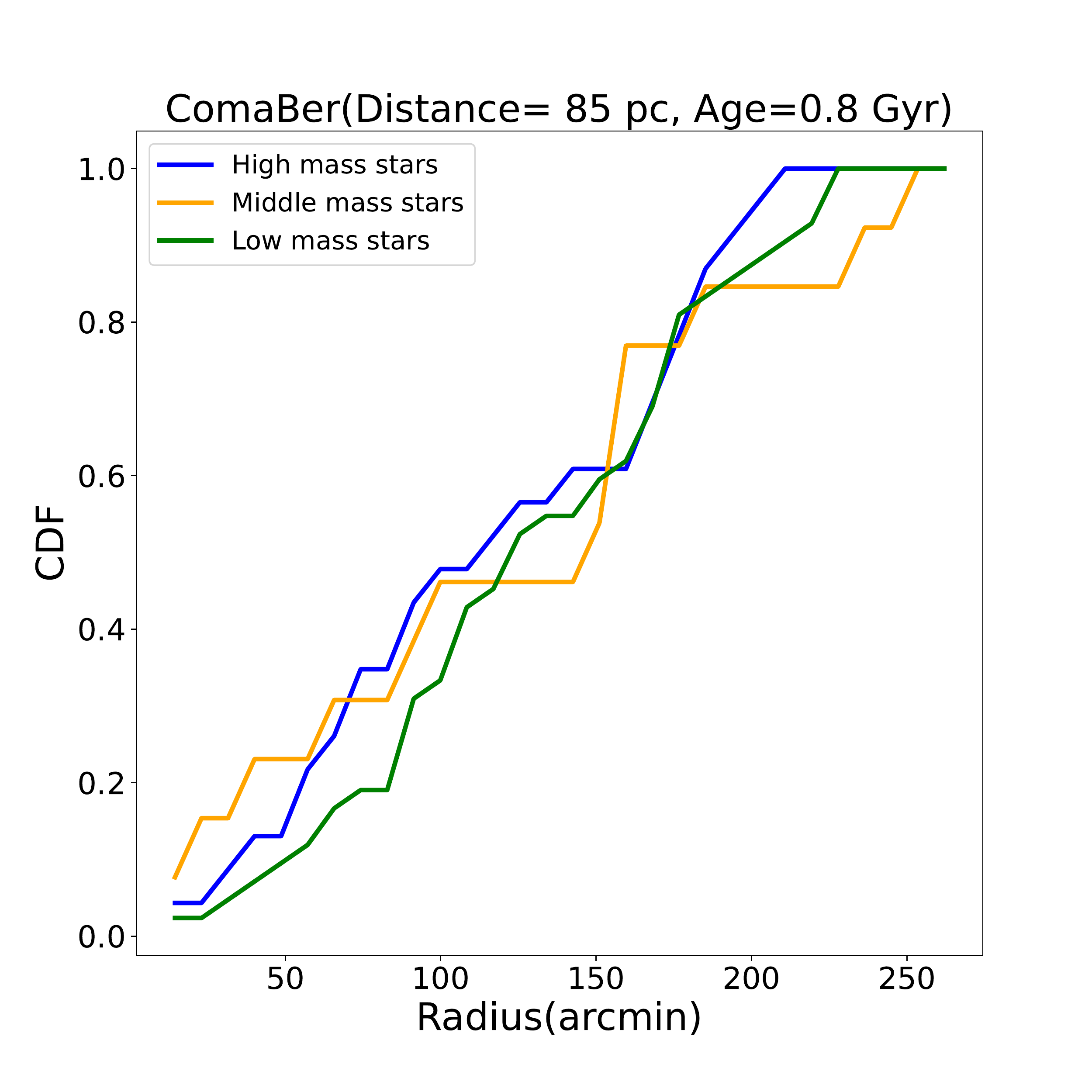}
        \end{subfigure}
        \begin{subfigure}{0.44\textwidth}
                \centering

                \includegraphics[width=\textwidth]{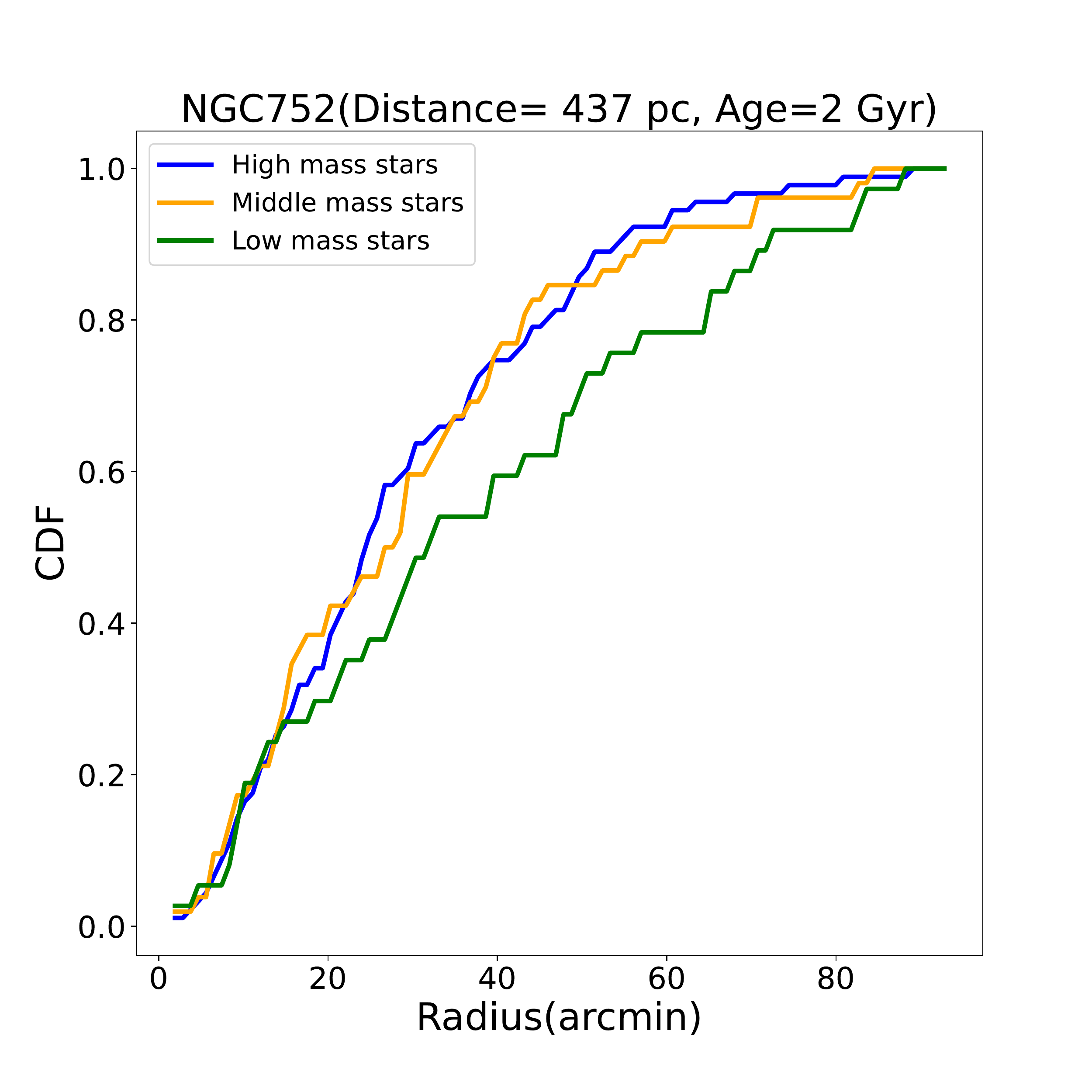}

        \end{subfigure}
        \begin{subfigure}{0.44\textwidth}
                \centering

                \includegraphics[width=\textwidth]{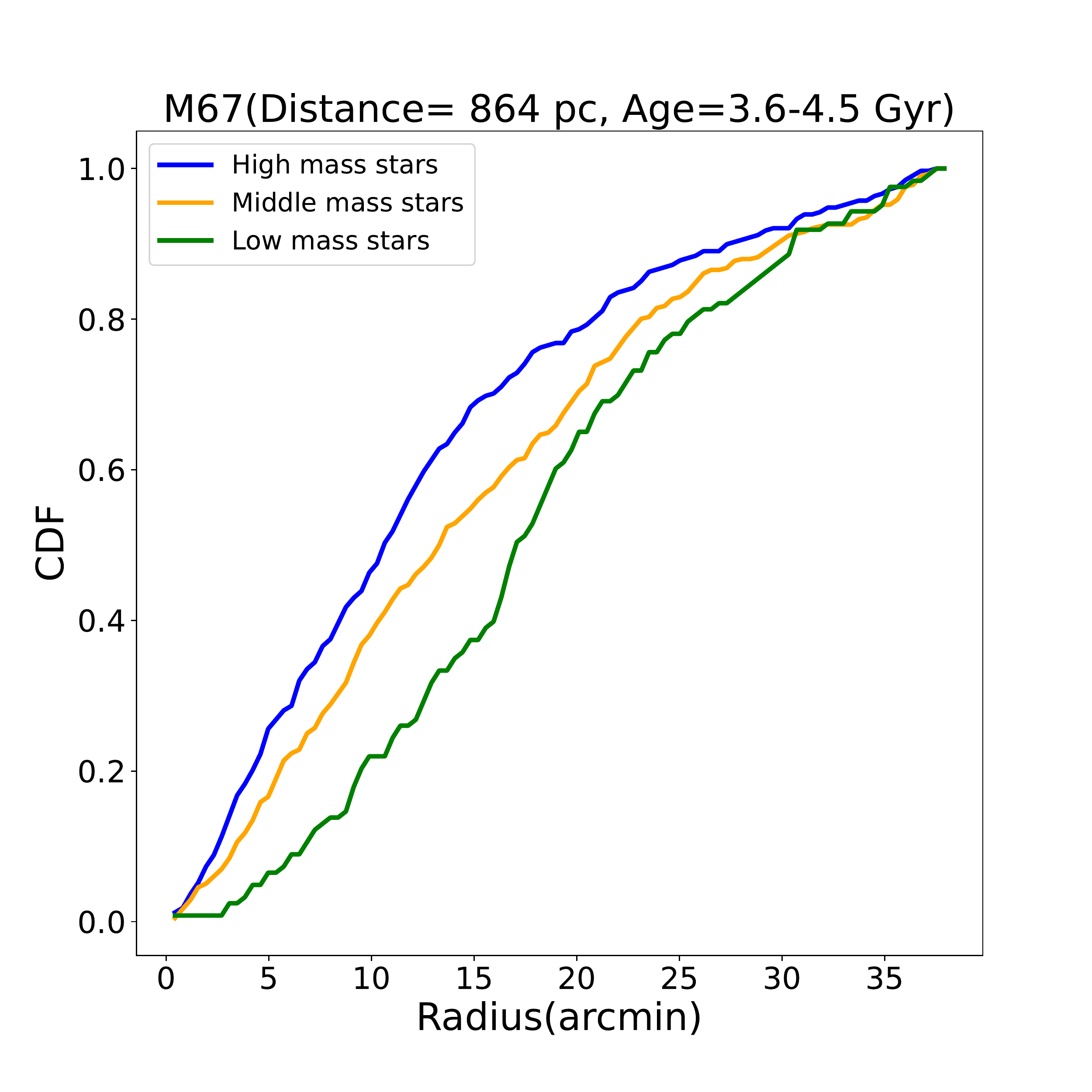}

        \end{subfigure}
        \begin{subfigure}{0.44\textwidth}
                \centering

                \includegraphics[width=\textwidth]{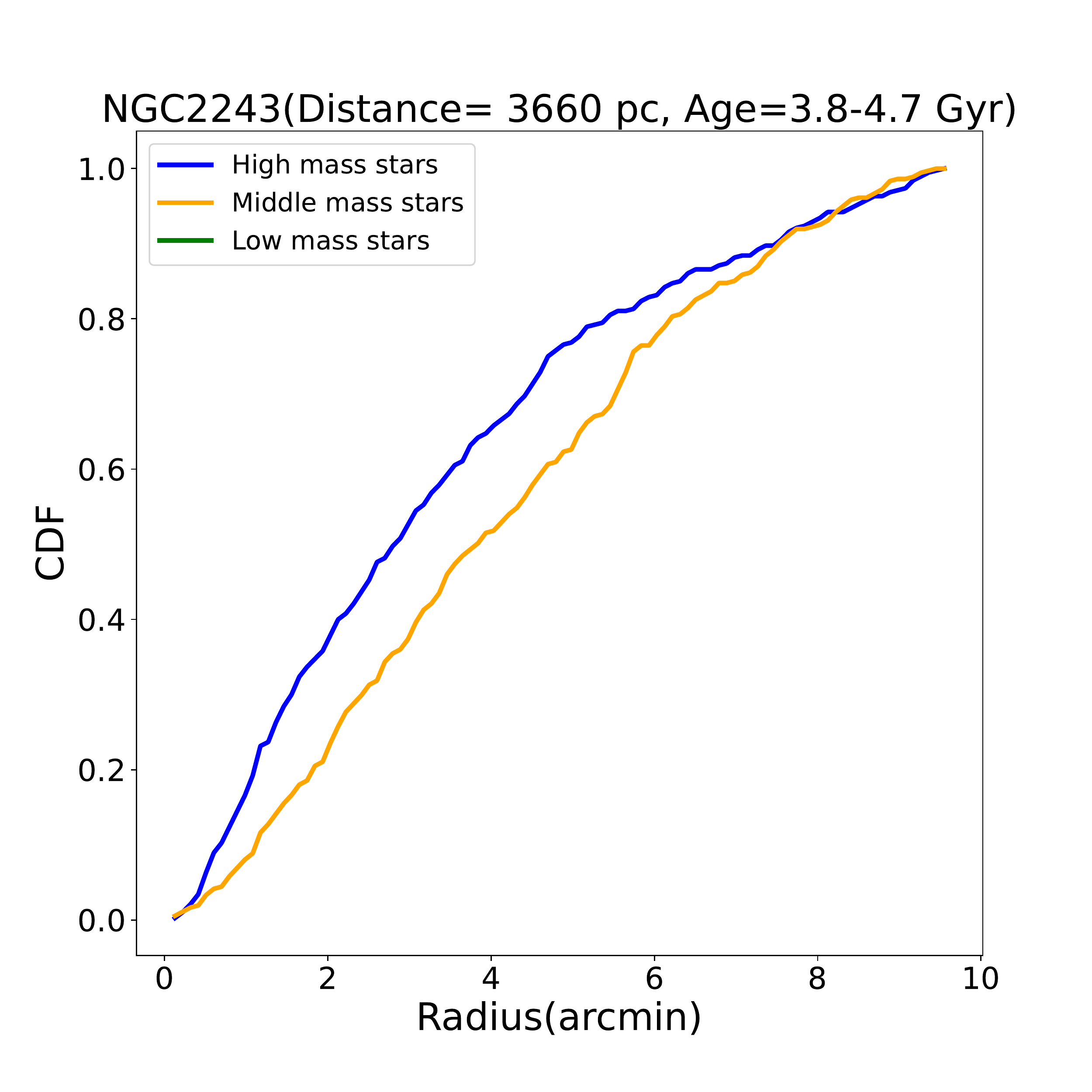}

        \end{subfigure}
  \caption{Radial cumulative distribution function, blue line shows massive stars, orange line shows middle mass stars and green line shows low mass stars.}
  \label{Mass segregation.fig}
\end{figure}
\newpage
\twocolumn
\noindent 
As shown in Fig~\ref{Mass segregation.fig}, M38 does not show clear evidence of mass segregation that could be due to its younger age.
On the other hand, in older clusters such as Coma Ber, NGC752 and M67 with the age of 0.8-3.5 Gyr we found clear indications of mass segregation as the CDF of low-mass stars differs from the CDF of high-mass stars.\\
Also, we found mass segregation in NGC2243, M67, and NGC2099 as the CDF of middle-mass stars differs from the CDF of high-mass stars, in NGC2243 and M67 more than NGC2099 based on their age. 
Since we can see that the older clusters are more segregated in mass, then its possible origin could  be dynamic evolution and not primordial mass segregation.\\

\section{CONCLUSION}\label{con.Data}

We used a proper combination of two unsupervised machine learning methods, DBSCAN, and GMM, which is a powerful tool to identify the most probable members of twelve open clusters M38, NGC2099, Coma Ber, NGC752, M67, NGC2243, Alessi01, Bochum04, M34, M35, M41, and M48.\\
For Comparison to other investigations based on machine learning algorithms and Gaia data, we applied our method in Gaia DR2. In addition, we found that more stars are seen outside the tidal radius for older clusters.\\
We also obtained the physical parameters of clusters such as the core and tidal radii and showed that the results are in good agreement with the previous works. In addition, we checked the distribution of low- and high-mass stars as a function of distance from the cluster center and found a clear indication of mass segregation in some clusters of our sample. Finally, by checking the CMD of the clusters,
we were able to identify some candidates for blue stragglers, white dwarfs, and RGB stars in some clusters.\\
Having the present-day structural properties for these clusters such as core and tidal radius and mass segregation, what remains to be seen is what this means for how the clusters started. Are all open clusters formed with the same initial conditions or do these initial conditions also depend on their birth environment? Did the clusters start mass segregated or are they present-day observations compatible with no primordial mass segregation? Determining the starting conditions at birth under which these clusters are formed and evolved can be possible by means of the N-body simulations.\\
\section*{DATA AVAILABILITY}
The data used in this work are available at \url{https://gea.esac.esa.int/archive/} and we are ready to share our data in this work in each section on request.

\bibliographystyle{mnras}
\bibliography{ref}

\onecolumn
\begin{figure}
\ContinuedFloat*
\centering
\captionsetup[subfigure]{labelformat=empty}
\begin{subfigure}{0.44\textwidth}
        \centering
           \includegraphics[width=\textwidth]{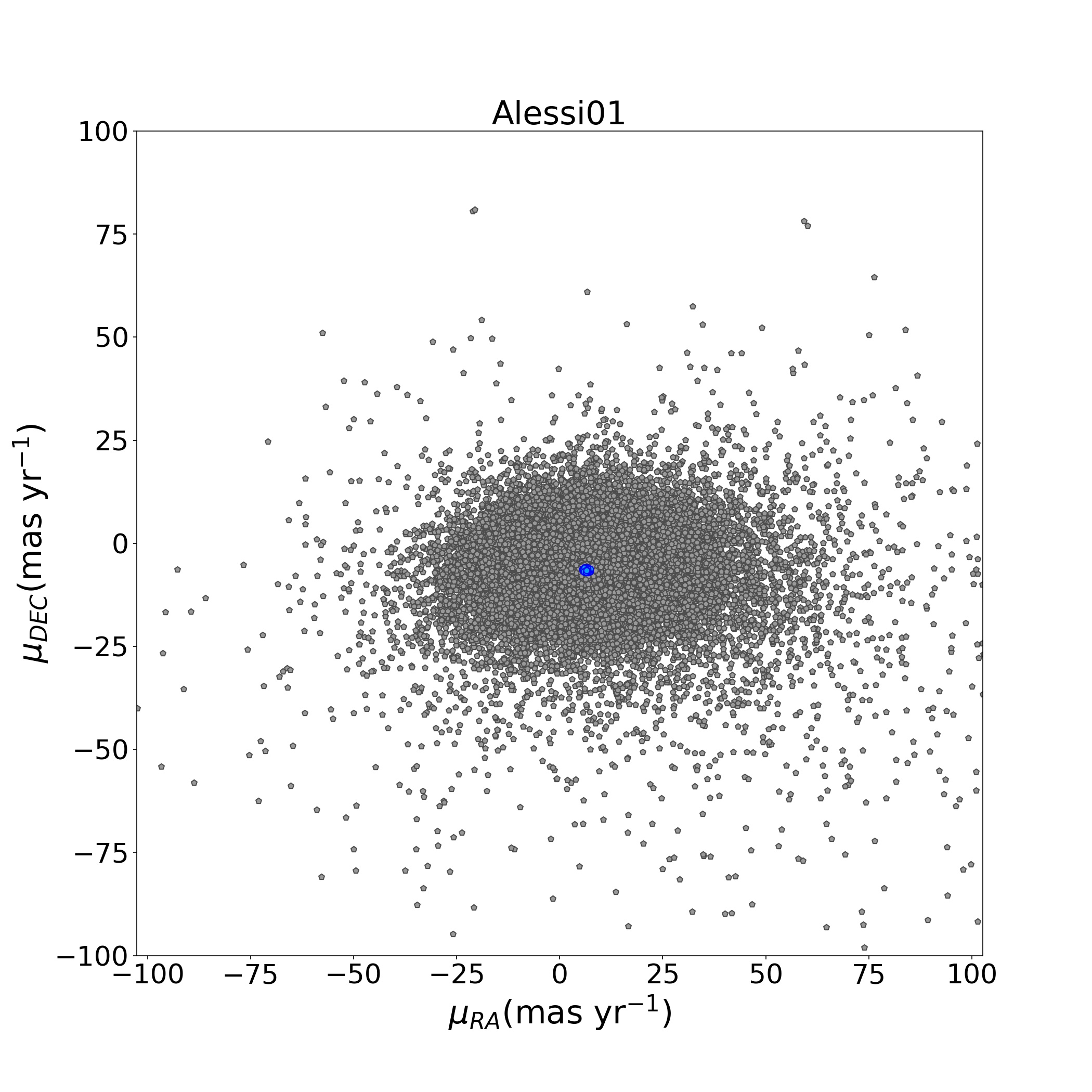}

        \end{subfigure}
        \begin{subfigure}{0.44\textwidth}

                \centering
                \includegraphics[width=\textwidth]{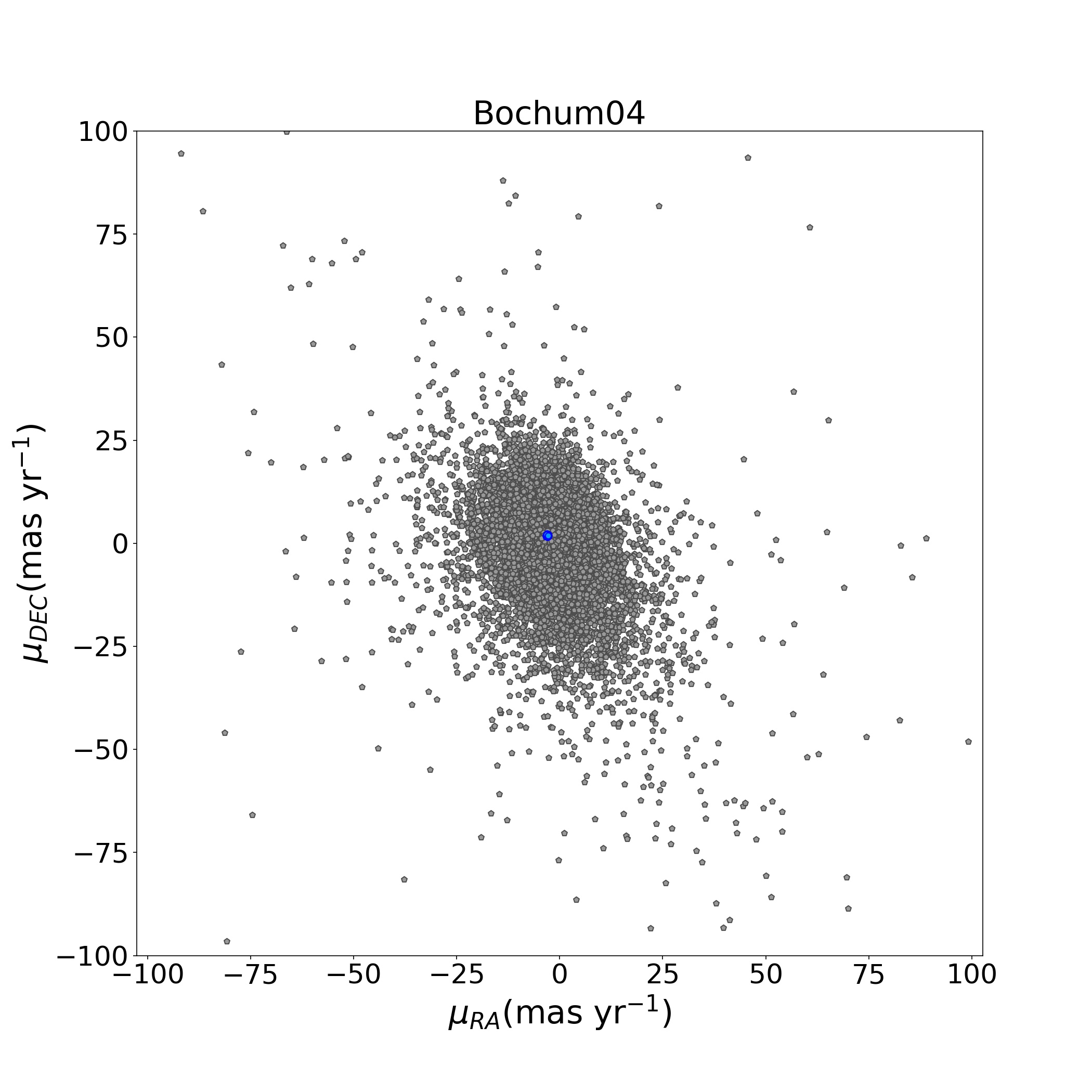}

        \end{subfigure}
        \begin{subfigure}{0.44\textwidth}
                \centering
           \includegraphics[width=\textwidth]{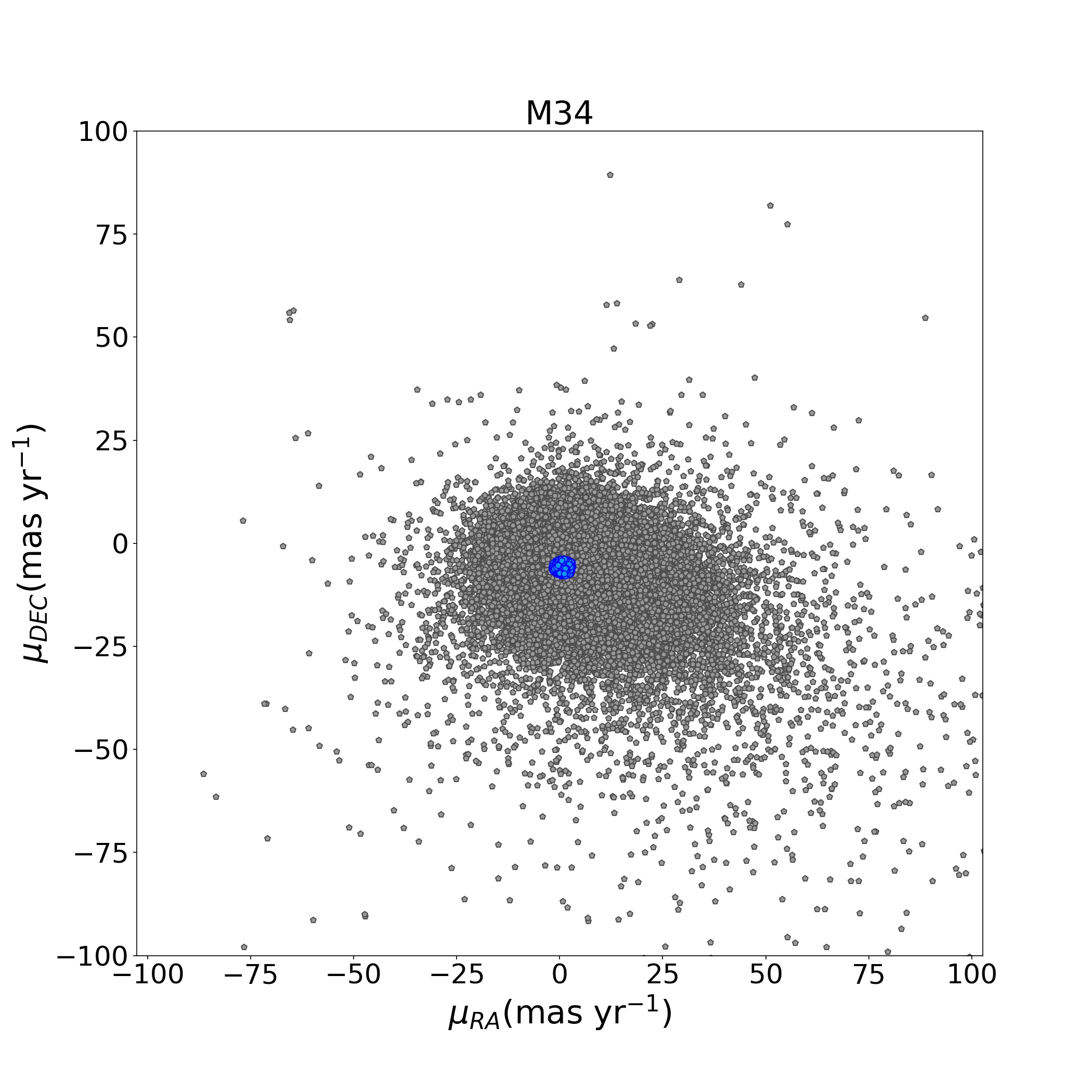}

        \end{subfigure}
        \begin{subfigure}{0.44\textwidth}
                \centering

                \includegraphics[width=\textwidth]{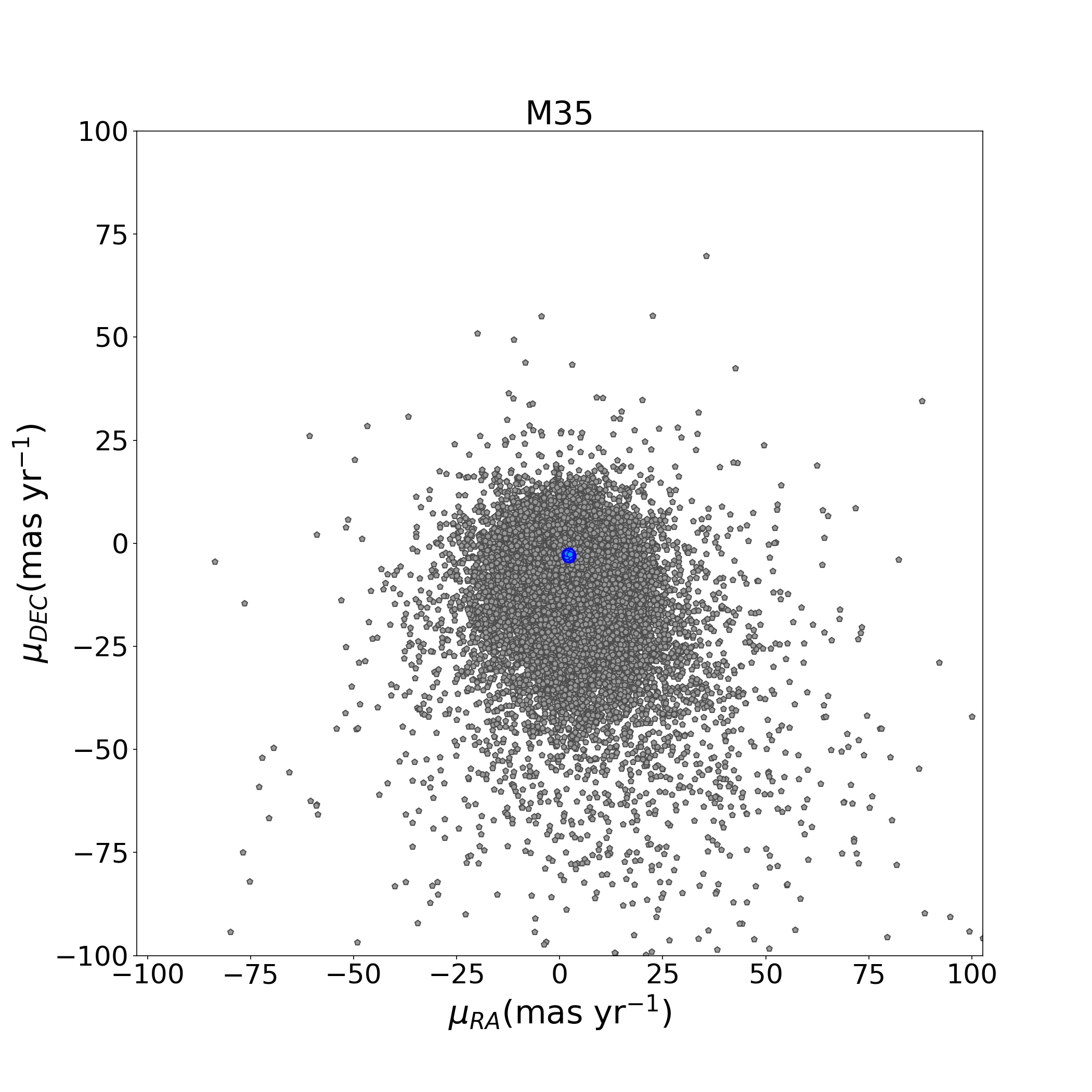}

        \end{subfigure}
        \begin{subfigure}{0.44\textwidth}
                \centering

                \includegraphics[width=\textwidth]{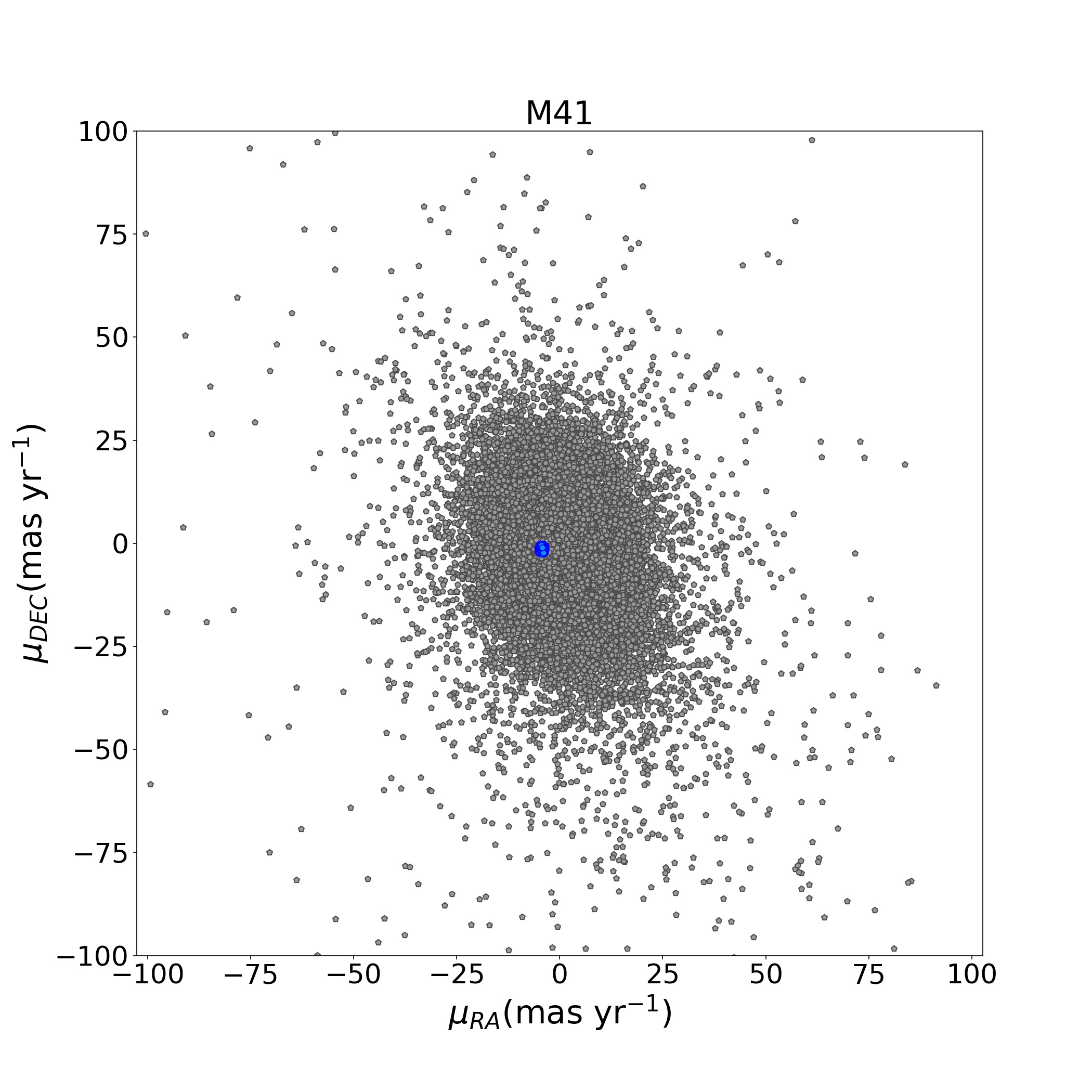}

        \end{subfigure}
        \begin{subfigure}{0.44\textwidth}
                \centering

                \includegraphics[width=\textwidth]{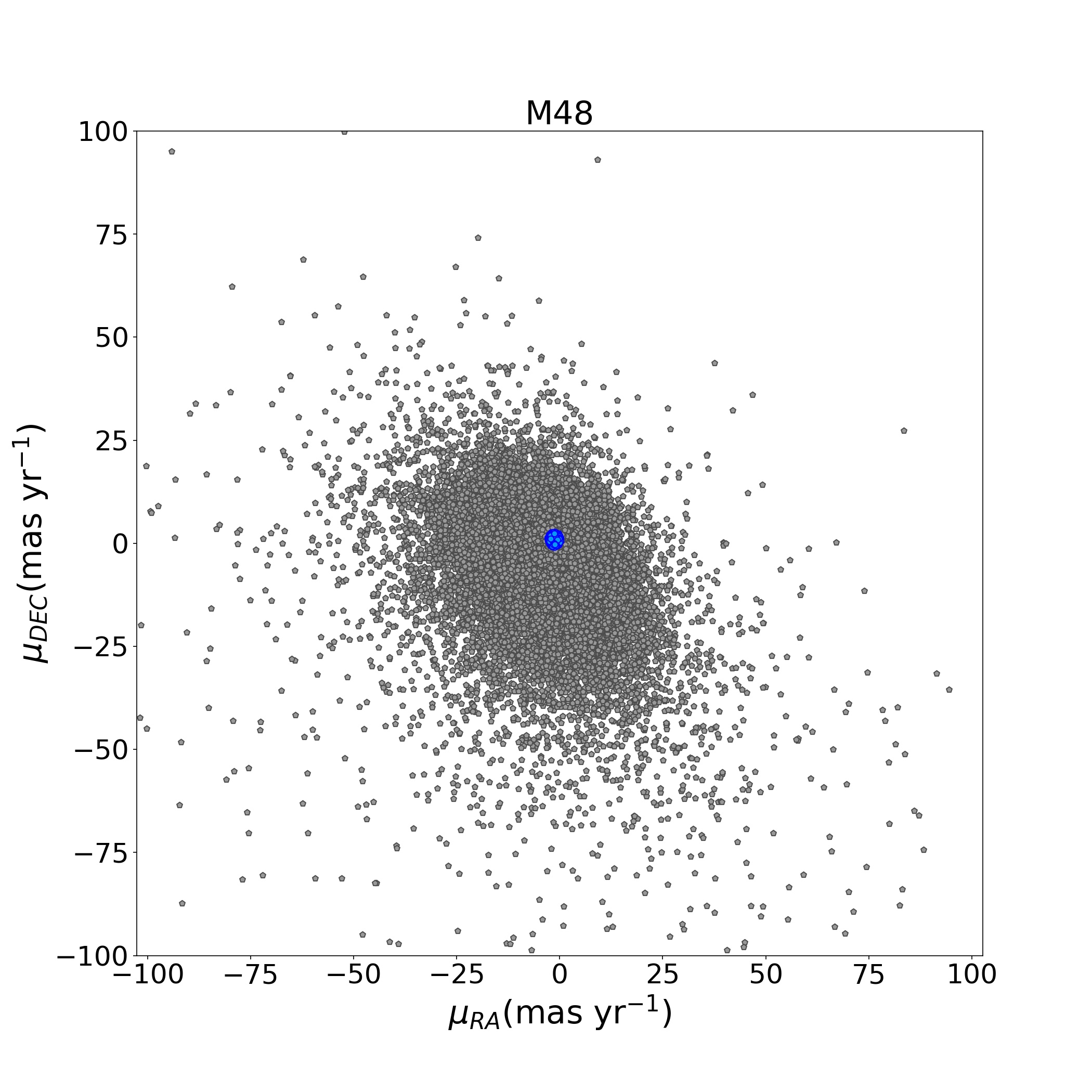}

        \end{subfigure}

  \caption{Proper motion of the cluster member star candidates from the field stars by DBSCAN, vertical axis is $pmRA$ and the horizontal axis is $pmDEC$, grey dots show the field stars and blue dots show stars that were selected by DBSCAN.}
  \label{a.proper motion of dbscan.fig}
\end{figure}

\begin{figure}
\ContinuedFloat*
  \centering
  \captionsetup[subfigure]{labelformat=empty}
        \begin{subfigure}{0.43\textwidth}
        \centering

                \includegraphics[width=\textwidth]{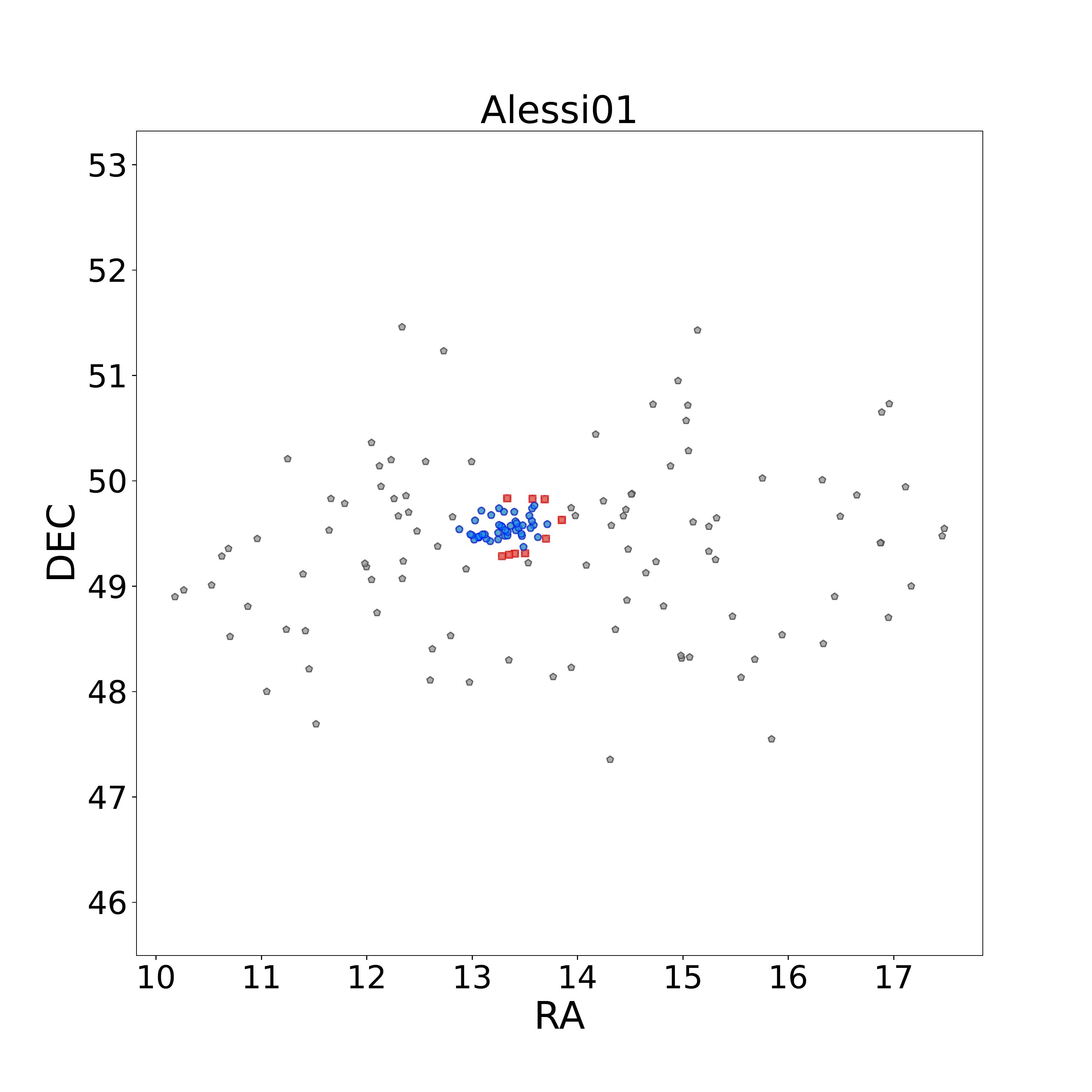}
        \end{subfigure}
        \begin{subfigure}{0.43\textwidth}

                \centering
                \includegraphics[width=\textwidth]{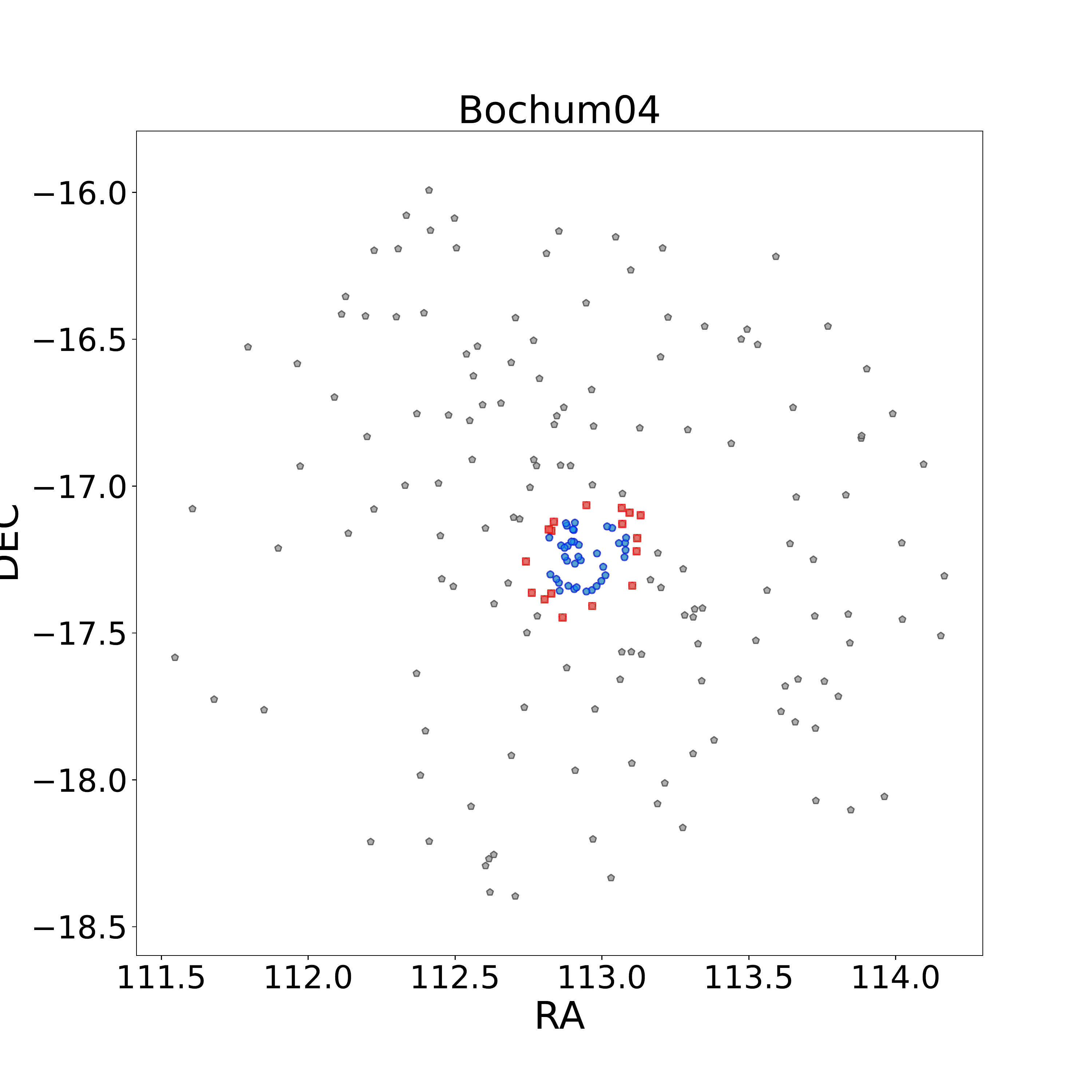}

        \end{subfigure}

  \begin{subfigure}{0.43\textwidth}
                \centering

                \includegraphics[width=\textwidth]{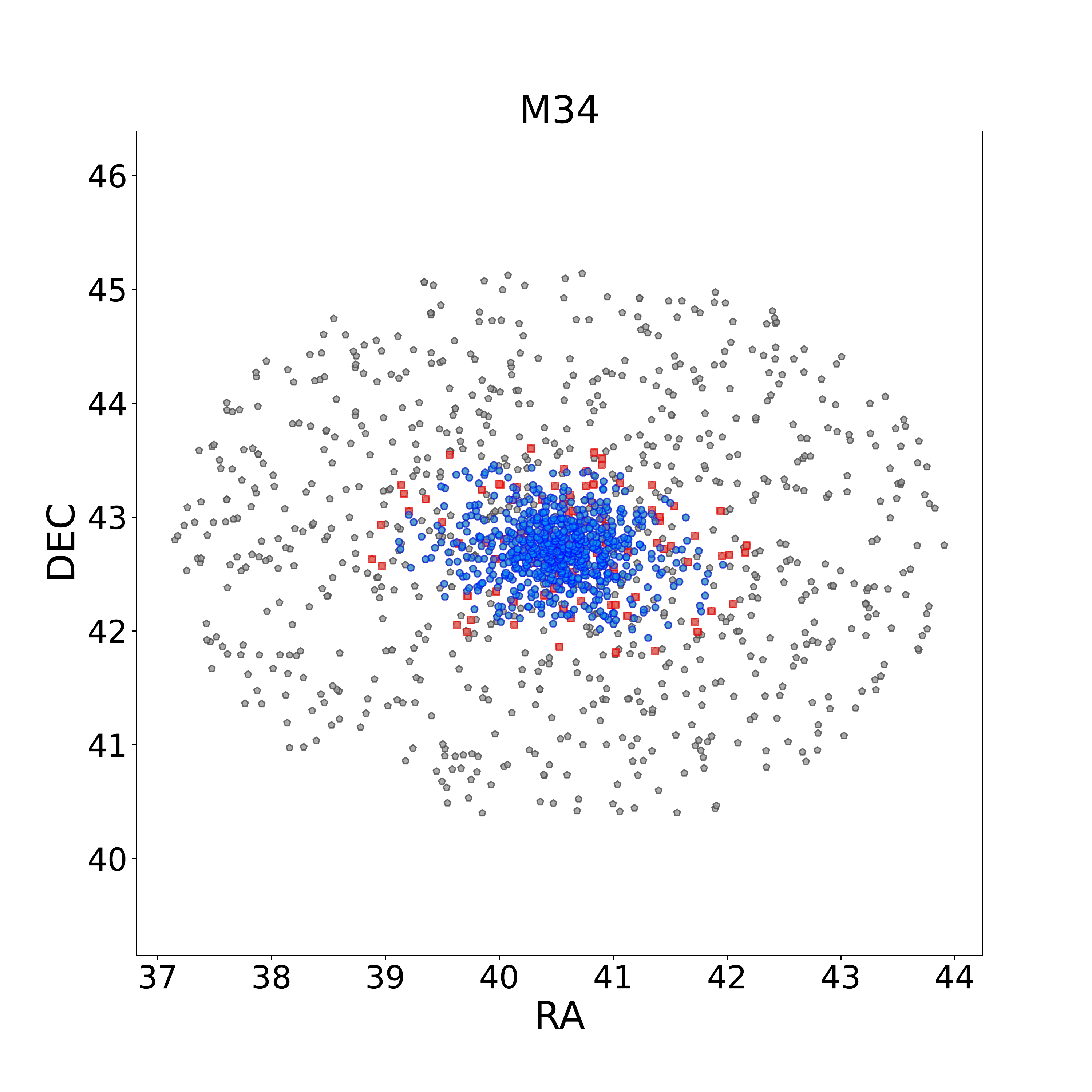}
        \end{subfigure}
        \begin{subfigure}{0.43\textwidth}
                \centering

                \includegraphics[width=\textwidth]{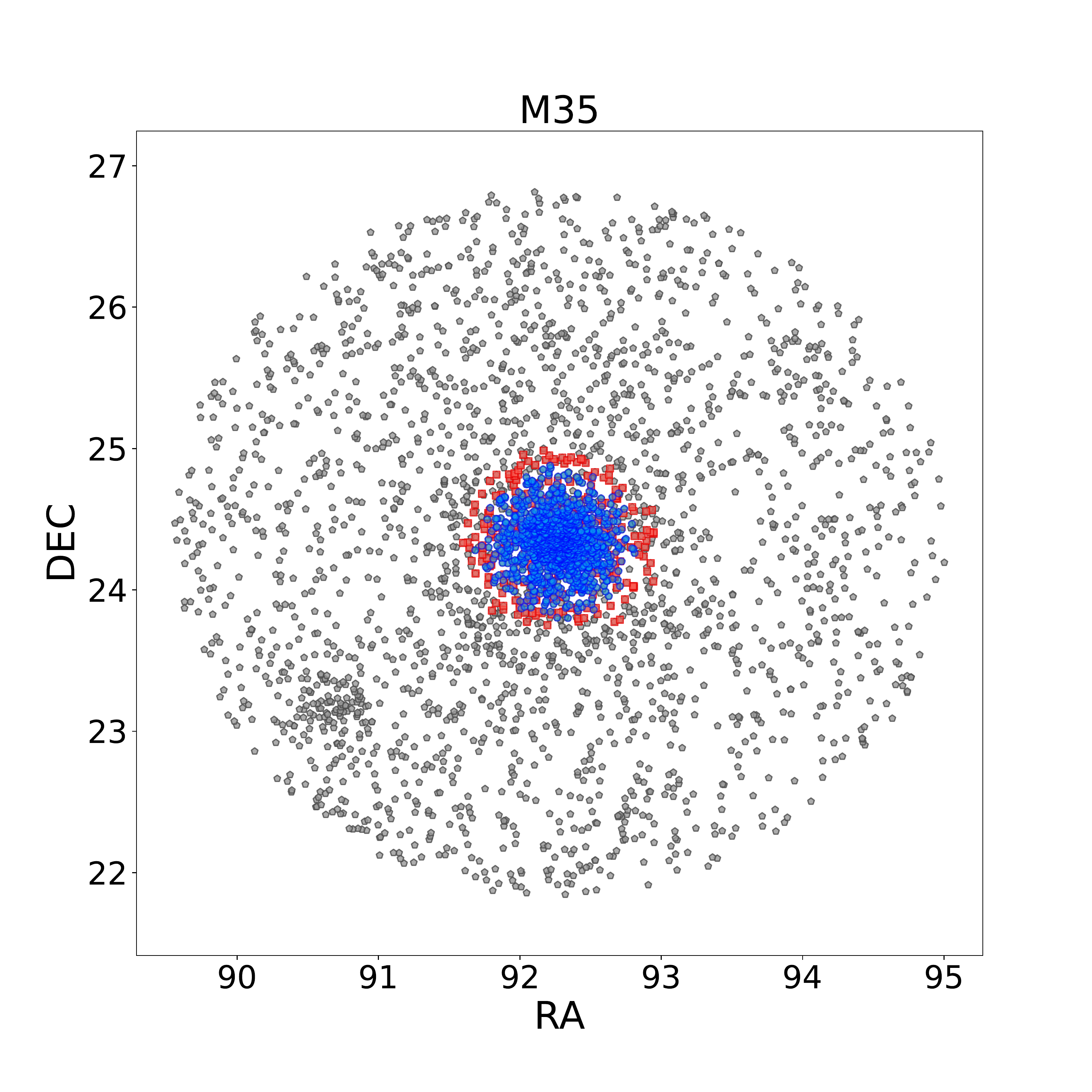}

        \end{subfigure}
        \begin{subfigure}{0.43\textwidth}
                \centering

                \includegraphics[width=\textwidth]{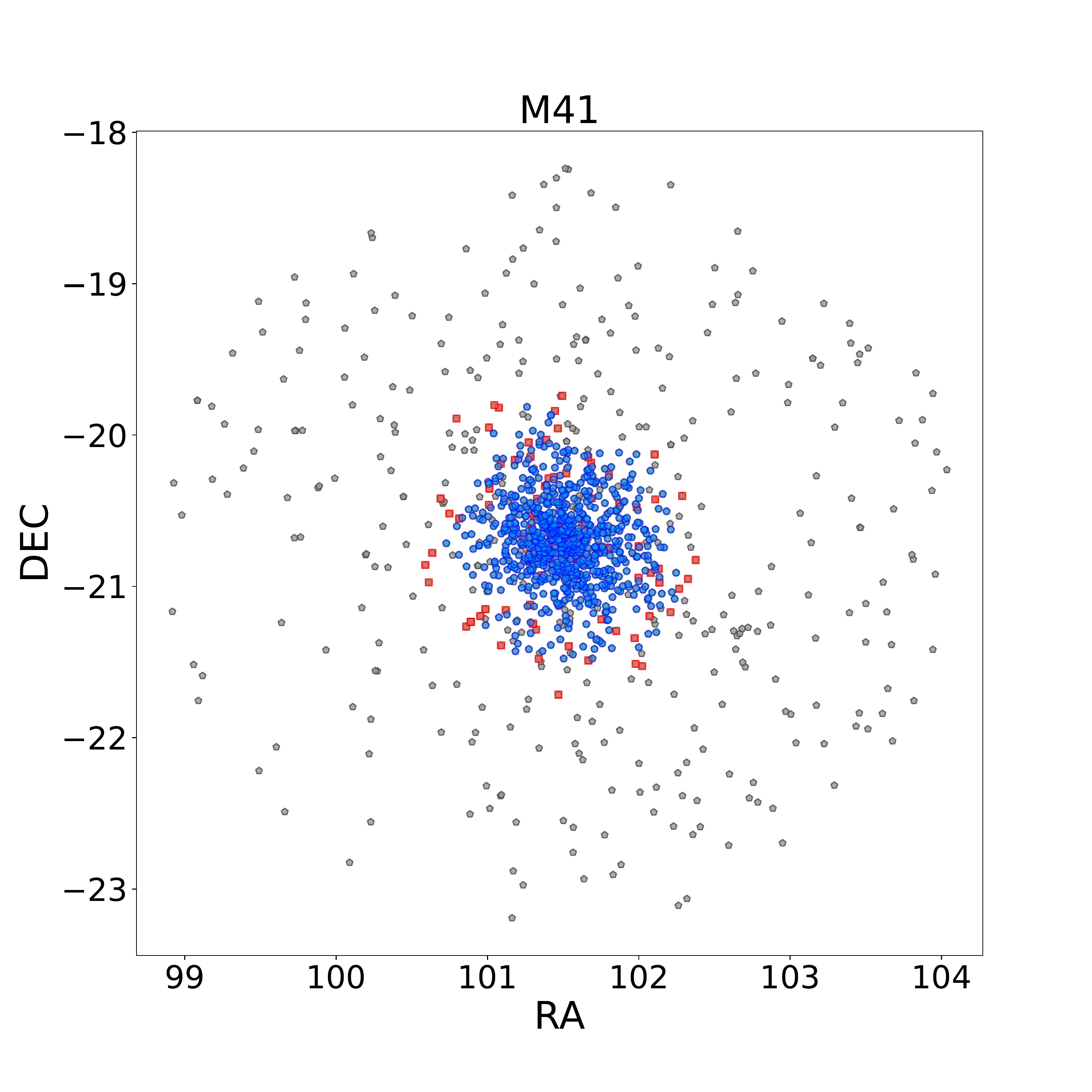}

        \end{subfigure}
        \begin{subfigure}{0.43\textwidth}
                \centering

                \includegraphics[width=\textwidth]{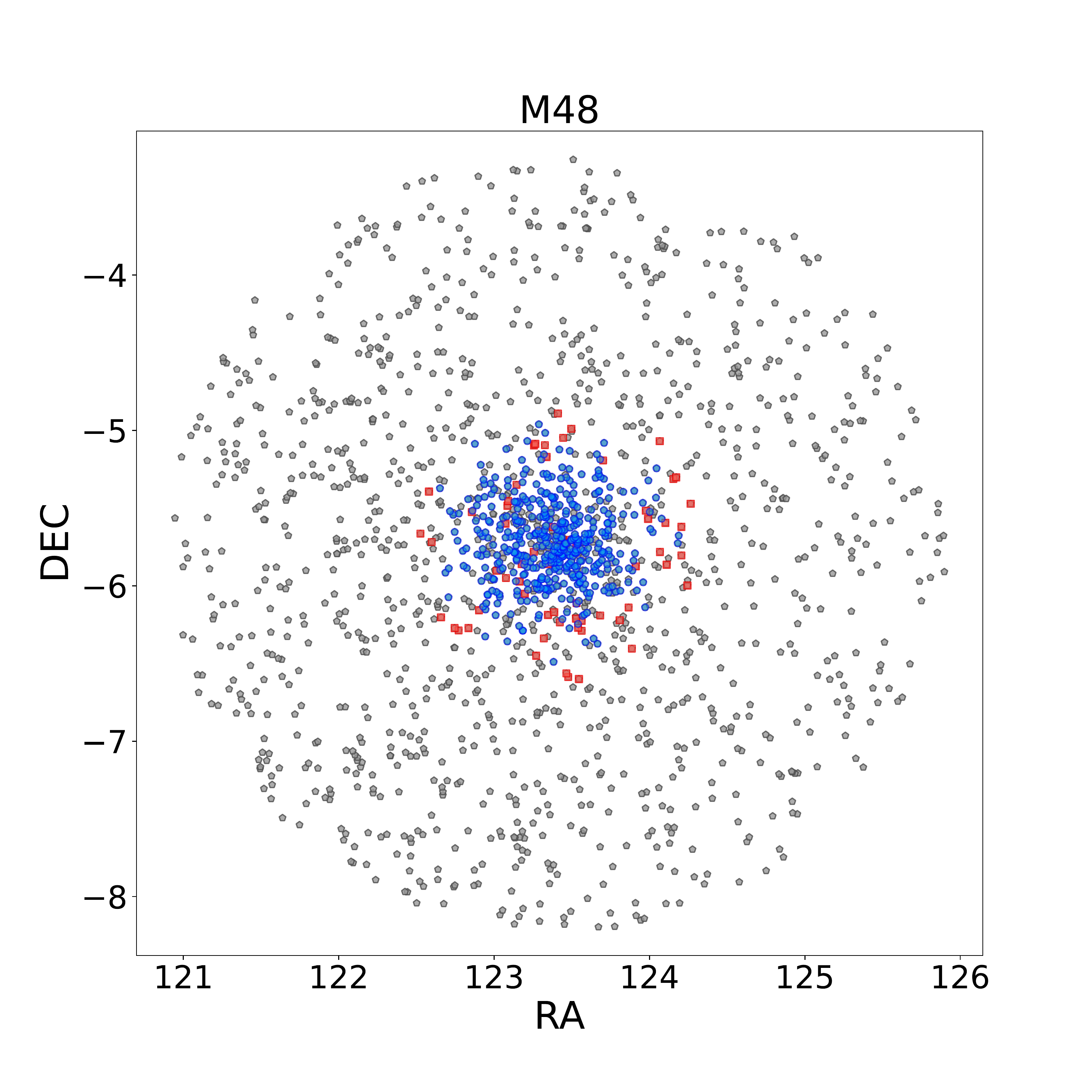}

        \end{subfigure}
  \caption{Position of stars selected by DBSCAN and GMM, grey dots show selected stars by DBSCAN but are not selected by GMM algorithm. Red dots show stars that were selected by GMM algorithm and have probability of membership between $>0.5$ and $<0.8$ and blue dots show stars that were selected by GMM algorithm and have probability $>0.8$.}
  \label{a.position of dbscan and GMM.fig.}
\end{figure}

\begin{figure}
\ContinuedFloat*
  \centering
  \captionsetup[subfigure]{labelformat=empty}
        \begin{subfigure}{0.43\textwidth}
        \centering

                \includegraphics[width=\textwidth]{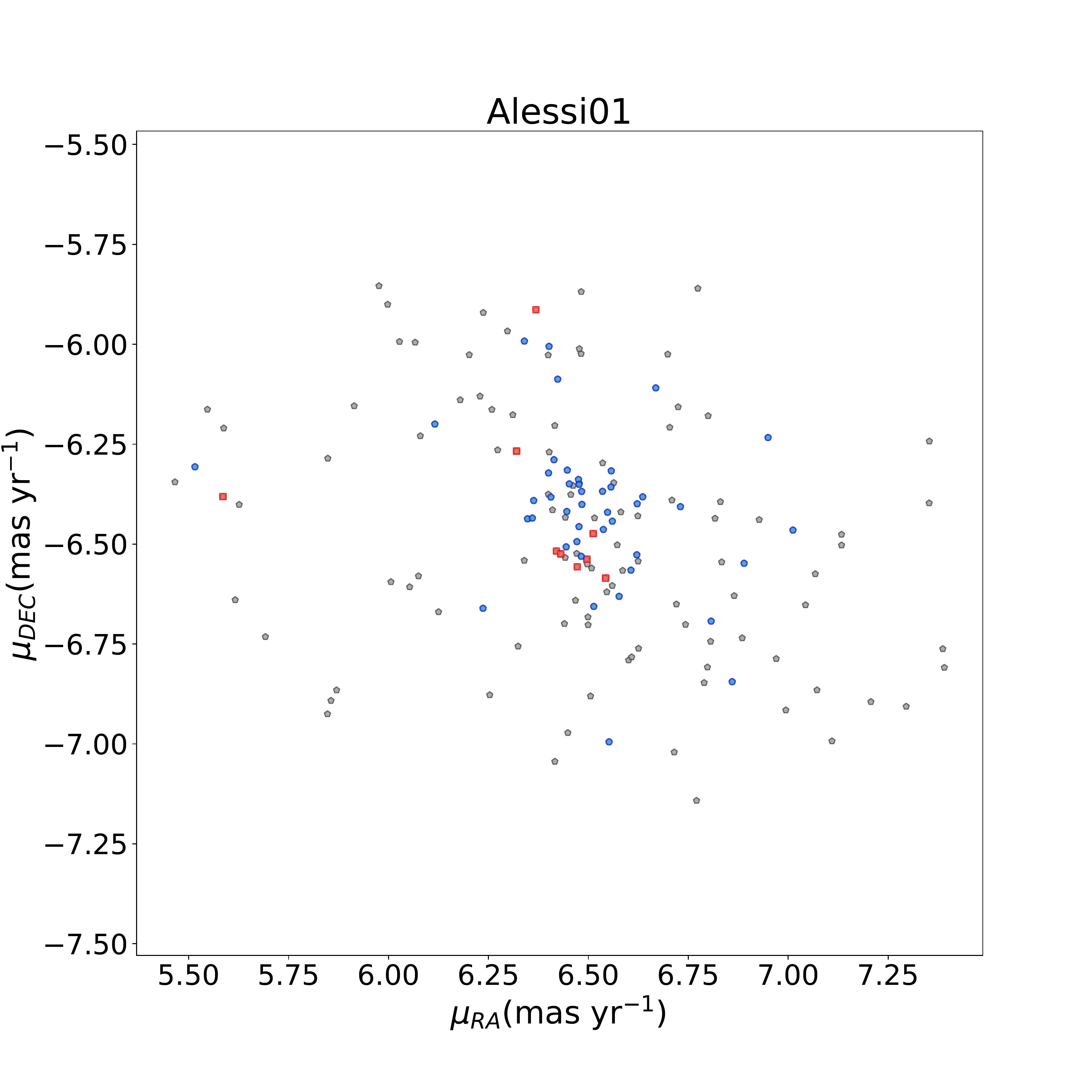}
        \end{subfigure}
        \begin{subfigure}{0.43\textwidth}

                \centering
                \includegraphics[width=\textwidth]{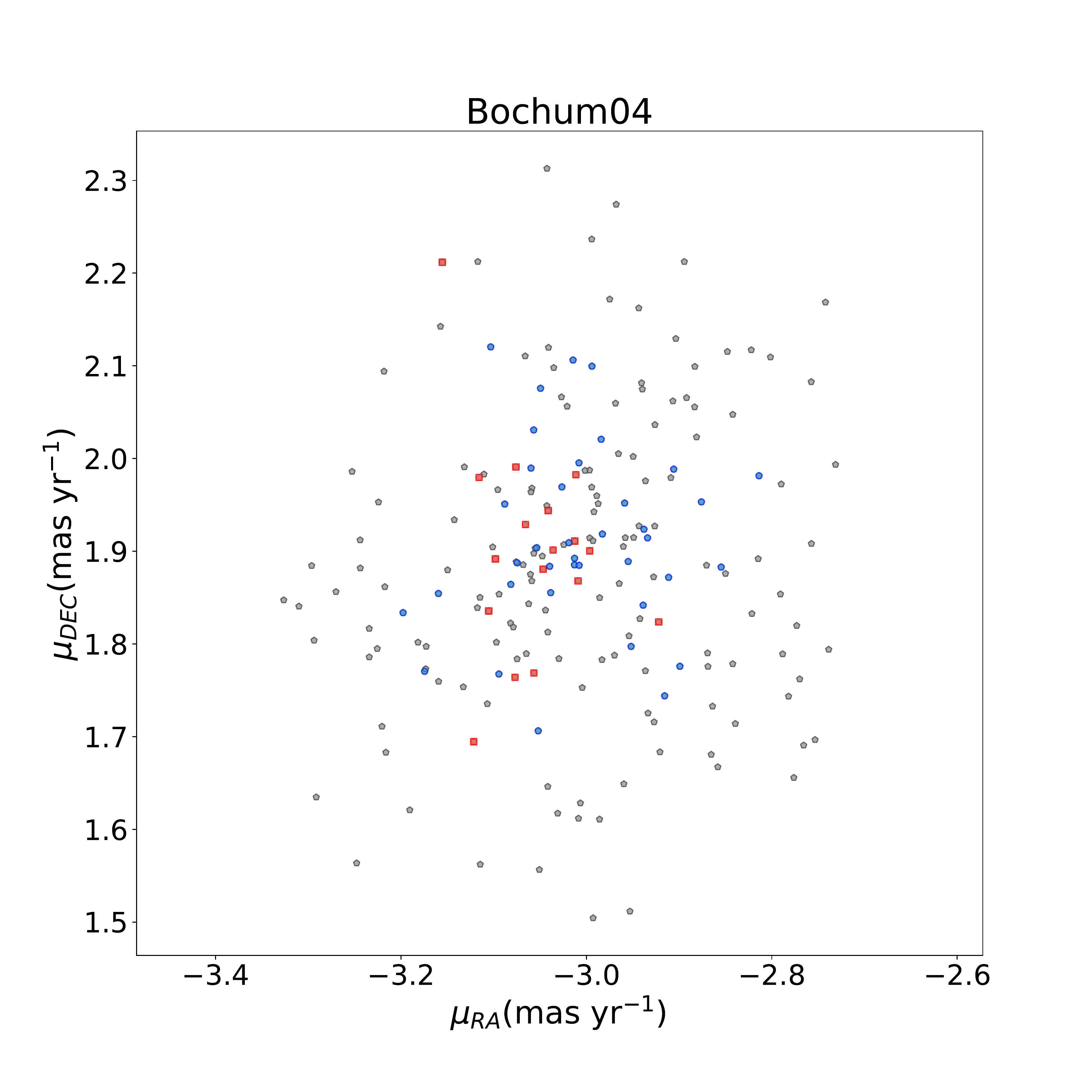}

        \end{subfigure}
  \begin{subfigure}{0.43\textwidth}
                \centering

                \includegraphics[width=\textwidth]{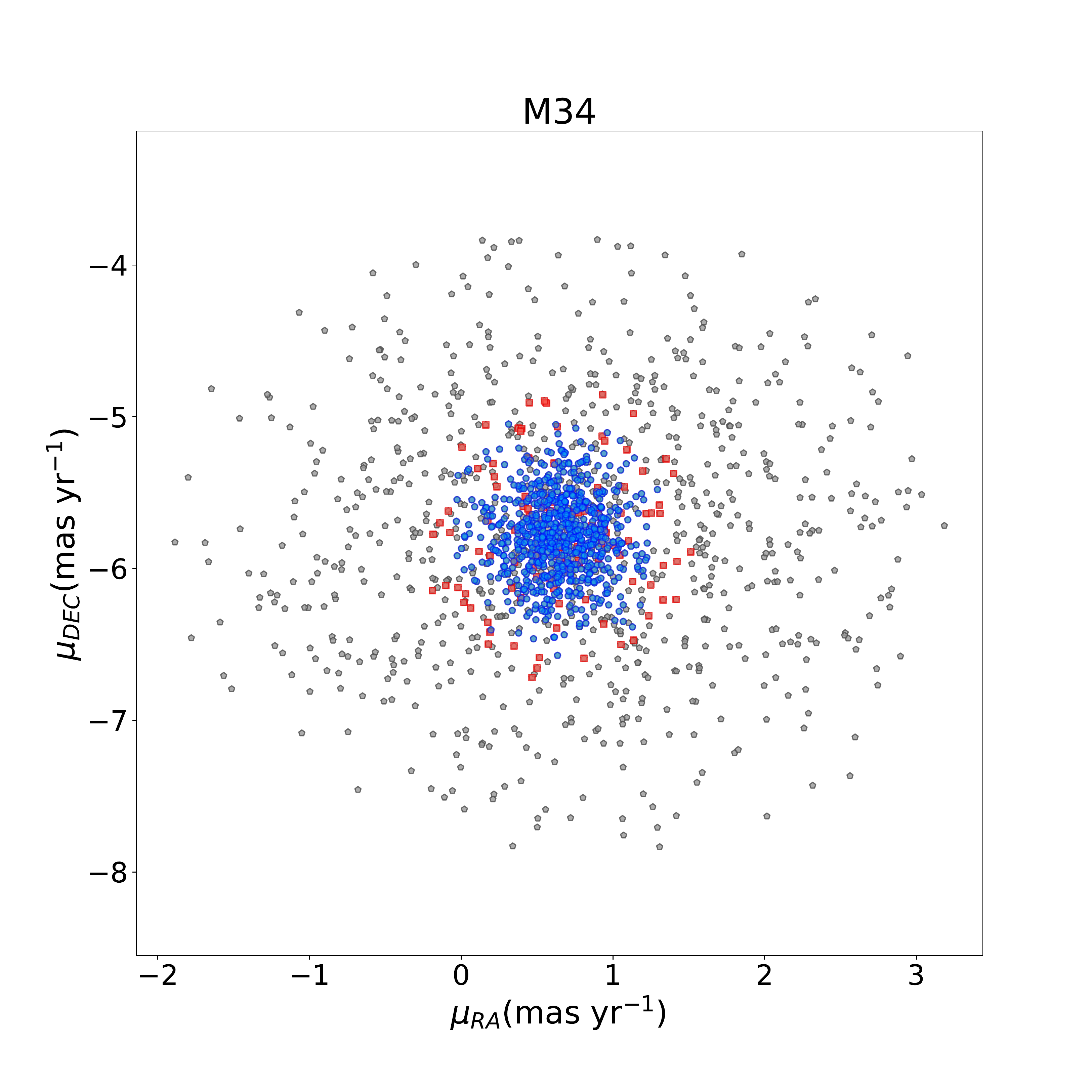}
        \end{subfigure}
        \begin{subfigure}{0.43\textwidth}
                \centering

                \includegraphics[width=\textwidth]{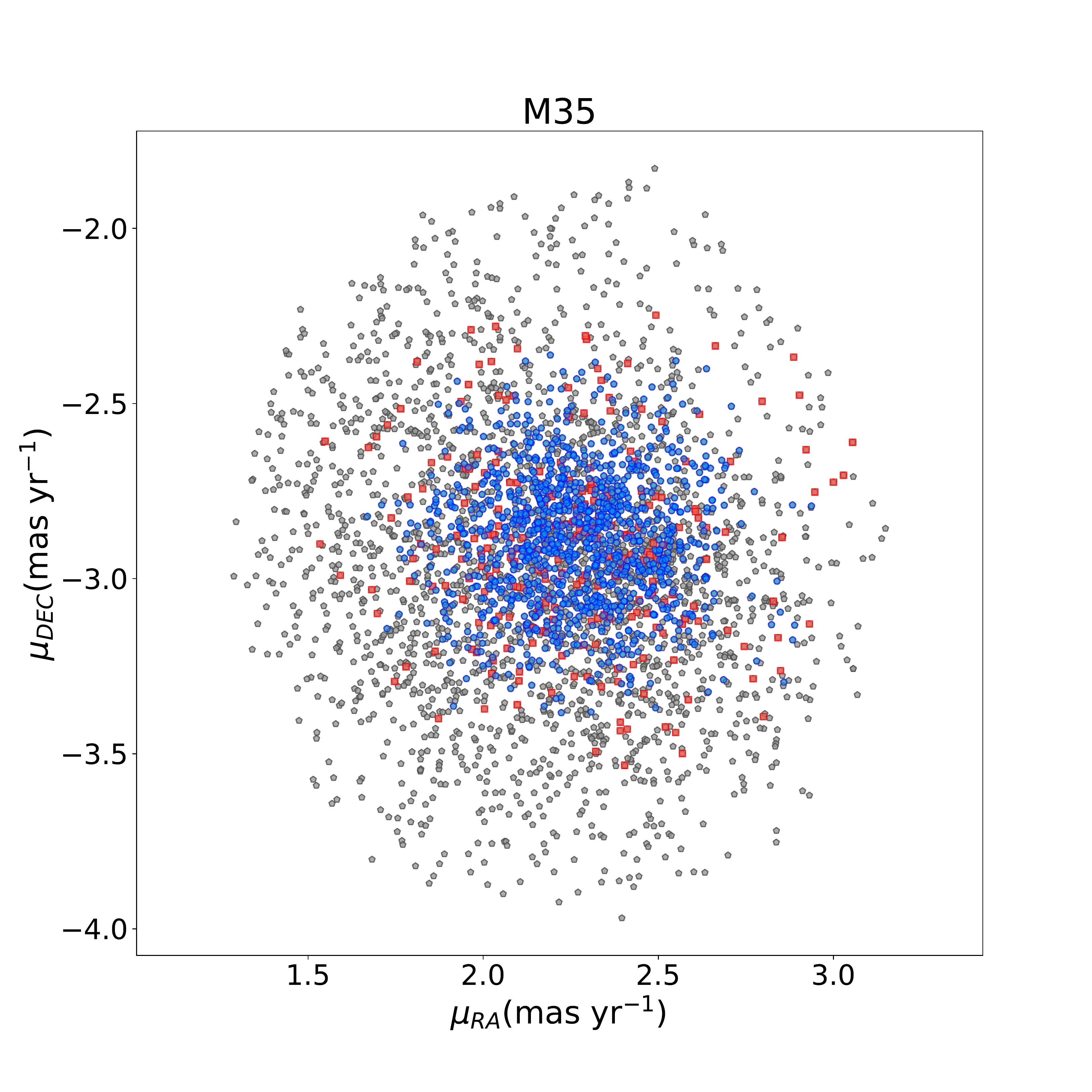}

        \end{subfigure}
        \begin{subfigure}{0.43\textwidth}
                \centering

                \includegraphics[width=\textwidth]{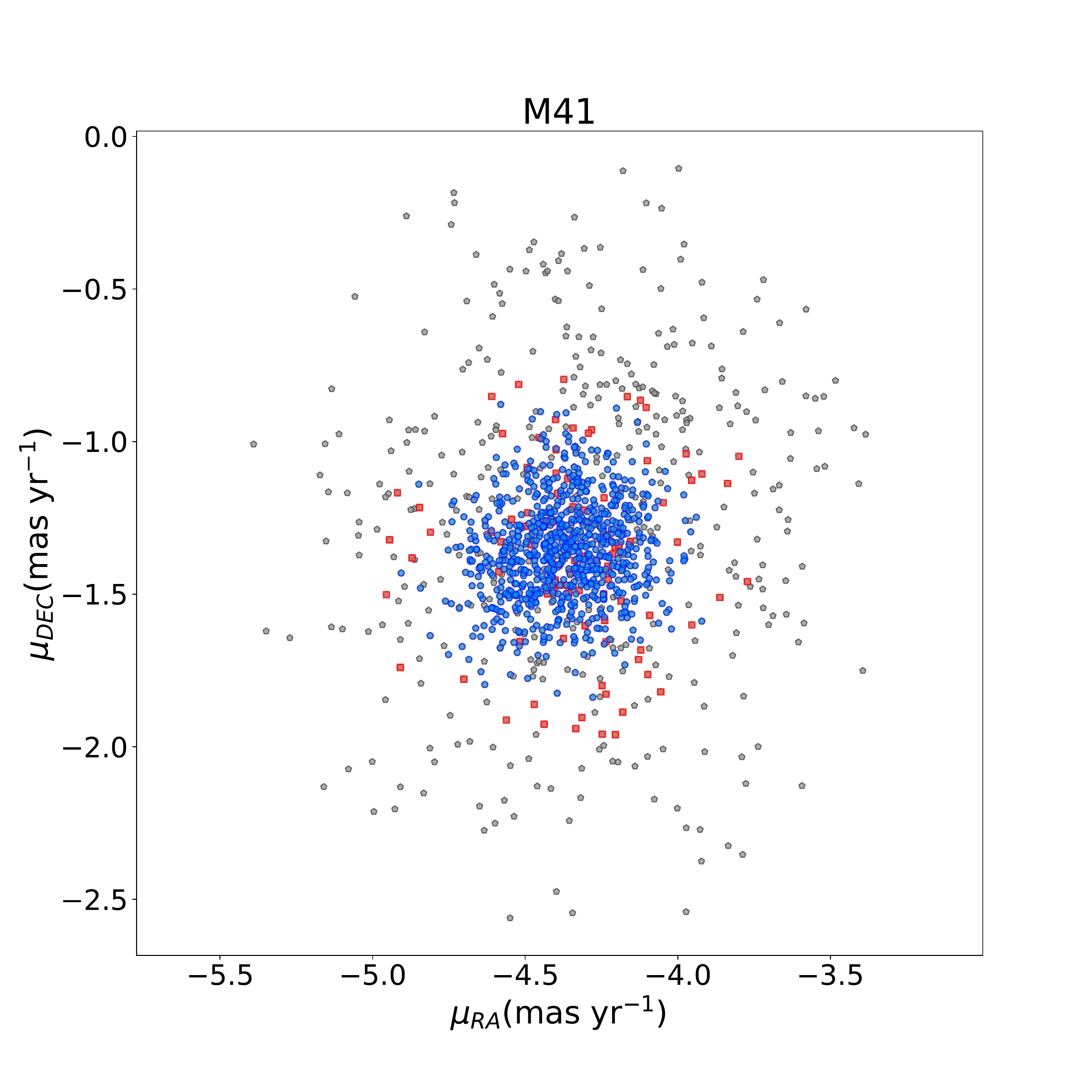}

        \end{subfigure}
        \begin{subfigure}{0.43\textwidth}
                \centering

                \includegraphics[width=\textwidth]{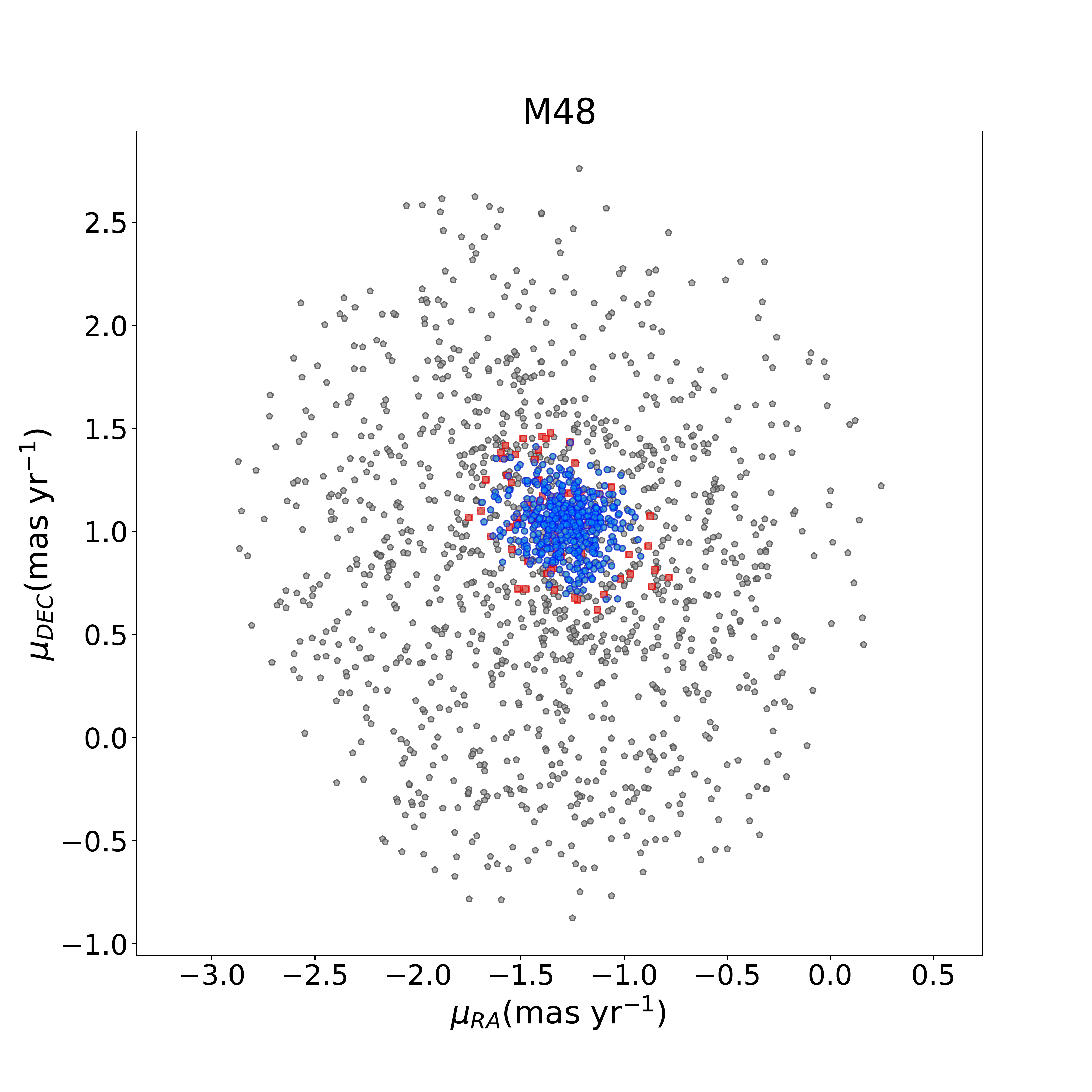}

        \end{subfigure}
  \caption{Proper motion of stars selected by DBSCAN and GMM, grey dots show selected stars by DBSCAN but are not selected by GMM algorithm. Red dots show stars that were selected by GMM algorithm and have probability of membership between $>0.5$ and $<0.8$ and blue dots show show stars that were selected by GMM algorithm and have probability $>0.8$.}
  \label{a.proper motion of dbscan and GMM.fig}
\end{figure}

\begin{figure}
\ContinuedFloat*
  \centering
  \captionsetup[subfigure]{labelformat=empty}
        \begin{subfigure}{0.43\textwidth}
        \centering

                \includegraphics[width=\textwidth]{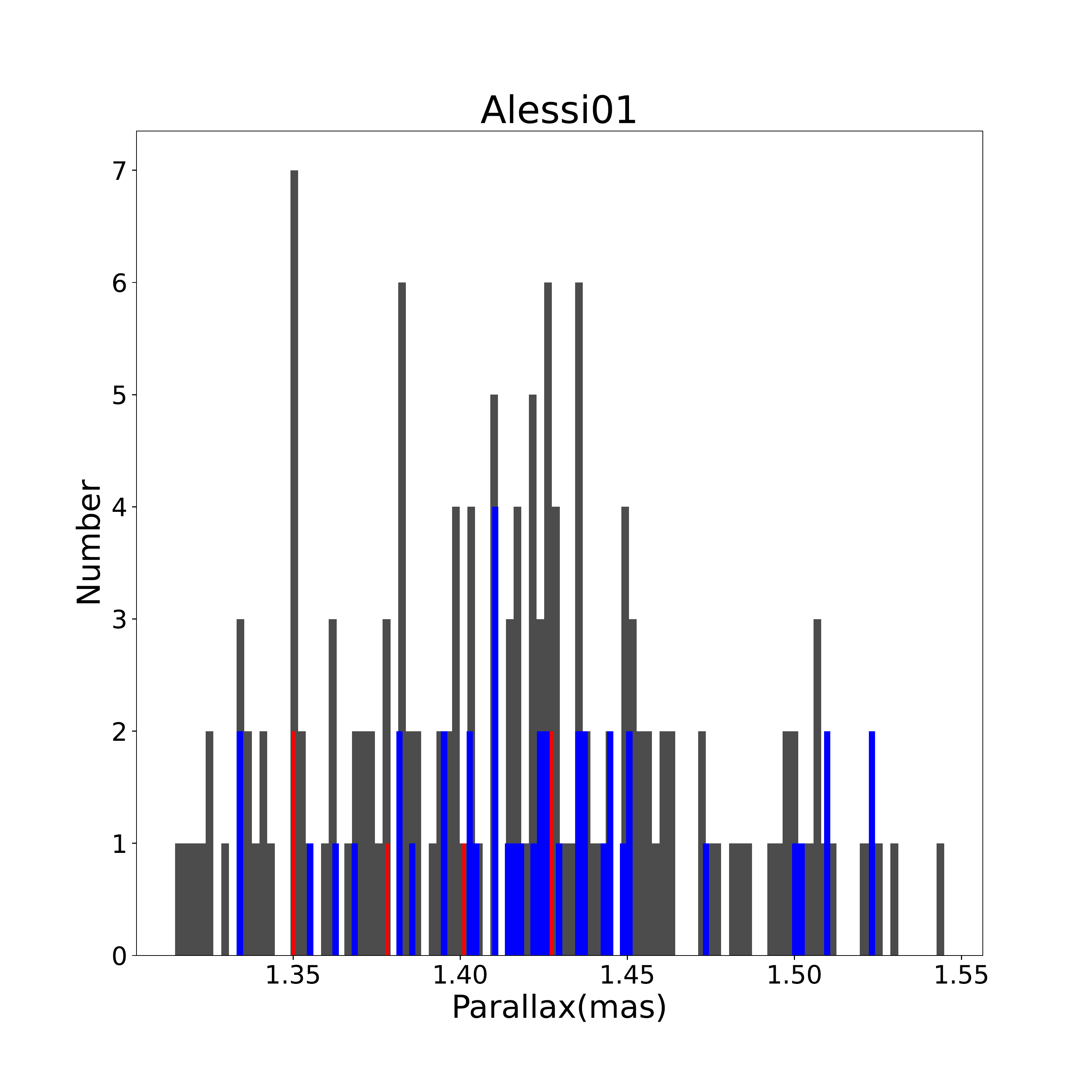}
        \end{subfigure}
        \begin{subfigure}{0.43\textwidth}

                \centering
                \includegraphics[width=\textwidth]{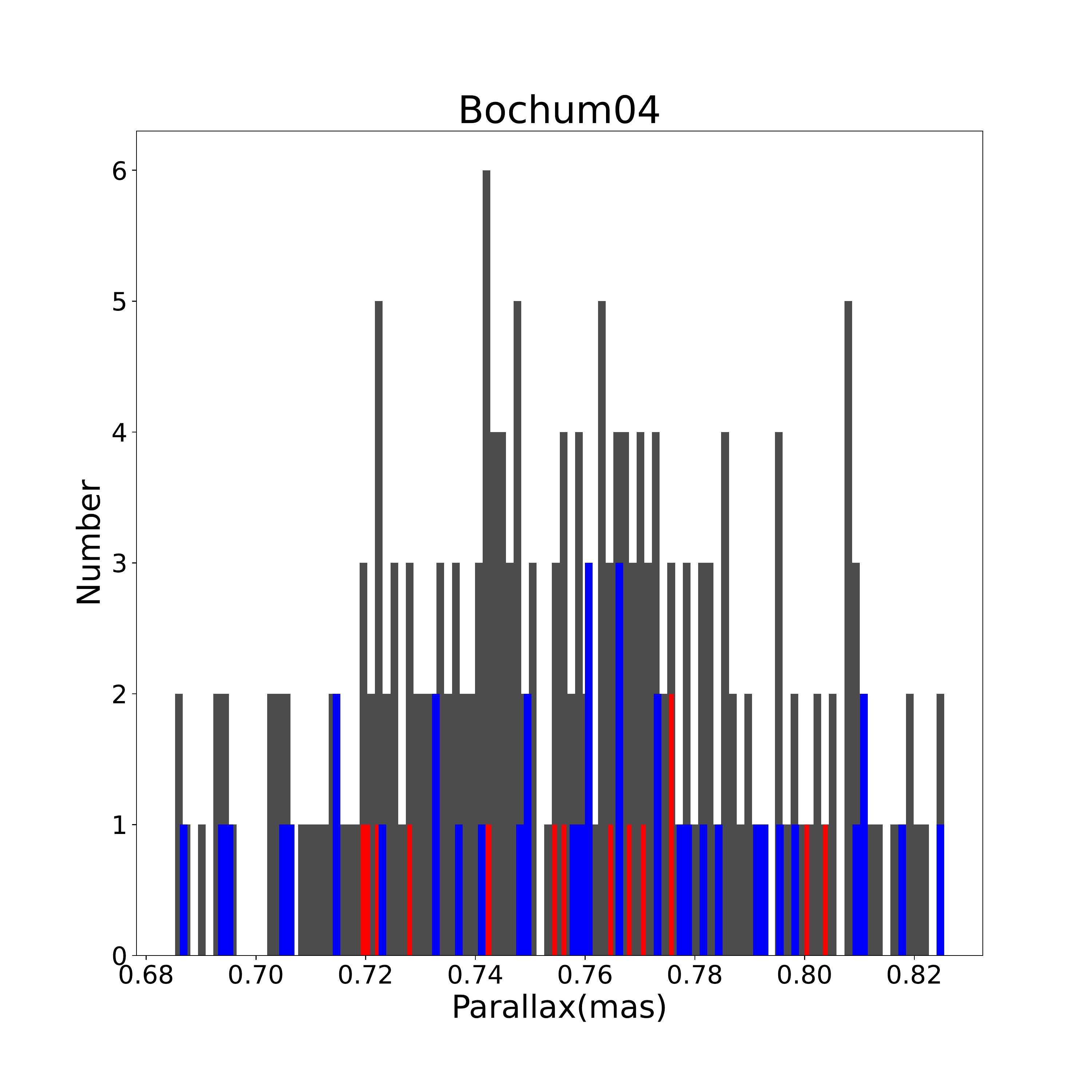}

        \end{subfigure}
        \begin{subfigure}{0.43\textwidth}
                \centering

                \includegraphics[width=\textwidth]{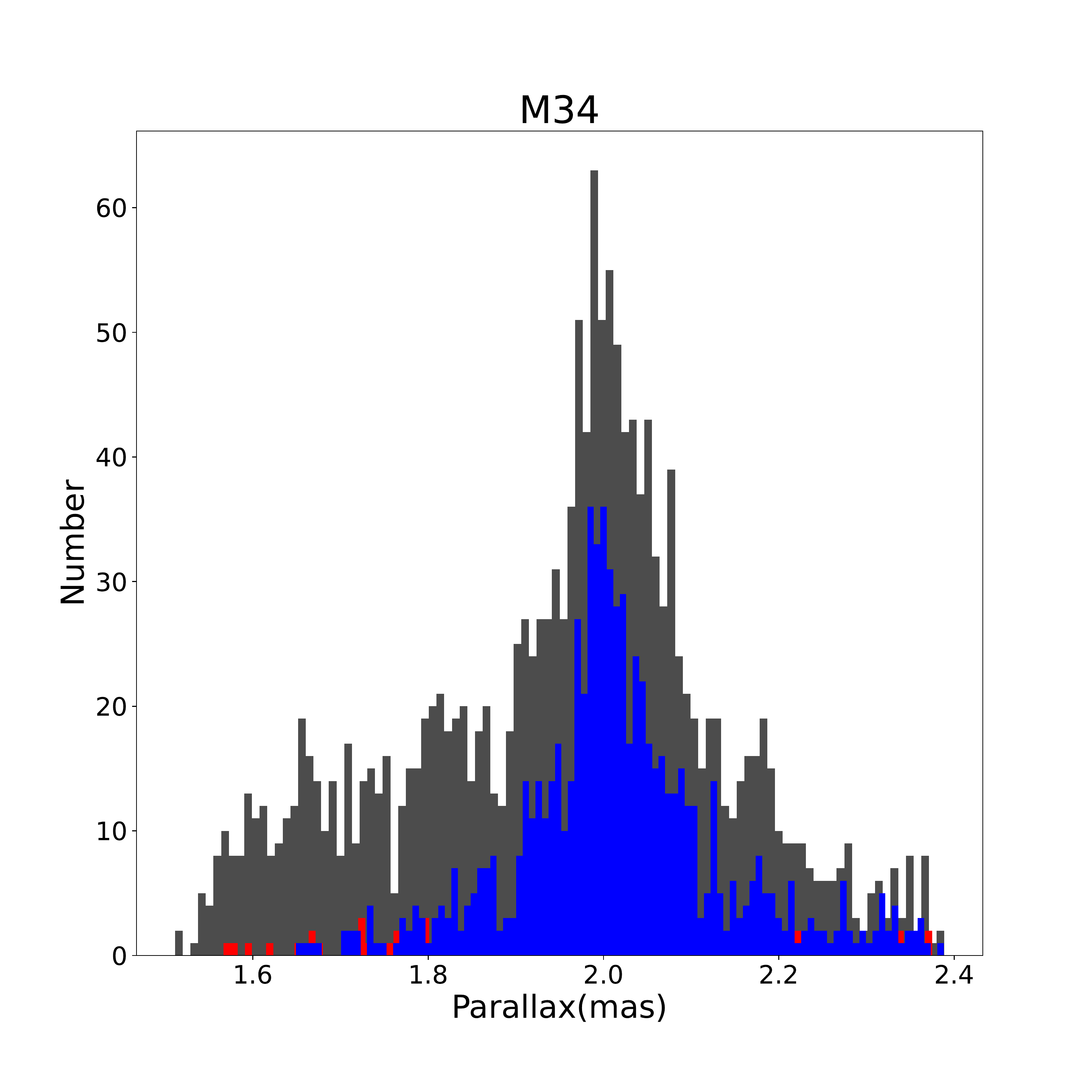}
        \end{subfigure}
        \begin{subfigure}{0.43\textwidth}
                \centering

                \includegraphics[width=\textwidth]{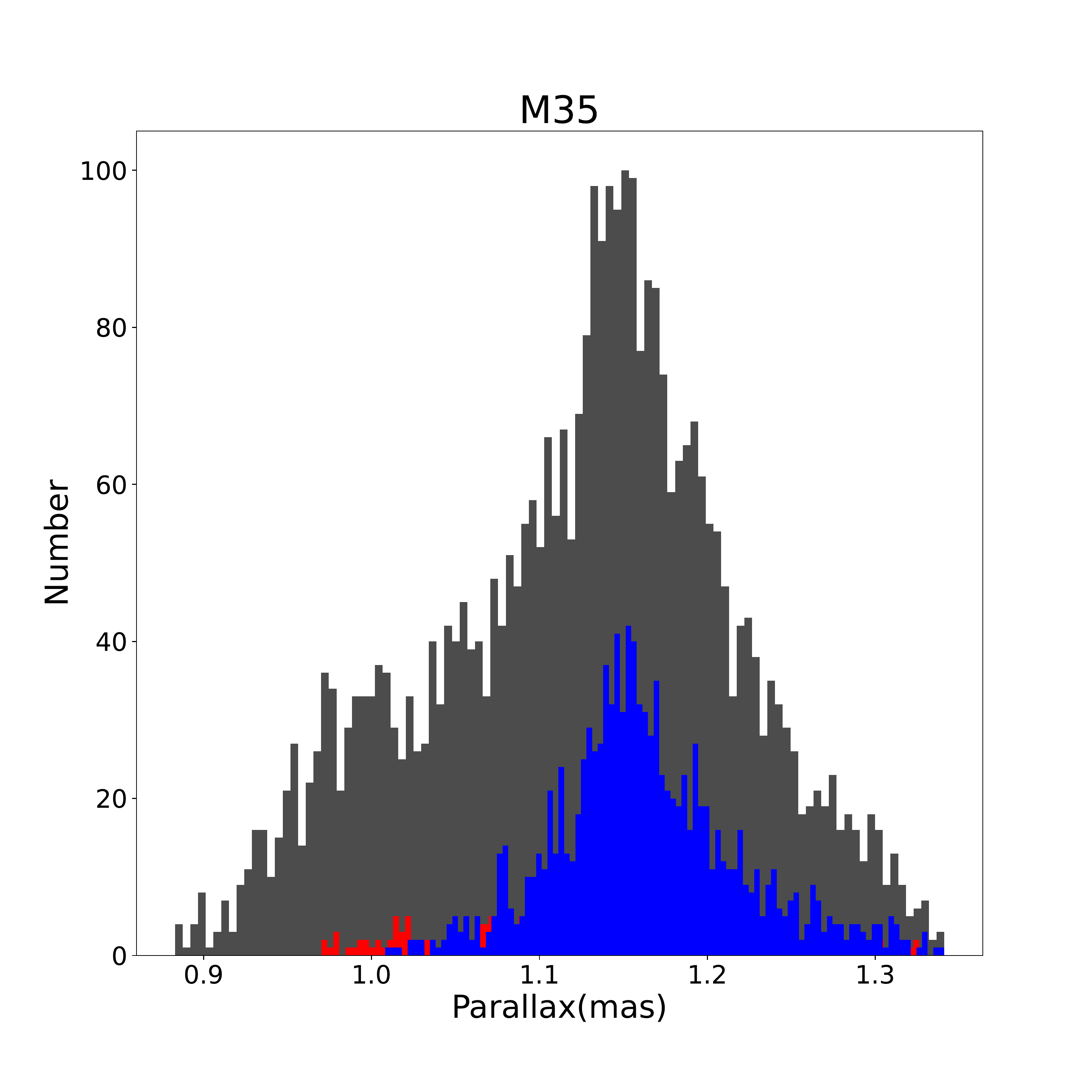}

        \end{subfigure}
        \begin{subfigure}{0.43\textwidth}
                \centering

                \includegraphics[width=\textwidth]{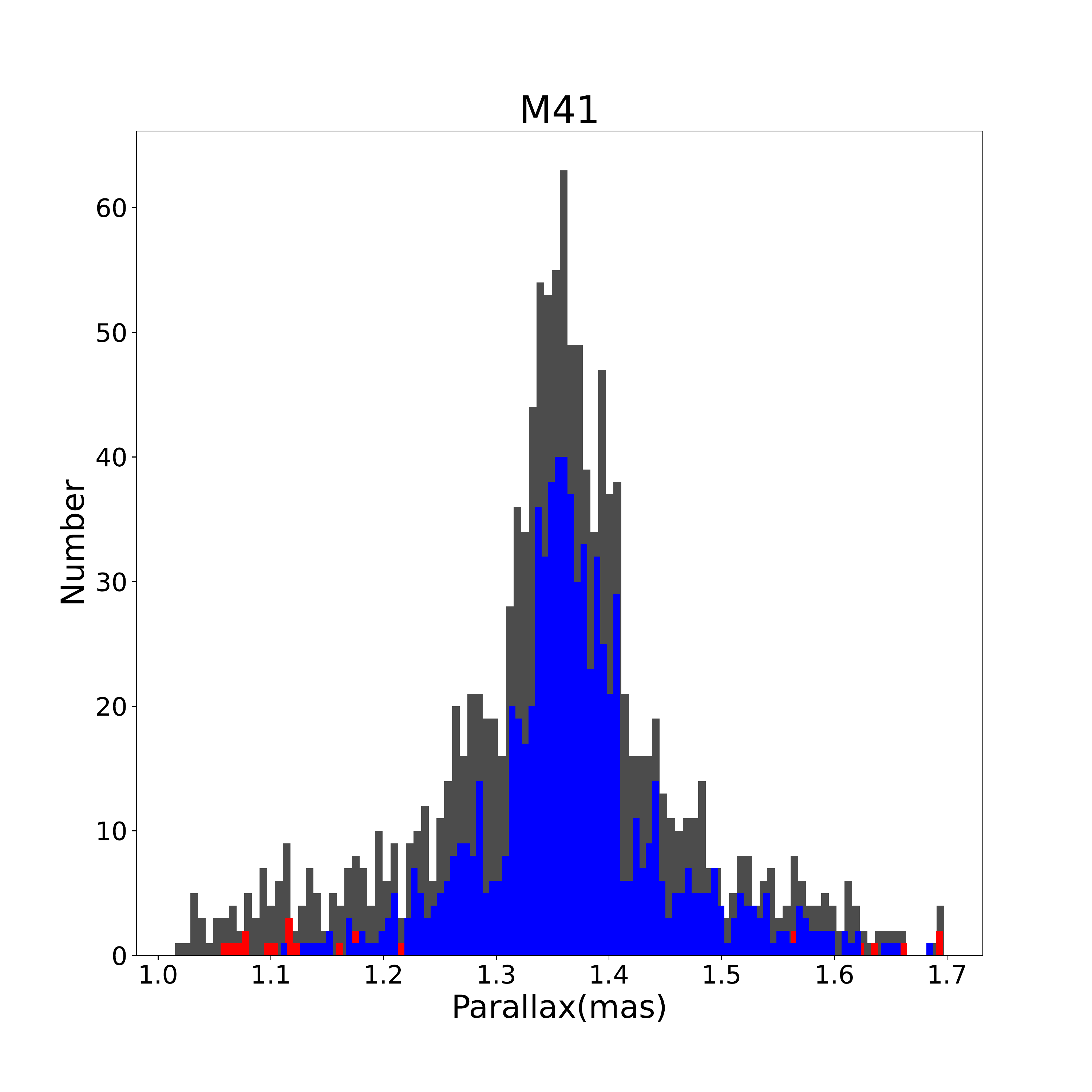}

        \end{subfigure}
        \begin{subfigure}{0.43\textwidth}
                \centering

                \includegraphics[width=\textwidth]{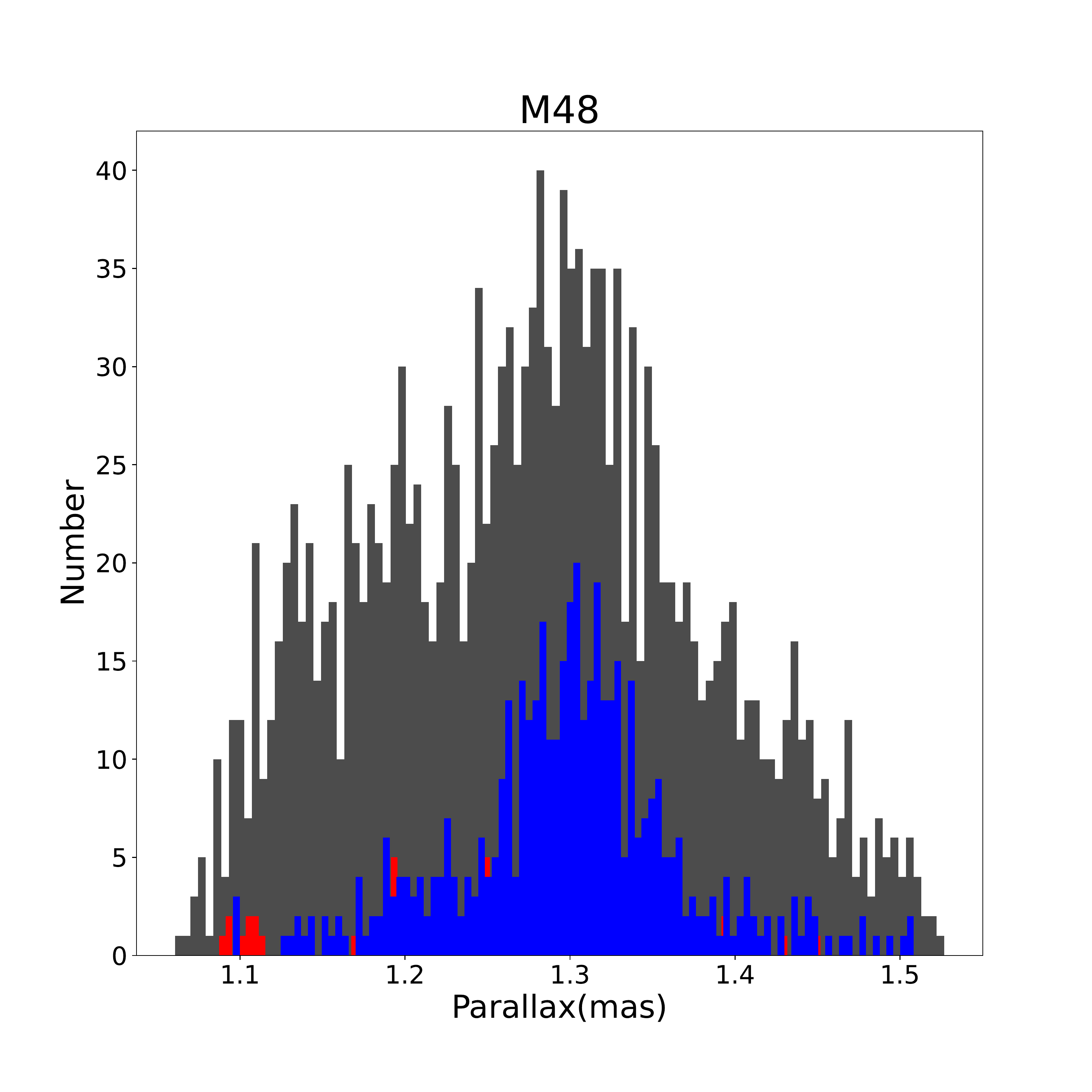}

        \end{subfigure}
  \caption{Parallax of cluster member star candidates by DBSCAN and GMM. Grey lines show selected stars by DBSCAN but were not selected by GMM algorithm. Red lines show stars that were selected by GMM algorithm and have probability of membership between $>0.5$ and $<0.8$ and blue lines show stars that were selected by GMM algorithm and have probability $>0.8$.}
  \label{a.parallax of dbscan and GMM.fig}
\end{figure}

\begin{figure}
\ContinuedFloat*
  \centering
  \captionsetup[subfigure]{labelformat=empty}
        \begin{subfigure}{0.43\textwidth}
        \centering

                \includegraphics[width=\textwidth]{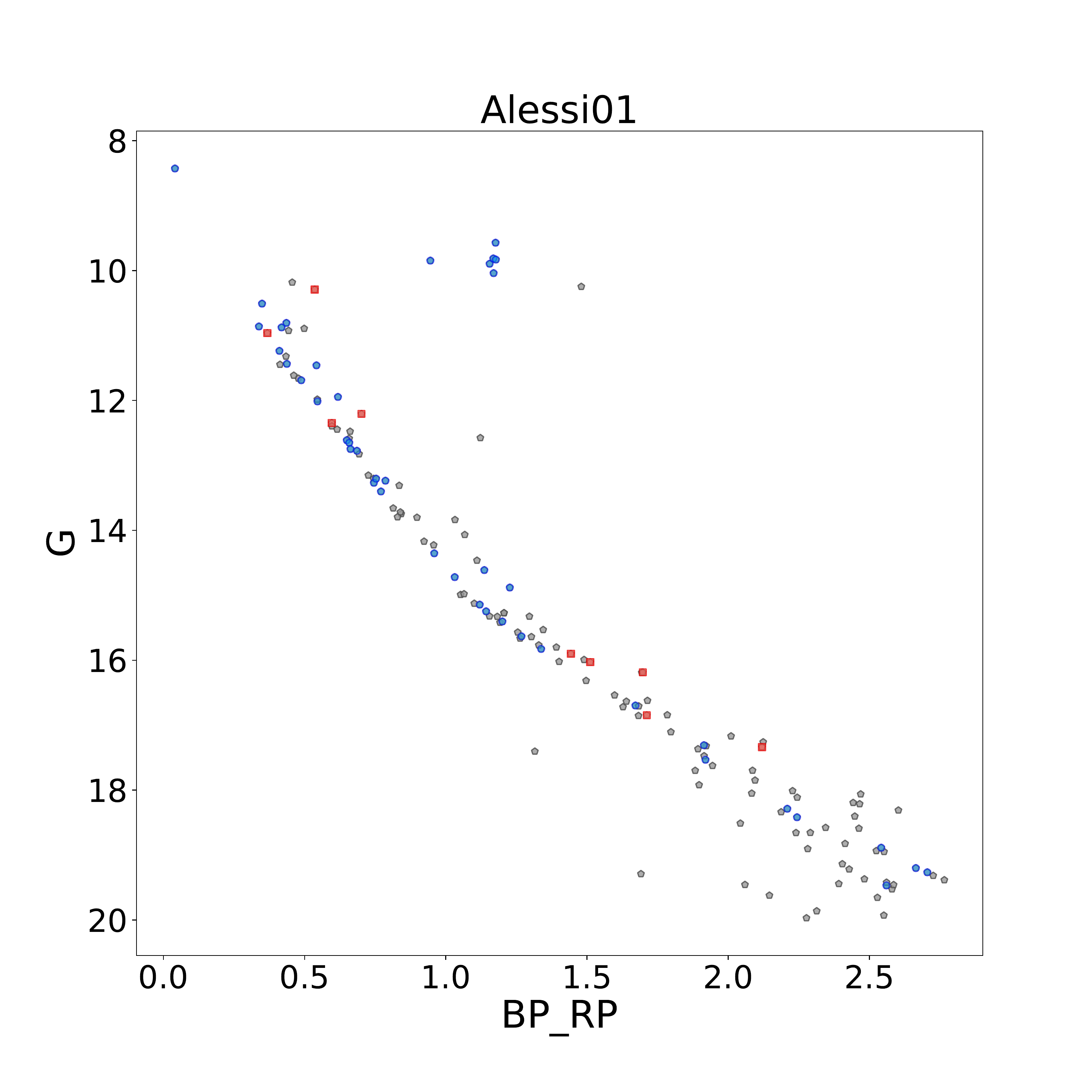}
        \end{subfigure}
        \begin{subfigure}{0.43\textwidth}

                \centering
                \includegraphics[width=\textwidth]{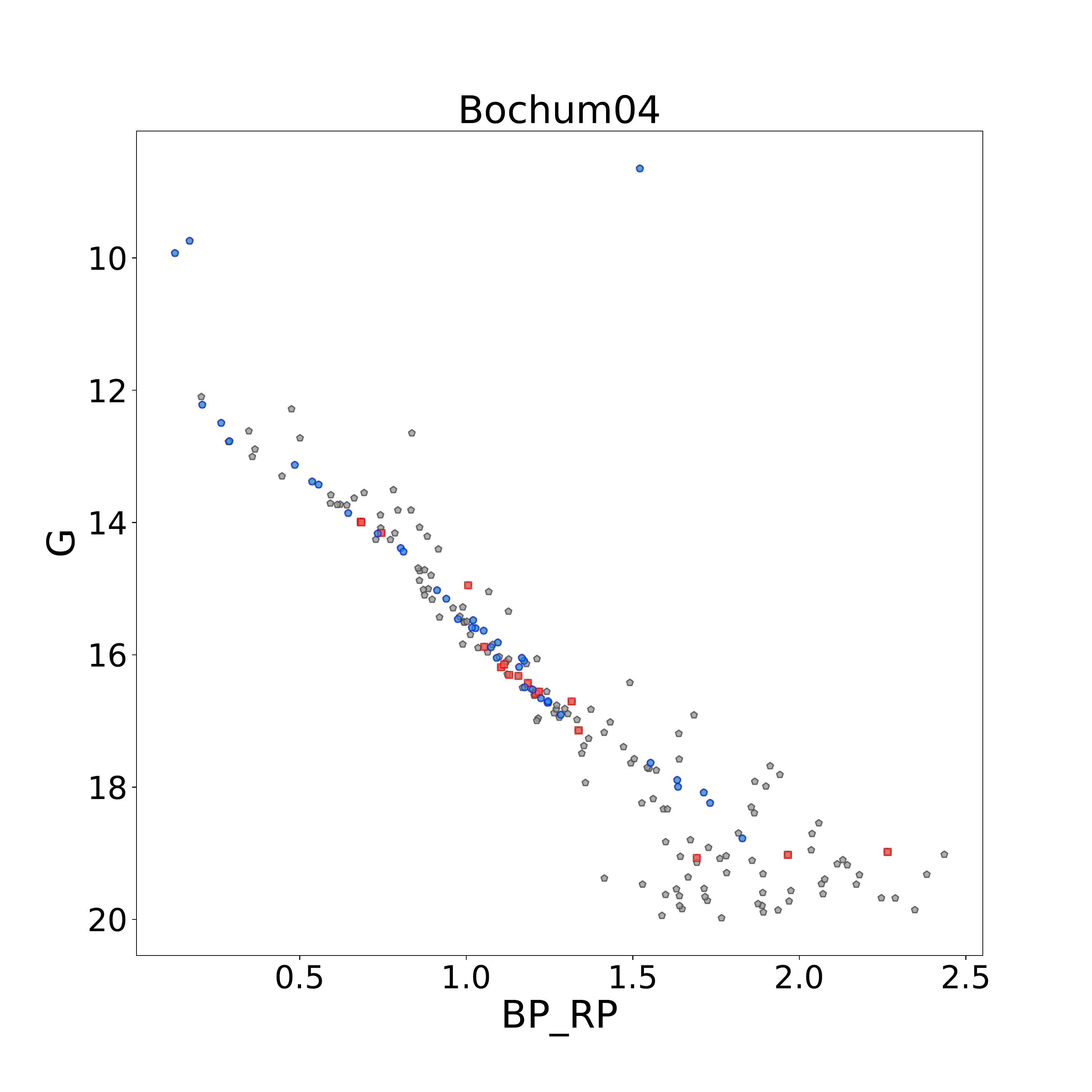}

        \end{subfigure}
        \begin{subfigure}{0.43\textwidth}
                \centering

                \includegraphics[width=\textwidth]{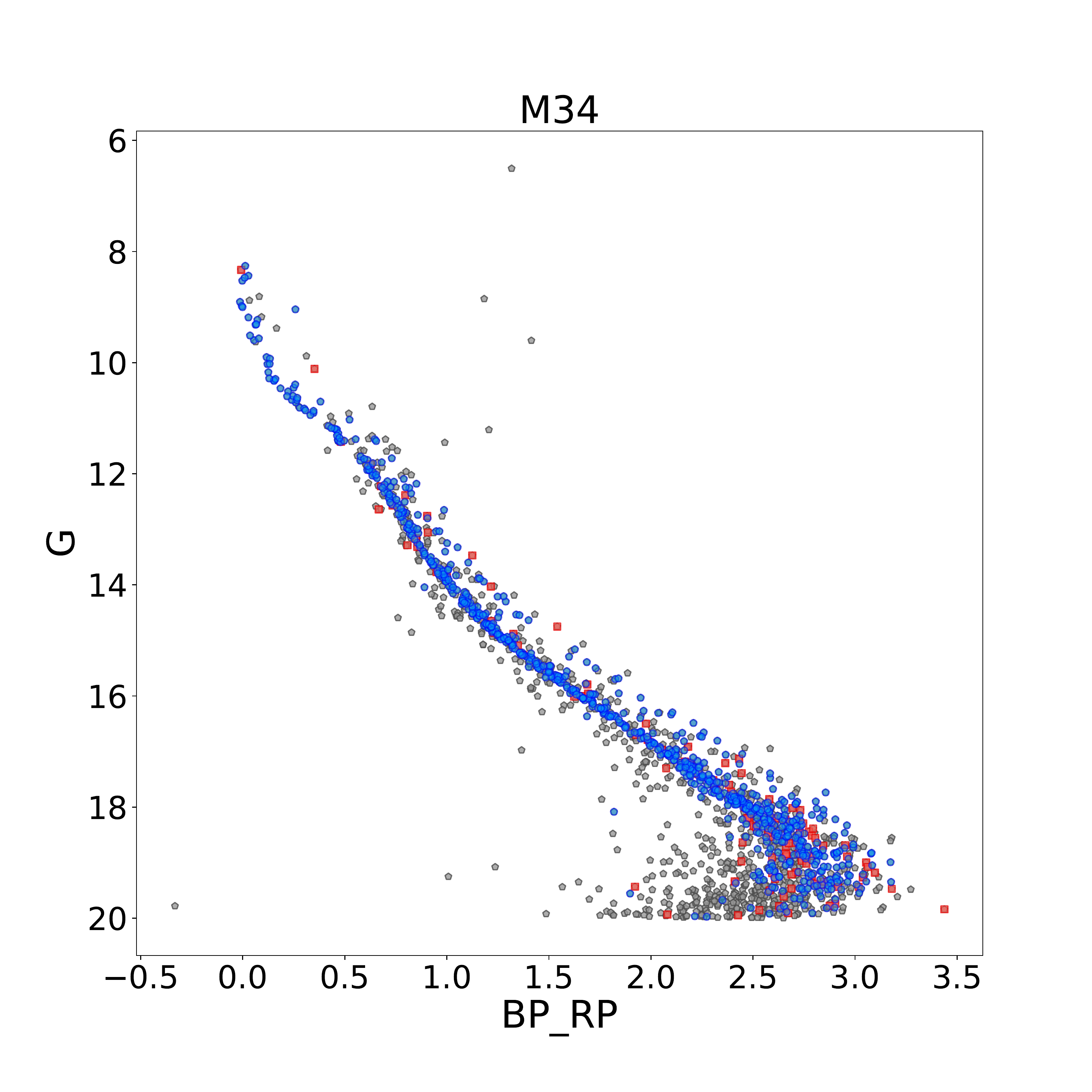}
        \end{subfigure}
        \begin{subfigure}{0.43\textwidth}
                \centering

                \includegraphics[width=\textwidth]{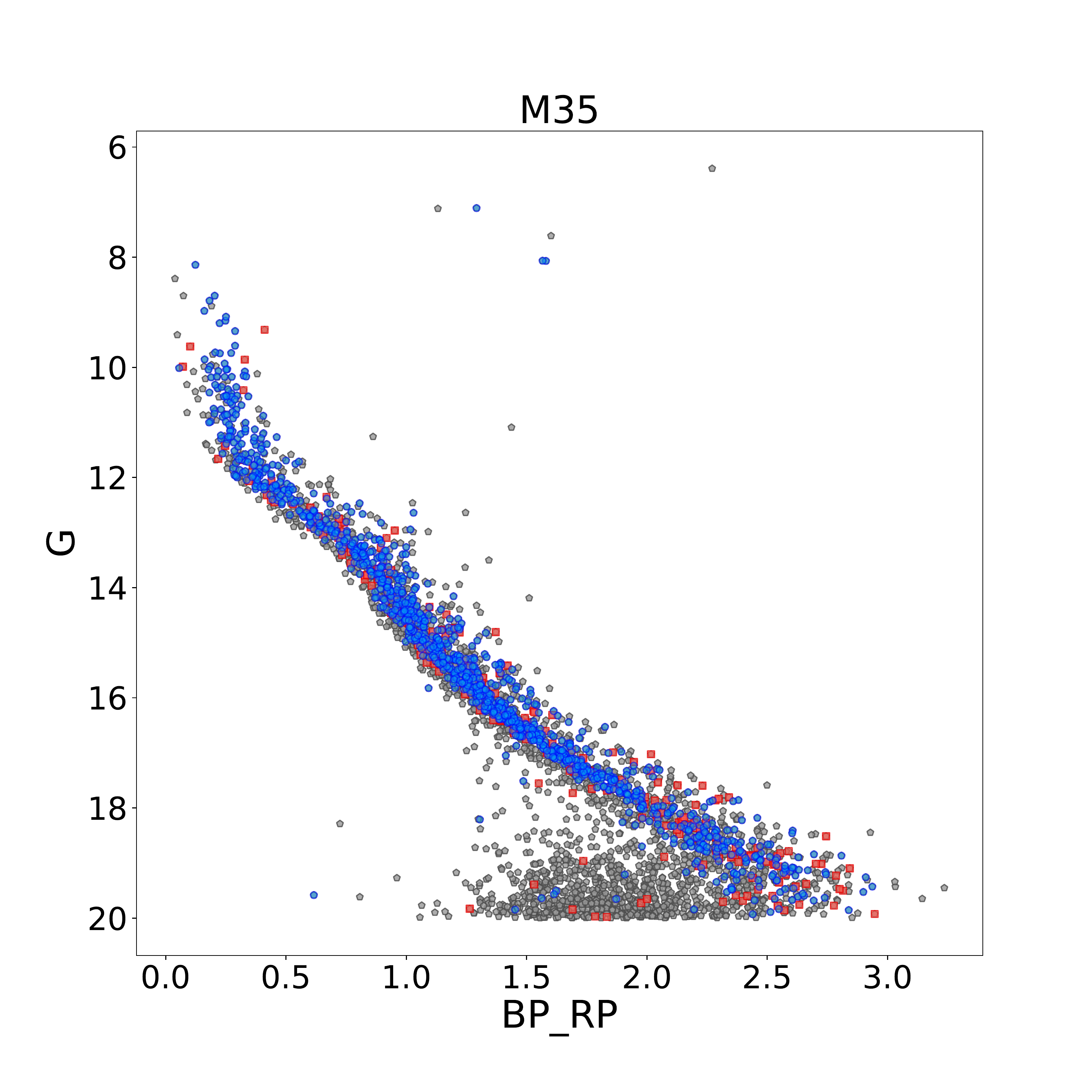}

        \end{subfigure}
        \begin{subfigure}{0.43\textwidth}
                \centering

                \includegraphics[width=\textwidth]{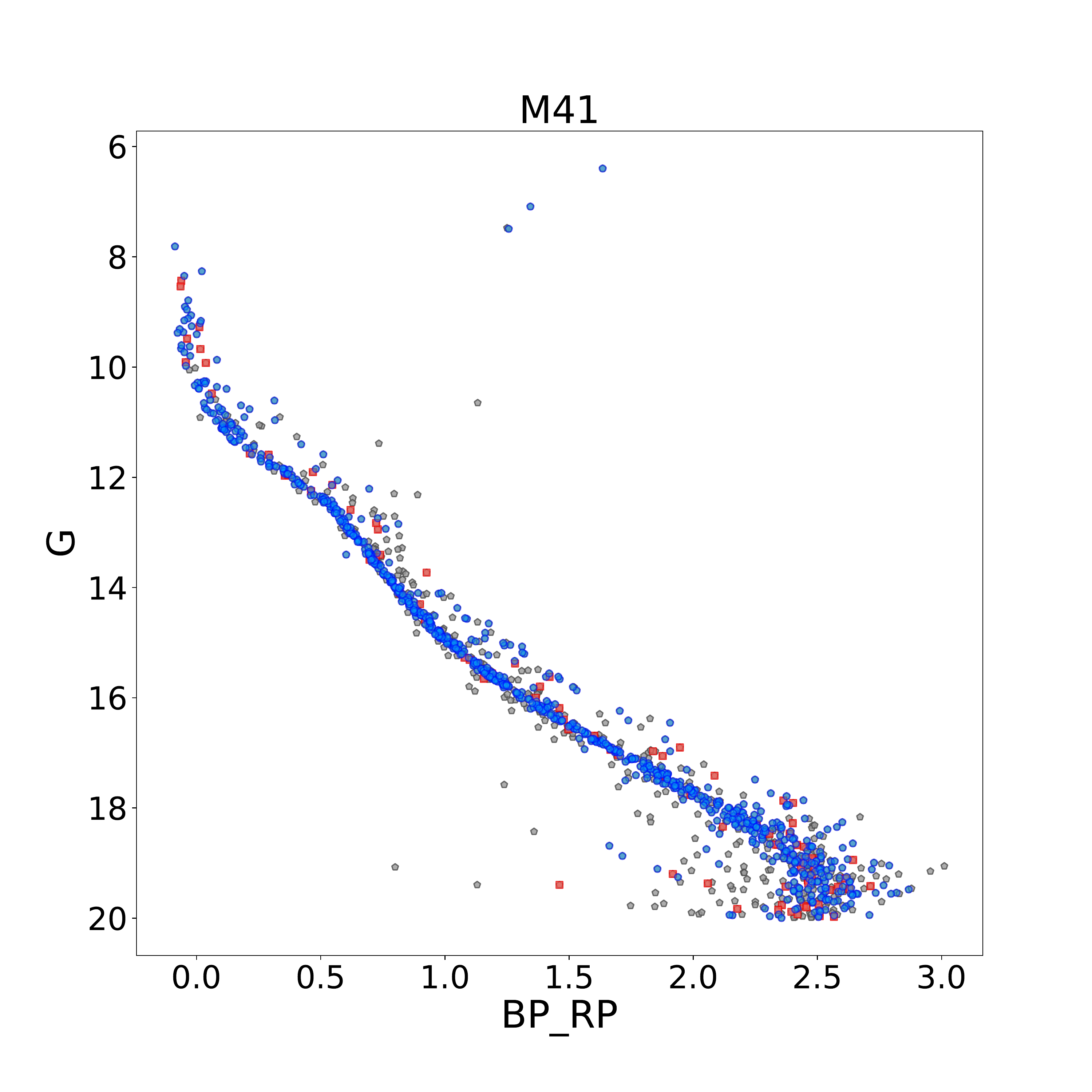}

        \end{subfigure}
        \begin{subfigure}{0.43\textwidth}
                \centering

                \includegraphics[width=\textwidth]{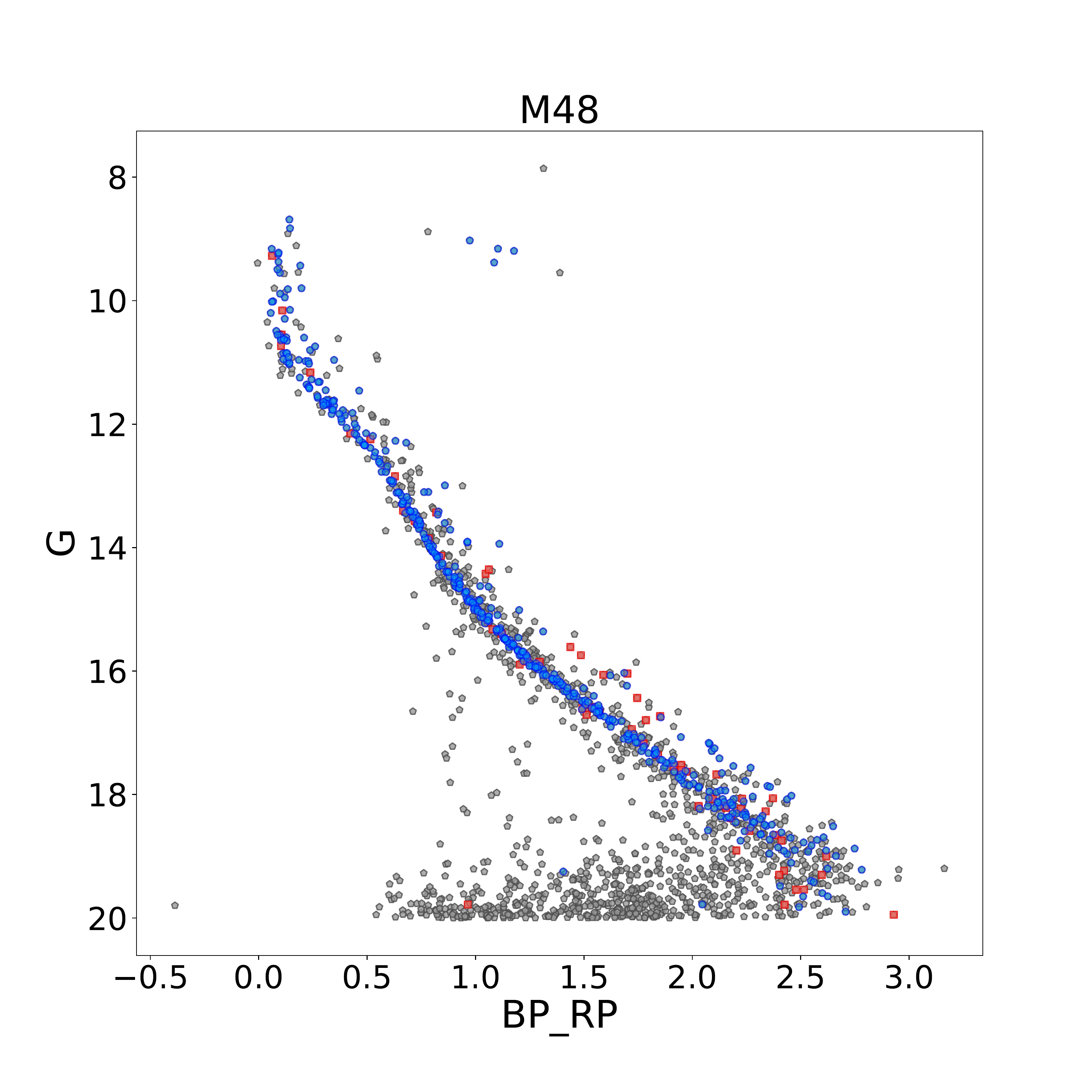}

        \end{subfigure}
  \caption{CMD of cluster member stars candidate from field stars by DBSCAN and GMM, grey dots show selected stars by DBSCAN that were not selected by GMM algorithm, red dots show stars that were selected by GMM algorithm and have a probability of membership $<0.8$ and blue dots show cluster members, the vertical axis is G magnitude, and the horizontal axis is $B-R$ that shows colors indicate.}
  \label{a.CMD of dbscan and GMM.fig}
\end{figure}

\begin{figure}[h!]
\ContinuedFloat*
  \centering
  \captionsetup[subfigure]{labelformat=empty}
        \begin{subfigure}{0.45\textwidth}
                \centering

                \includegraphics[width=\textwidth]{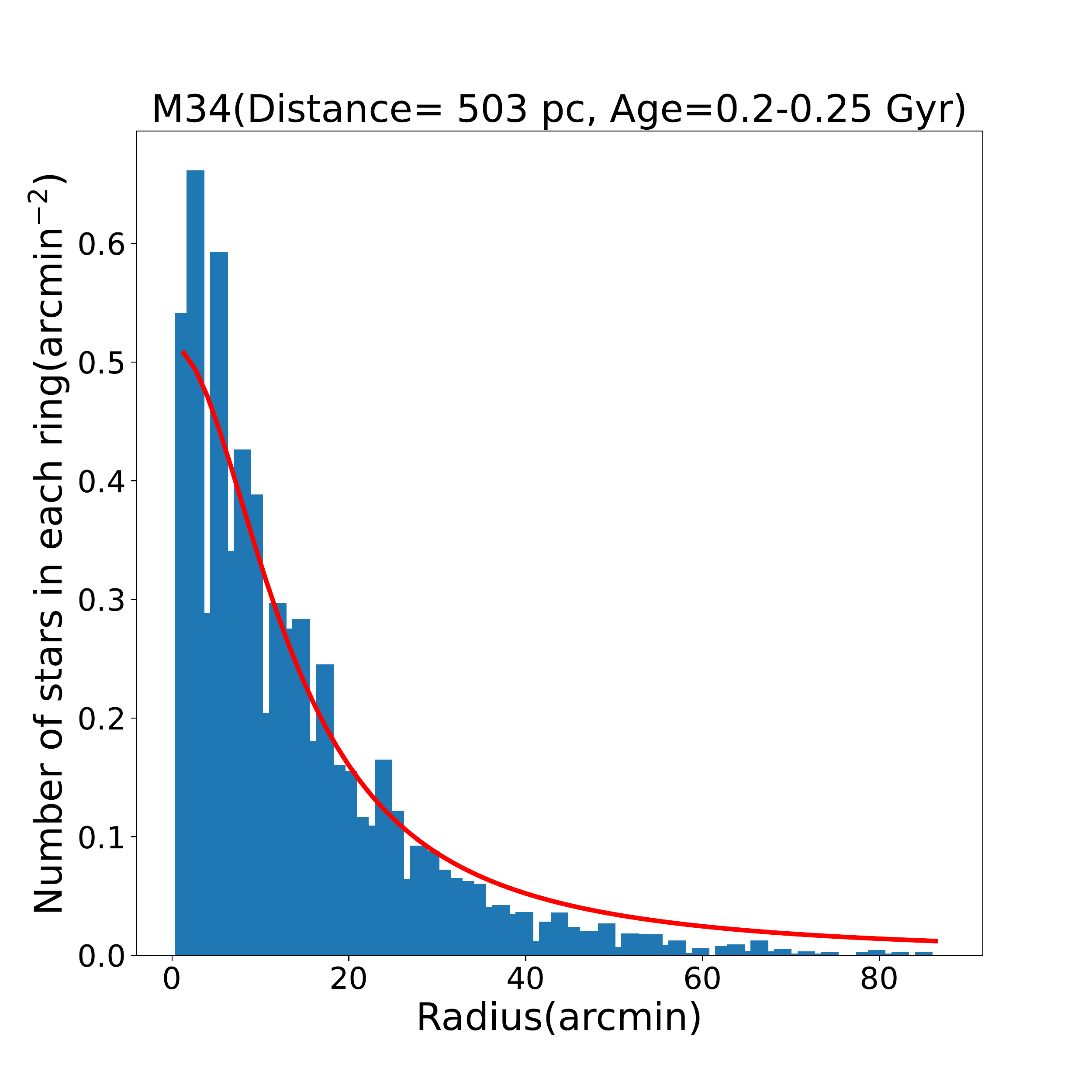}
        \end{subfigure}
        \begin{subfigure}{0.45\textwidth}
                \centering

                \includegraphics[width=\textwidth]{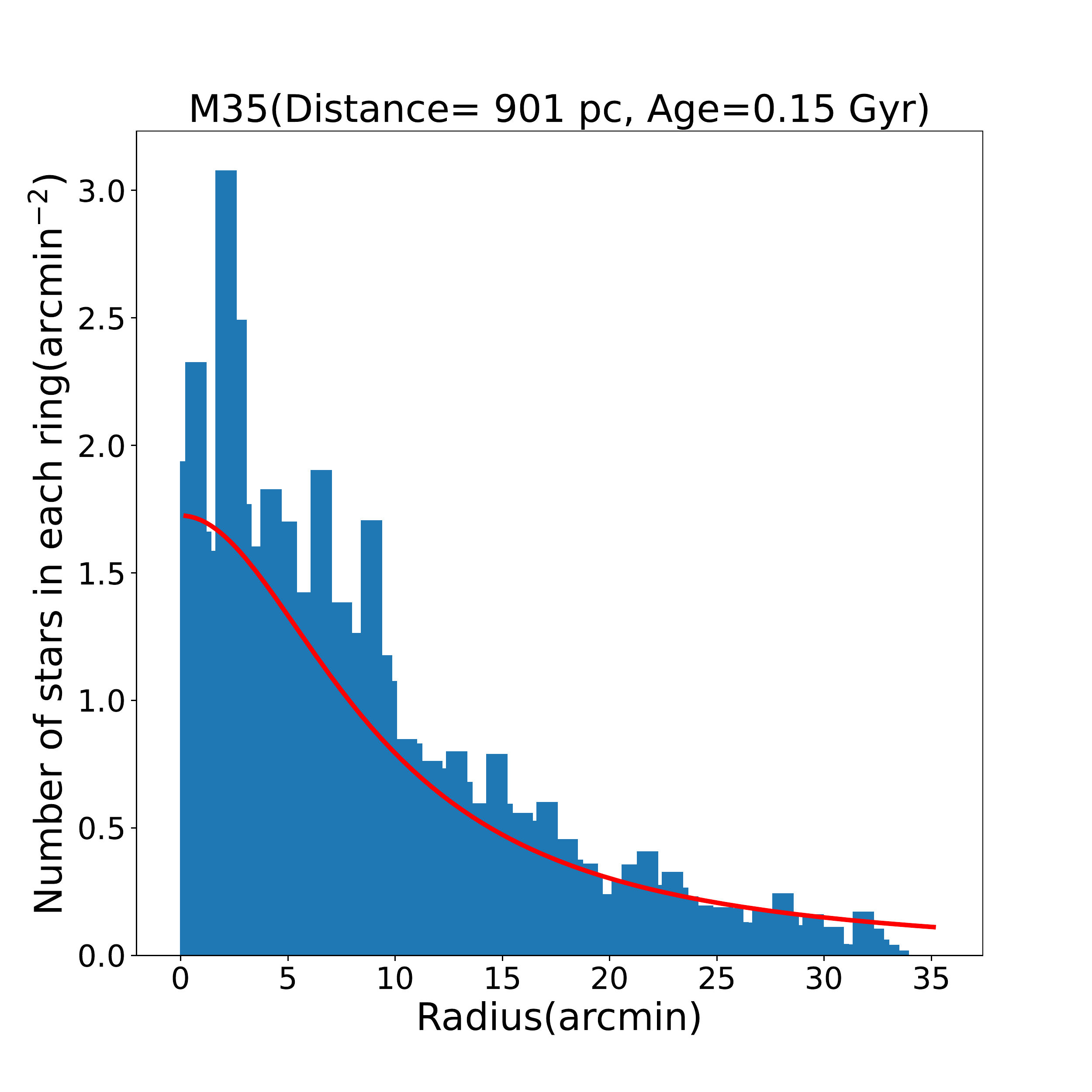}

        \end{subfigure}
        \begin{subfigure}{0.45\textwidth}
                \centering

                \includegraphics[width=\textwidth]{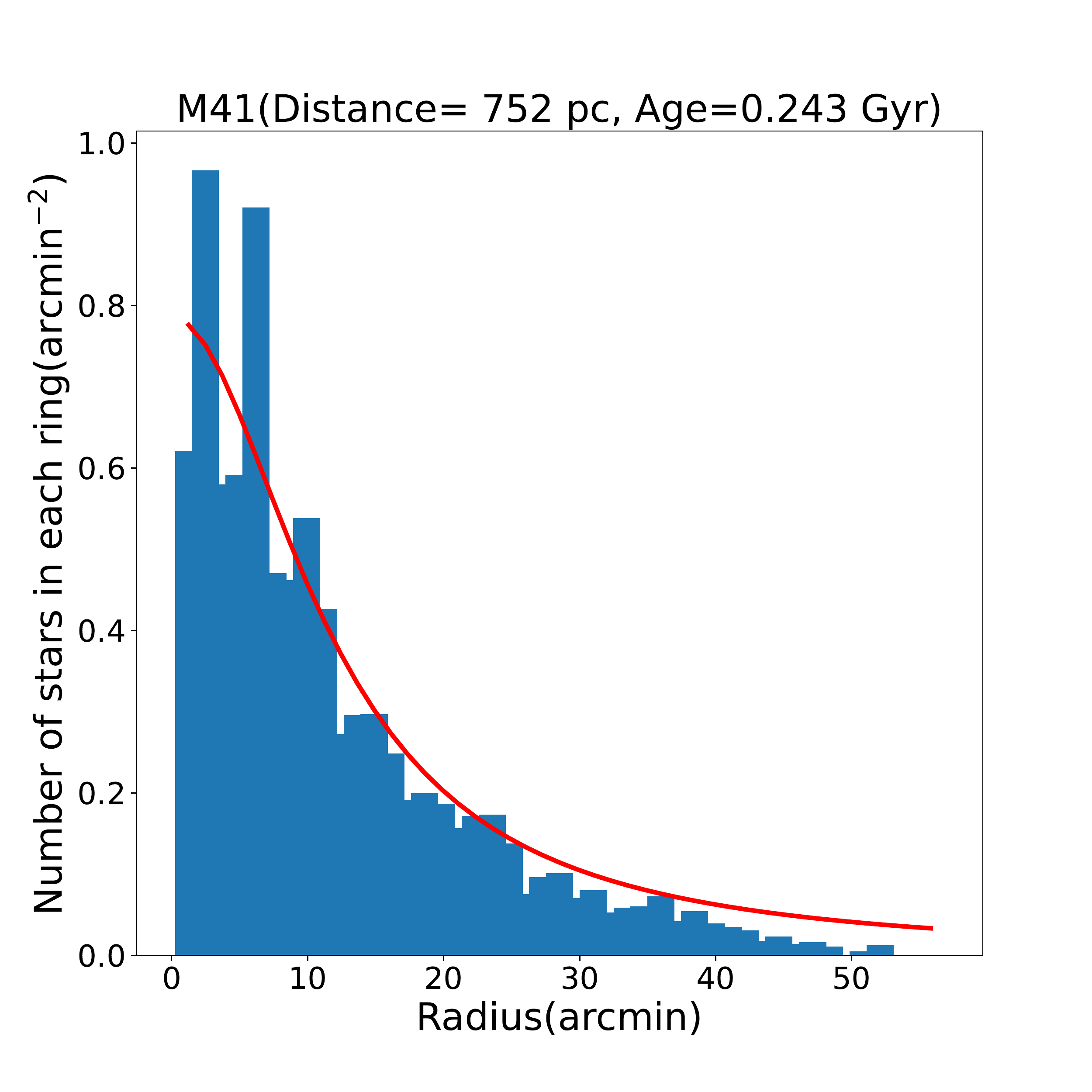}

        \end{subfigure}
        \begin{subfigure}{0.45\textwidth}
                \centering

                \includegraphics[width=\textwidth]{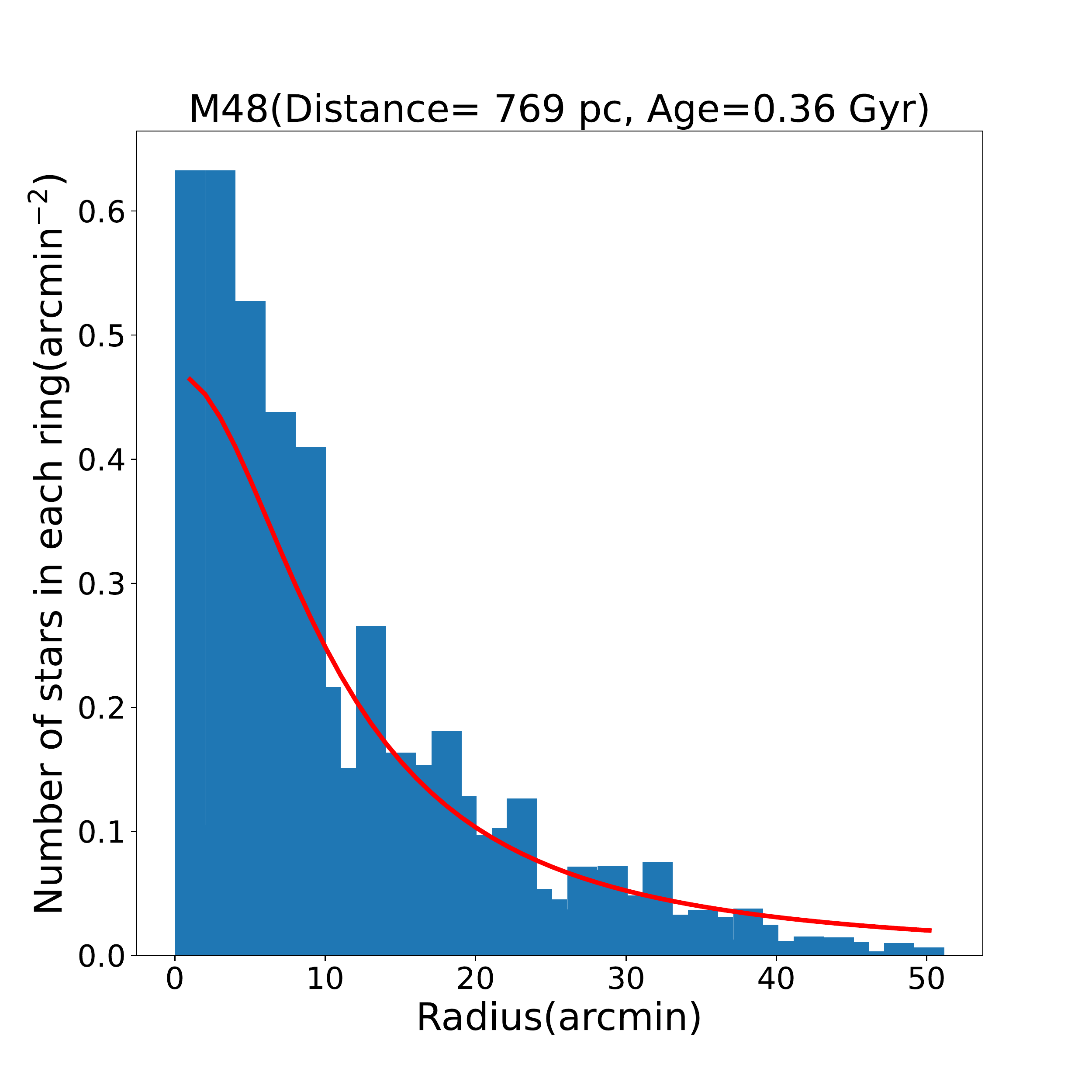}

        \end{subfigure}
  \caption{King profile fit on the distribution of stars in the cluster, the vertical axis is the density of stars number and the horizontal axis is radius.}
  \label{a.stars distribotion.fig}
\end{figure}
\begin{figure}
\ContinuedFloat*
  \centering
  \captionsetup[subfigure]{labelformat=empty}
        \begin{subfigure}{0.45\textwidth}
                \centering

                \includegraphics[width=\textwidth]{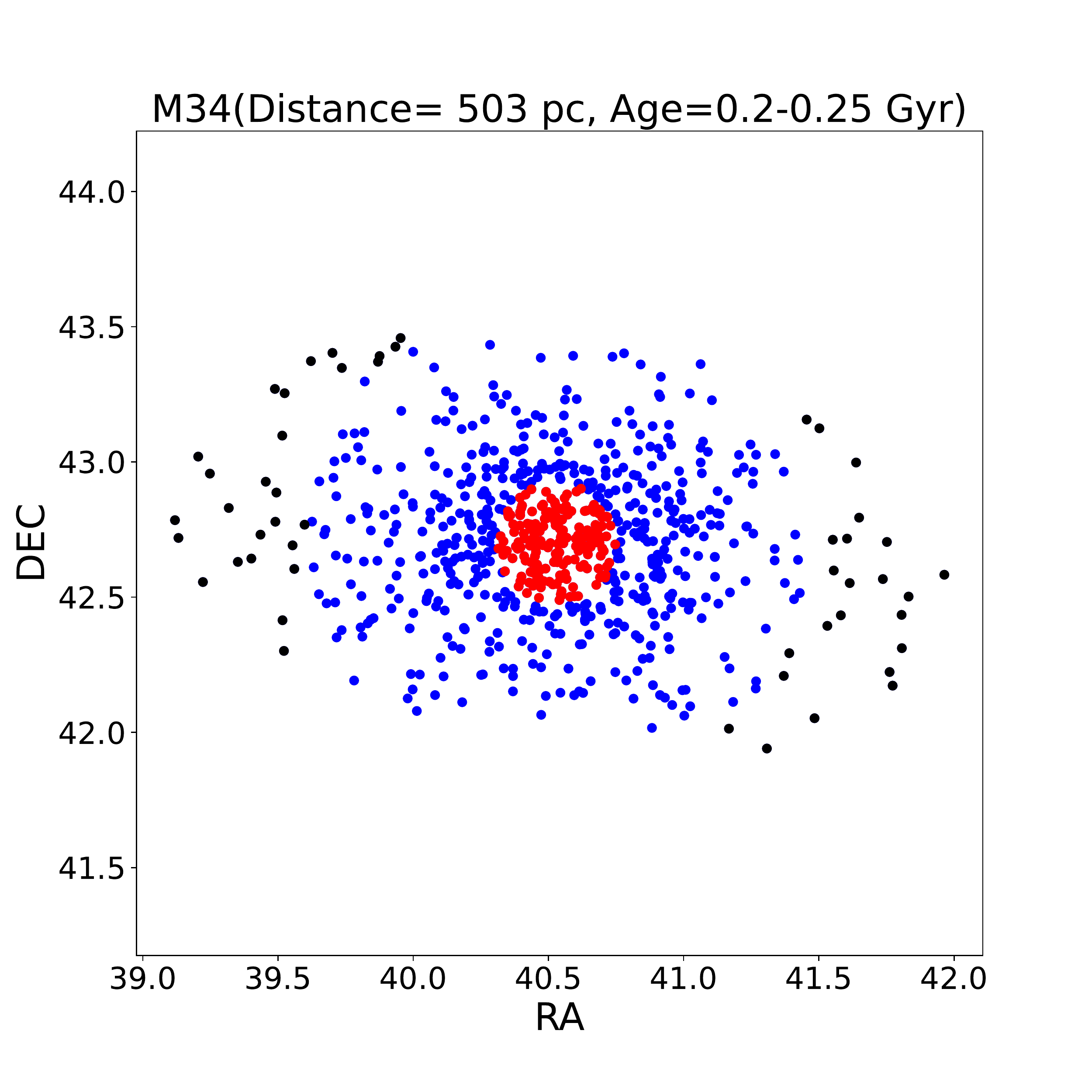}
        \end{subfigure}
        \begin{subfigure}{0.45\textwidth}
                \centering

                \includegraphics[width=\textwidth]{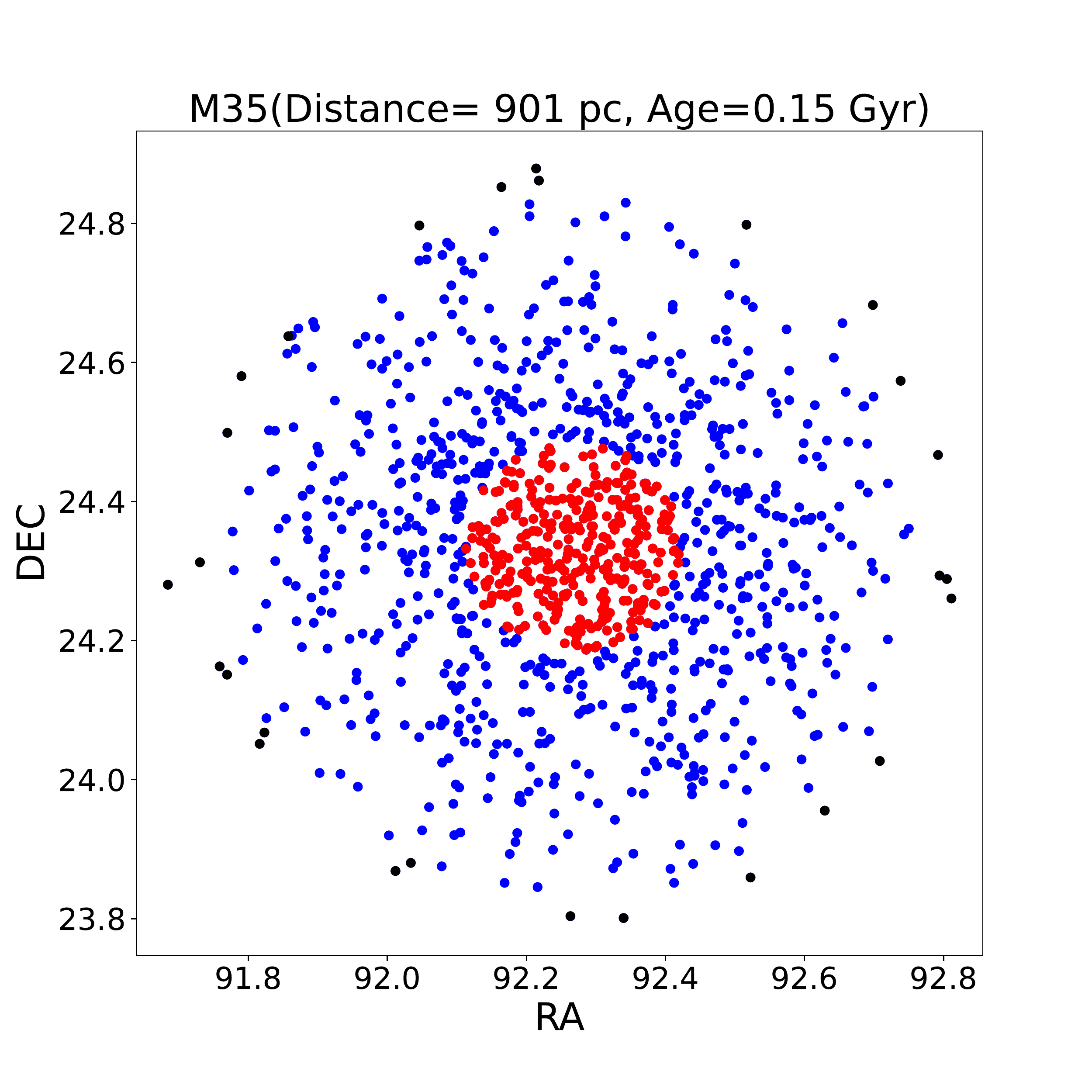}

        \end{subfigure}
        \begin{subfigure}{0.45\textwidth}
                \centering

                \includegraphics[width=\textwidth]{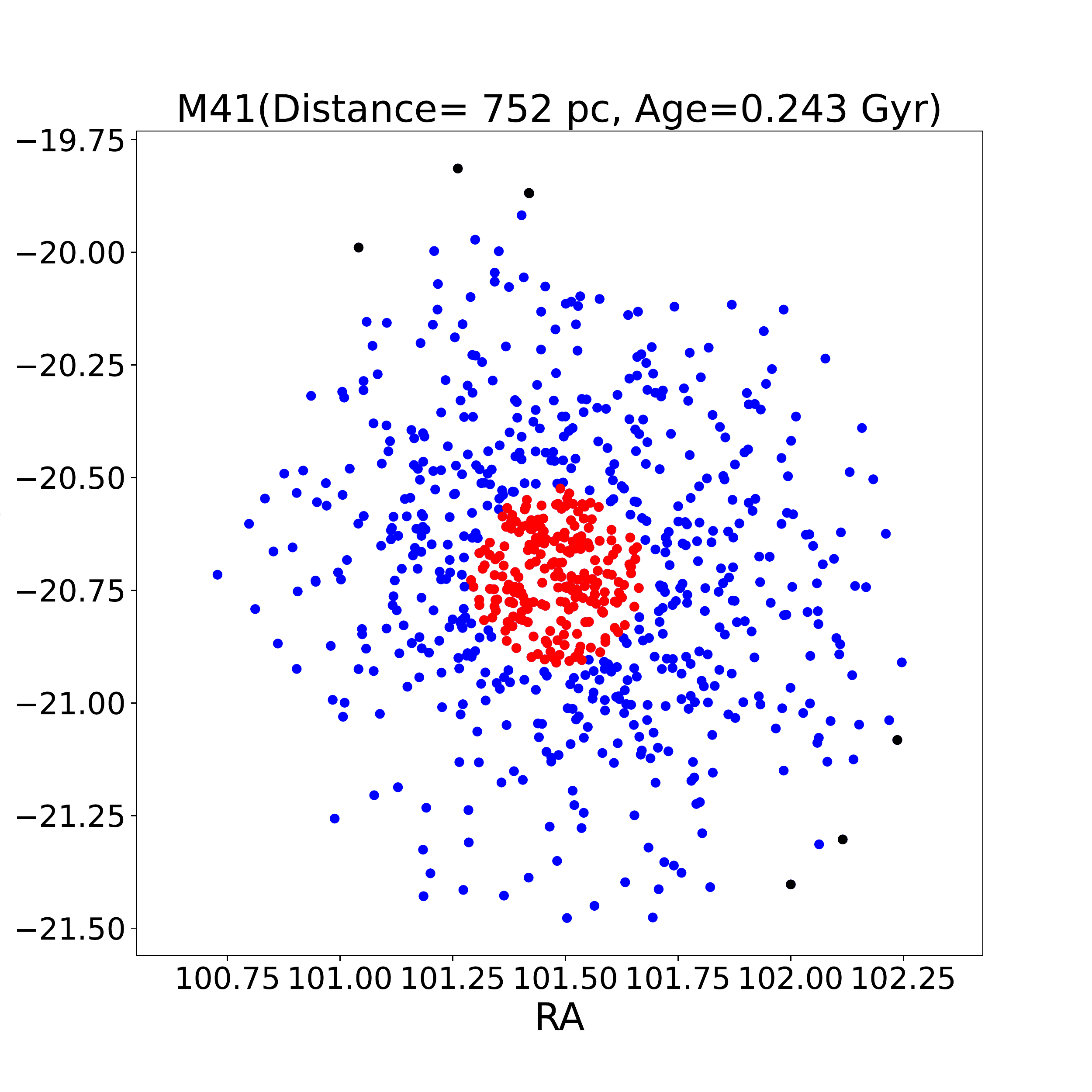}

        \end{subfigure}
        \begin{subfigure}{0.45\textwidth}
                \centering

                \includegraphics[width=\textwidth]{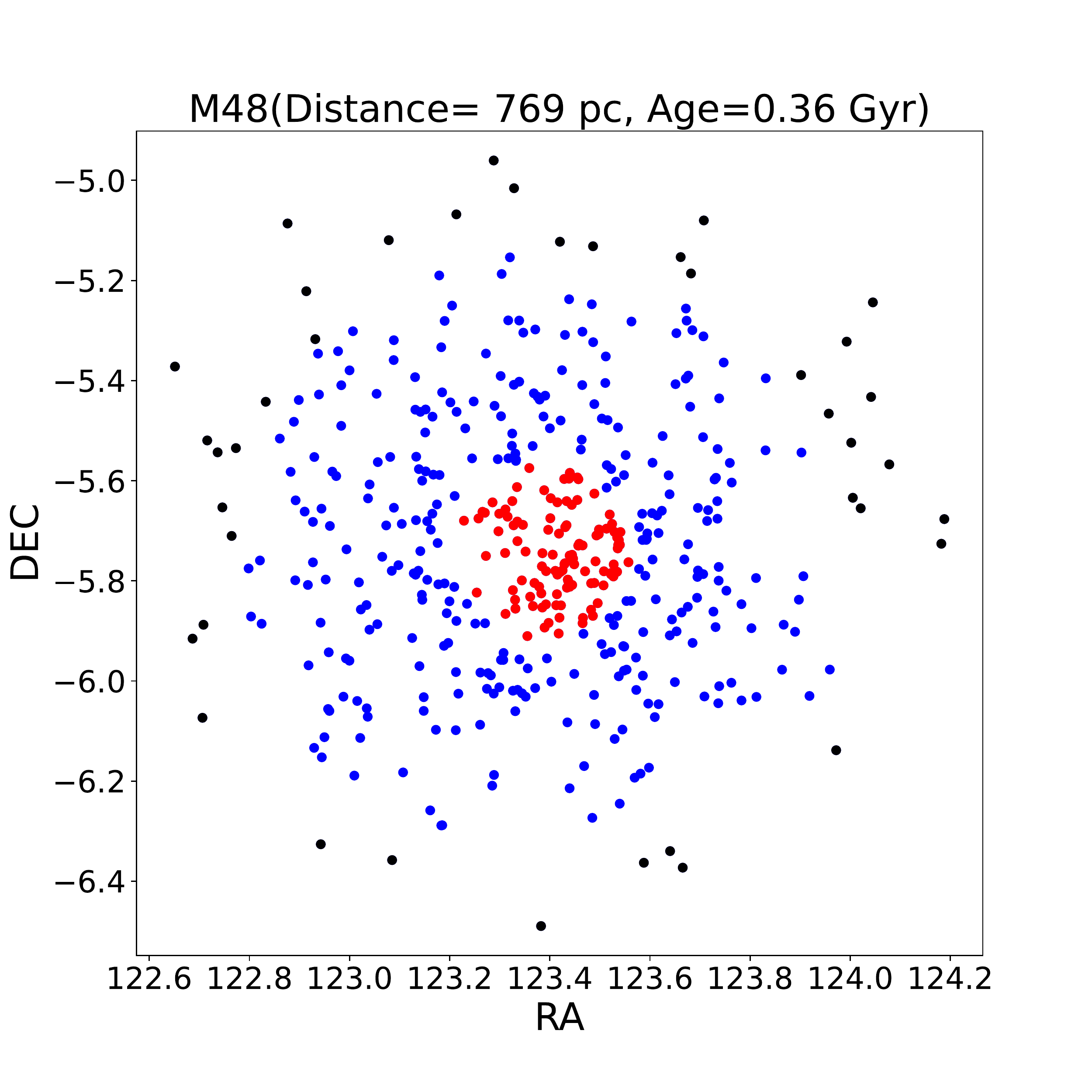}

        \end{subfigure}
  \caption{Position of cluster stars. Red dots show stars within the cluster core, and black dots show stars that are distributed outside the tidal radius.}
  \label{a.position from fields.fig}
\end{figure}
\begin{figure}[h!]
\ContinuedFloat*
  \centering
  \captionsetup[subfigure]{labelformat=empty}
        \begin{subfigure}{0.45\textwidth}
                \centering

                \includegraphics[width=\textwidth]{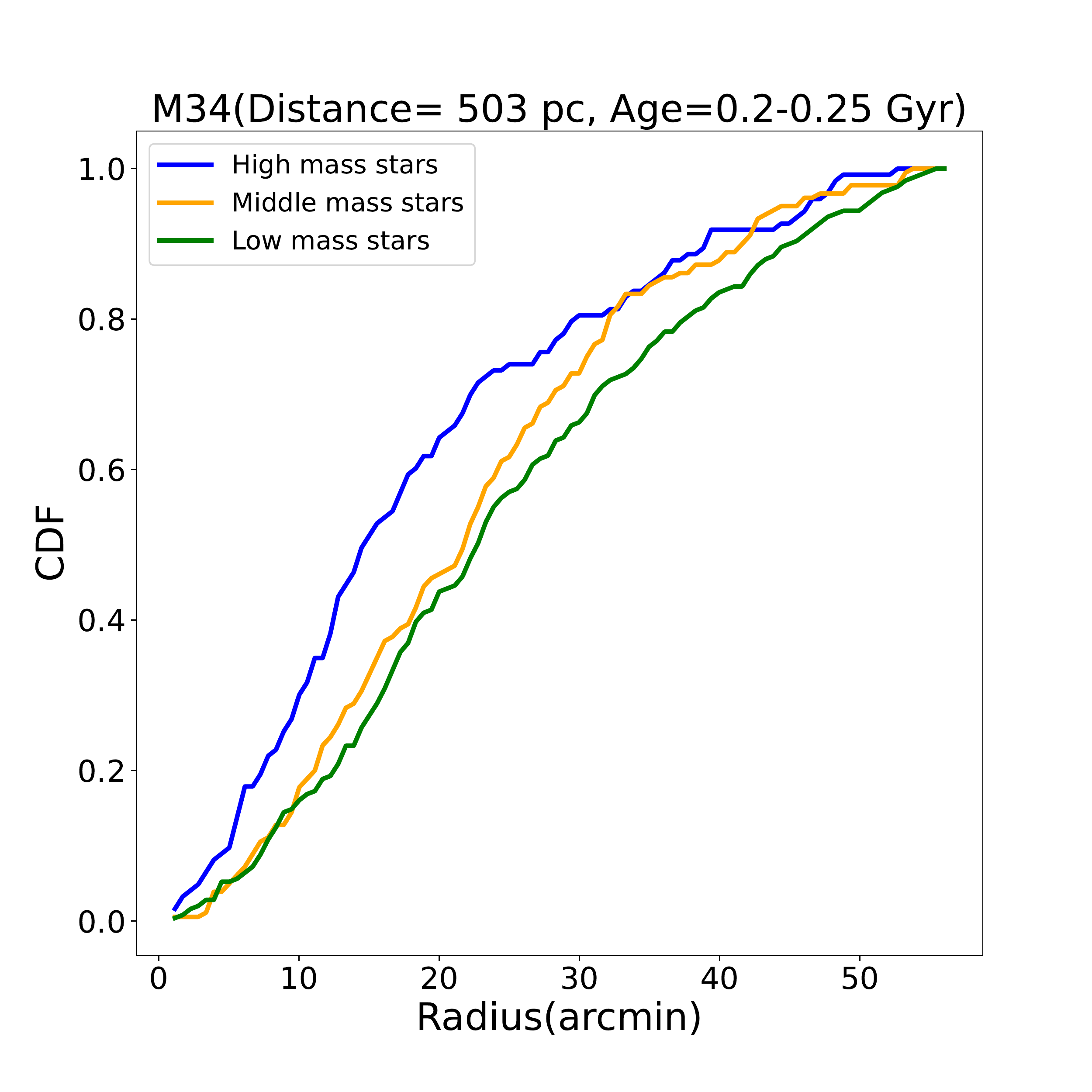}
        \end{subfigure}
        \begin{subfigure}{0.45\textwidth}
                \centering

                \includegraphics[width=\textwidth]{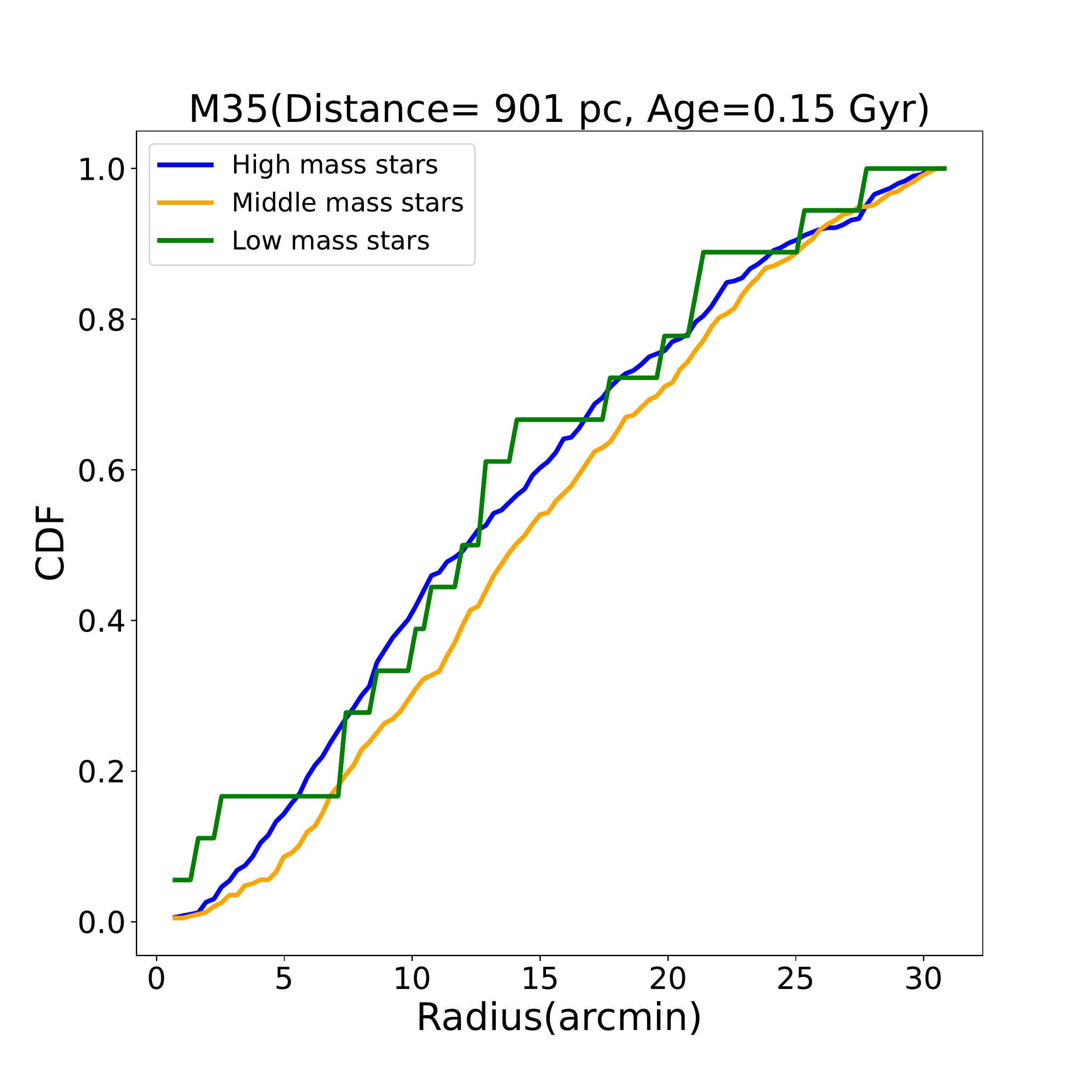}

        \end{subfigure}
        \begin{subfigure}{0.45\textwidth}
                \centering

                \includegraphics[width=\textwidth]{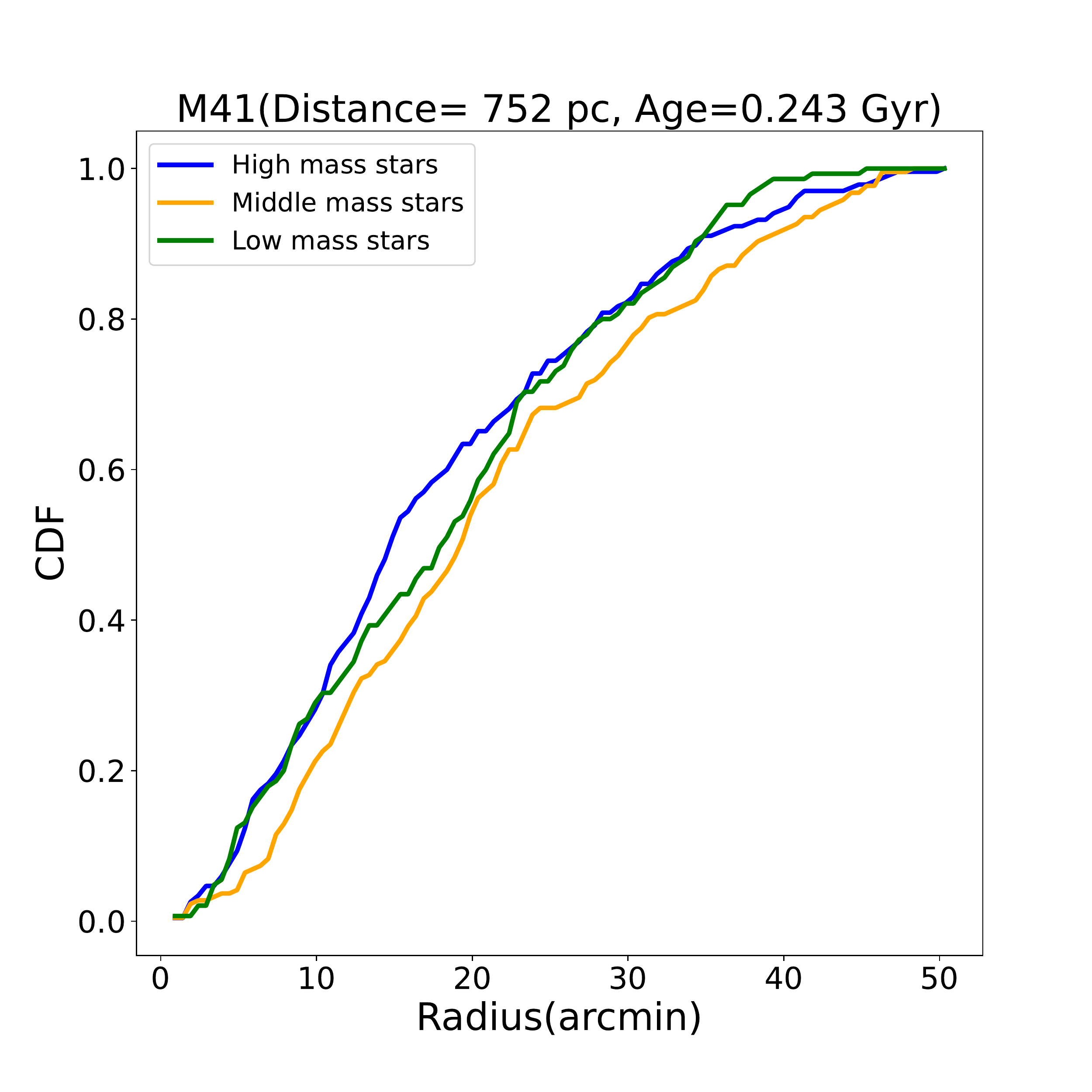}

        \end{subfigure}
        \begin{subfigure}{0.45\textwidth}
                \centering

                \includegraphics[width=\textwidth]{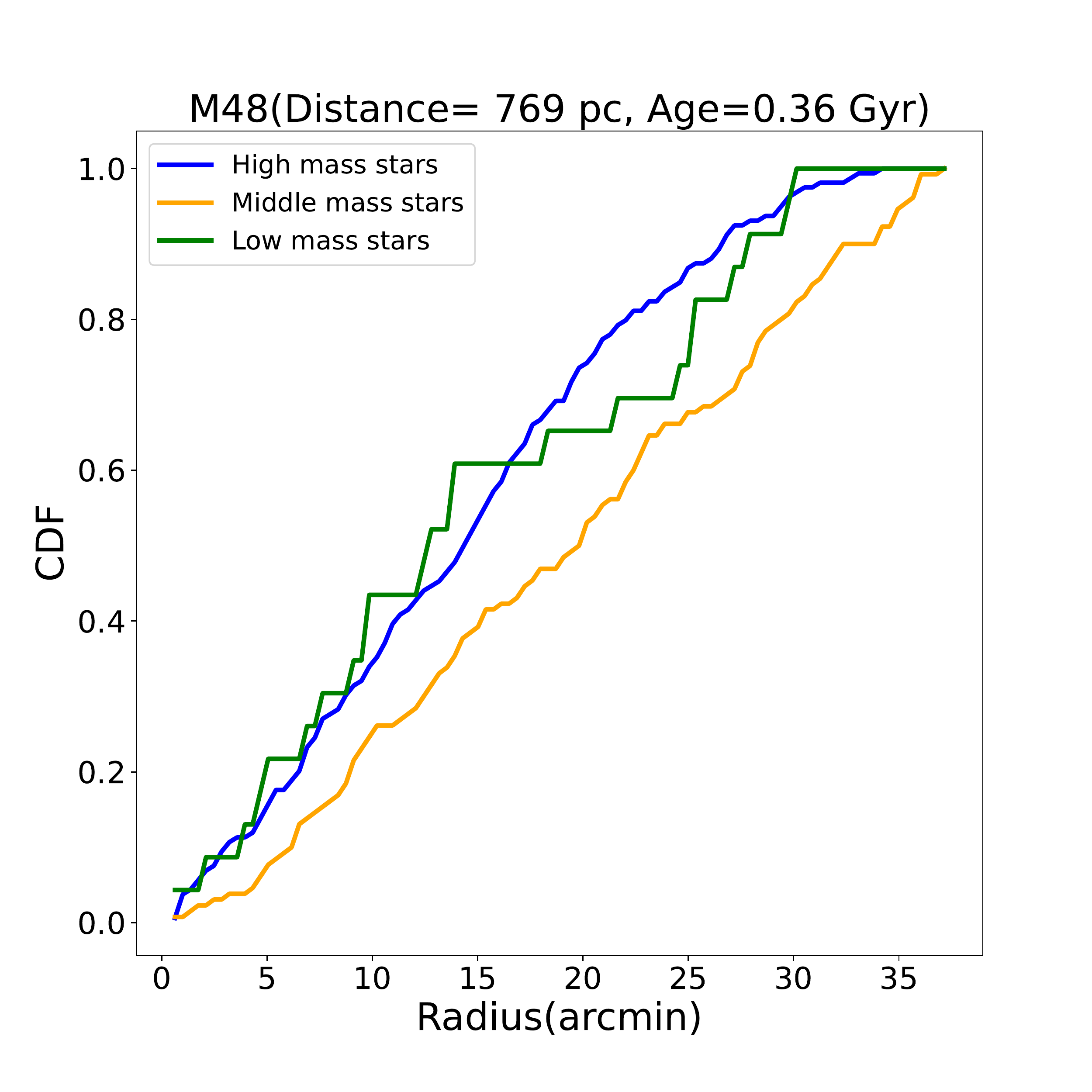}

        \end{subfigure}
  \caption{Radial cumulative distribution function, blue line shows massive stars, orange line shows middle mass stars and green line shows low mass stars.}
  \label{a.Mass segregation.fig}
\end{figure}

\end{document}